# Formation of dust grains in the winds of cool giants

von
Dipl. Phys. Carsten Dominik
aus Berlin

vom Fachbereich 4 (Physik)
der Technischen Universität Berlin
zur Erlangung des akademischen Grades

Doktor der Naturwissenschaften (Dr. rer. nat.)

genehmigte Dissertation

Berlin 1992
D 83

# Formation of dust grains in the winds of cool giants

von
Dipl. Phys. Carsten Dominik
aus Berlin

Promotionsausschuß:
Vorsitzender : Prof. Dr. G. von Oppen
Berichter :     Prof. Dr. E. Sedlmayr
                Prof. Dr. K. Beuermann

Tag der mündlichen Prüfung : 26. 2. 1992

*Silizium gibt's wie Sand am Meer ...*

(N. Bliefert)

# Zusammenfassung

Kleine Staubteilchen mit einem Durchmesser von in der Regel weniger als $1\mu m$ stellen eine nahezu allgegenwärtige Komponente des interstellaren Mediums dar. Obwohl diese Komponente nur einen Massenbruchteil von etwa 1% ausmacht, kommt ihr doch besondere Bedeutung zu. Zum einen bestehen ca. 98% der interstellaren Materie (ISM) aus Wasserstoff und Helium, so daß also die Hälfte des verbleibenden Materials (d. h. aller Elemente schwerer als Helium) in den Festkörperpartikeln gebunden ist. Zum anderen bestimmen die Festkörper wegen ihrer großen Wirkungsquerschnitte für die Wechselwirkung mit Licht wesentlich das Absorptions- und Emissionsverhalten der ISM (sofern sie Staubteilchen enthält). Darüber hinaus stellt die Oberfläche der Staubteilchen eine Komponente dar, die die chemische Entwicklung der ISM entscheidend beeinflussen kann: auf einer solchen Oberfläche können Reaktionen stattfinden, die in der Gasphase nicht möglich wären.

Auf diese Weise haben die Staubteilchen stets großen Einfluß auf die Prozesse in der ISM.

Nun zeigen jedoch einfache Abschätzungen, daß die Staubteilchen kaum in der ISM entstanden sein können. Die Temperaturen sind dort meist so niedrig und die Dichten so gering, daß die Energiebarriere auf dem Weg vom Atom zum festen Körper nicht effektiv überschritten werden kann. Eine Möglichkeit zur Darstellung einer solchen Abschätzung wird im Anhang dieser Arbeit vorgestellt und diskutiert.

Wo entstehen also die Staubkörner, die im interstellaren Medium so häufig sind? Sowohl theoretische Untersuchungen als auch Beobachtungen weisen deutlich darauf hin, daß die Bedingungen für die Entstehung von Staub in der Regel dort günstig sind, wo Sterne in Massenverlustereignissen einen Teil ihrer (zunächst staubfreien) Materie an die ISM zurückgeben. Solche Ereignisse sind zum Beispiel Sternwinde, Supernovae und Novae. Die abgegebene Materie ist im Vergleich zur ISM sehr dicht und durchläuft auf dem Weg weg vom Stern einen Kühlprozeß, der durch die relevanten Bereiche in der $p - T$ Ebene führen kann.

Auf diesem Weg müssen zunächst Keime entstehen, die stabil sind und nicht mehr verdampfen. Diese Keime bilden dann den Ausgangspunkt für das Wachstum von größeren Staubteilchen. In allen bisherigen Untersuchungen wurde im wesentlichen nur die Bildung von reinen Staubkörnern untersucht, die entweder ausschließlich aus den Atomen eines Elementes oder aber aus einem festen Monomer, einem Molekül gleicher Zusammensetzung wie der Festkörper, aufgebaut sind. Genau dies ist jedoch ein Fall, der in den meisten Situationen, in denen Staub gebildet wird, nicht zu erwarten ist. Vielmehr besteht ja die Gasphase aus einer Vielzahl von chemischen Elementen, die unterhalb bestimmter kritischer Temperaturen eher gemäß ihrer kinetis-chen Auftreffraten als nach bestimmten festen stoichiometrischen Verhältnissen auf die vorhandenen Keime aufwachsen werden.

In Kapitel 1 wird die sicherlich leistungsfähigste bekannte Methode zur Berechnung des Staubwachstums, das Momentenverfahren [Gail and Sedlmayr 1988], in einer Weise modifiziert, die die Behandlung eines solchen gemischten Wachstums ermöglicht, ohne die wichtigen Vorteile des Verfahrens (die hohe Effizienz) zu verlieren. Darüber hinaus werden Methoden angegeben, wie aus der Berechnung des Staubwachstums auch alle Informationen sowohl über die Größenverteilung der entstehenden Staubkörner als auch über ihre innere chemische Zusammensetzung erhalten werden können, ebenfalls ohne Einschränkung der Effizienz des Verfahrens. Die Anwendung dieses Verfahrens auf die Staubbildung in einer sauerstoffreichen Situation (zum Beispiel der Wind eines M-Riesen) in Kapitel 4 zeigt die Funktionsfähigkeit der Methode unter der Voraussetzung, daß die benötigten physikalischen Daten (hauptsächlich die Reaktionskonstanten zwischen den Staubkörnern und den aufwachsenden Molekülen) bekannt sind. Dies ist derzeit allerdings noch problematisch.

Ein bedeutsames Faktum in der Theorie der Staubentstehung ist, daß der Staub nicht nur passiv aus der Gasphase auskondensiert, sondern, insbesondere aufgrund seines großen Absorptionsquerschnitts, aktiv seine Entstehungsumgebung mitgestaltet. Zum Beispiel können die Staubkörner bei der Absorption von Sternenlicht den Impuls der Strahlung aufnehmen, ihn durch Stöße wieder an das umgebende Gas abgeben und auf diese Weise erhebliche hydrodynamische Effekte verursachen, die wiederum die Entstehungsbedingungen des Staubes bestimmen. Das vielleicht deutlichste Beispiel hierfür sind die Winde kühler Riesen und Überriesen, die möglicherweise allein durch den Strahlungsdruck auf Staub in Gang gesetzt oder durch diesen zumindest stark angefacht werden können. In langperiodisch veränderlichen Sternen können neuesten Resultaten zufolge sogar Stoßwellen durch den Strahlungsdruck auf Staub entstehen.

Aus diesem Grund wird in den Kapiteln 2 und 3 das Modell eines staubgetriebenen Windes behandelt. Eine ausführliche Diskussion der zwischen Gas und Staub auftretenden Driftbewegung zeigt, daß die Staubteilchen sich stets mit ihrer Gleichgewichtsdriftgeschwindigkeit bewegen und daher nach Abschluß der Beschleunigungsphase tatsächlich sämtlichen durch Absorption von Licht aufgenommenen Impuls an das Gas weitergeben.
Weiterhin wird eine reduzierte Gleichgewichtschemie vorgestellt, die mit sehr geringen Aufwand an Rechenzeit bereits sehr gute Ergebnisse für die Gleichgewichtsdichten der häufigsten Moleküle liefert.

Die Lösungen des den staubgetriebenen Wind beschreibenden Gleichungssystems werden in Kapitel 3 beschrieben und diskutiert. Dabei wird insbesondere anhand der Entwicklung der Größenverteilung der Staubkörner im Wind der Entstehungsprozeß der Staubkörner verfolgt und das sich dabei ergebende Potenzspektrum für die Teilchengrößen diskutiert. Es zeigt sich, daß die Ursache für die besondere Form des Spektrums gerade in der gegenseitigen Determination von hydrodynamischer Windstruktur einerseits und Staubbildung andererseits zu finden ist.

Mit staubgetriebenen Winden ist zumindest in dem hier diskutierten Fall von C-Sternen ein extrem hoher Massenverlust verbunden. Dieser Massenverlust impliziert Entwicklungszeitskalen, die vergleichbar mit oder sogar kürzer als die entsprechenden nuklearen Entwicklungszeiten der Sterne in diesen Phasen sein können. Daher sind Auswirkungen auf die Entwicklung der betroffenen Sterne zu erwarten. Aus diesem Grund wird (ebenfalls in Kapitel 3) die Abhängigkeit der Massenverlustrate von stellaren Parametern diskutiert. Dabei zeigt sich, daß diese zwar nicht sehr von den chemischen Häufigkeiten abhängt, dafür jedoch dramatisch mit zunehmender Leuchtkraft, abnehmender Photosphärentemperatur und abnehmender Sternmasse zunimmt. Selbst eine einfache Zeitentwicklung, bei der die Variationen von Leuchtkraft und Temperatur außer acht gelassen wurden, zeigt, daß nur die Abnahme der Sternmasse bereits zu einer Art Instabilität in der Entwicklung der Massenverlustrate führen kann. Daher bieten sich staubgetriebene Massenverluste als ein möglicher Mechanismus für den Übergang eines Sternes vom Roten Riesen zum Planetarischen Nebel an.

# Contents

# List of Figures

# List of Tables

# Introduction

About five billion years ago, our solar system was formed by a gravitational collapse of a dense region in the interstellar medium[1]. The region probably was part of a huge molecular cloud, one of the largest objects in the galaxy. The cloud was a rather cold ($T \simeq 50K$) and dense ($n \simeq 10^6 cm^{-3}$) accumulation of interstellar material, containing up to $10^6$ solar masses. Its chemical composition was what we are used to from our sun: 70% (by mass) hydrogen, 28% helium and 2% *heavy elements* (i.e. heavier than helium). However, this is almost the end of the common chemical properties. In the sun, all matter is present as atoms, ions and a few kinds of diatomic molecules. In the molecular cloud, hydrogen and helium were also present as atoms or $H_2$, but almost all the heavy elements were condensed to small solid particles, usually smaller than $1\mu m$. The only important exceptions were the very stable molecules $CO$ and $N_2$, and sulfur. Though these were relatively abundant molecules, still one half of the total mass of heavy elements was present in solid form.

The presence of these particles, today somewhat carelessly named *dust* by astronomers, had striking effects on the evolution of the collapsing region.

The collapse of the self-gravitating material set free gravitational energy which was partly converted into thermal energy. In order to proceed with the collapse it was important that this energy could be radiated away, but the cloud soon became opaque for the radiation emitted by excited atoms. However, it still was transparent for the far infrared radiation emitted by dust grains which received energy from the gas by collisions with atoms and molecules. So the collapse continued.

Most of the material fell down to the center forming the protostar, the progenitor of our sun. With time the protostar heated up. All solid particles inside the star evaporated. And the protostar began to glow.

Farther out the temperature was not high enough to evaporate the solid particles. Due to their excellent absorption capabilities all the light emitted from the star was absorbed and then reemitted at longer wavelengths. In this stage the system was an impressive infrared object if seen from outside.

The remaining fraction of the cloud formed a rotating disk around the core. Due to the friction between dust and gas, the motions of the dust perpendicular to the disk were broken and a very thin dust enriched disk evolved. In this disk gravitational instabilities (dust only!) or a slow growth by coagulation led to the formation of small bodies called planetesimals. Gravitation between these bodies further accelerated the process of formation of larger lumps ending in only few large planets and many smaller planetoids. The remaining material was blown out by the onseting strong wind of the protostar.

---

[1]The descriptions of the formation of the solar system and the planets are essentially taken from [Shu *et al.* 1987], [Wetherill 1980], [Pollack 1984], [Coradini *et al.* 1987], and [Weidenschilling *et al.* 1988]. It should be noted that, though state of the art, this picture is still changing.

On the planets conditions developed which are rather uncommon in the universe. Moderate temperatures and high densities provided an excellent basis for a complex chemistry. On the third innermost planet this processes culminated in the evolution of life.

The evolution of the whole system, especially the formation of the planets, was decisively influenced by the presence of solid particles in the presolar nebula. But why were these particles present in the molecular cloud? Where and under which condition were they formed. What are the reasons that the heavy elements were incorporated in the grains with an efficiency of 50%?

The first guess, that the grains originate from the interstellar medium itself, turns out to be wrong. Already simple estimates show that the timescale available in the interstellar medium is not sufficient in order to produce considerable amounts of dust[2]. Grains may well be processed in the ISM but the seed particles have to be formed elsewhere. It turns out that only a rather narrow region in the $p-T$ diagram allows the initial formation of solid particles and this part of the $p-T$ plain is passed mainly in the outflow of material from stars.

This point of view is consistent with many observations. In the environments of certain late type giant stars large amounts of dust can be seen. The dust shows itself by the typical infrared excess which is due to thermal emission of the dust. Since there is an outflow of material from the stars, the grains are obviously produced in the outflow: In the stars atmosphere the temperatures are to high for dust grains to exist. Thus they must be formed in the cooling outflow.

The formation of dust grains is first of all a chemical problem. Which are the possible condensates that will be formed under certain conditions? Which are the first chemical reactions to take place in the chain of reactions leading from atoms to macroscopic particle. Are the dust grains chemically processed by reactions with aggressive gas species? And so forth.

However, the formation of dust grains does not only happen passively in the outflows. The newly produced dust completely changes the physical properties of the whole system. Since the cross section of the particles for the interaction with photons is by orders of magnitude larger than that of the gas it dominates the absorption of the star's light and thereby plays an important role for the thermodynamic structure of the object. If the radiation field is not isotropic, the absorption of photons leads to a transfer of momentum from the radiation field to the dust particles which has hydrodynamic consequences.

---

[2] A new way to demonstrate this estimate is discussed in Appendix I

The appearance of the dust grains has striking effects on the chemistry. The growth of the grains leads to the consumption of certain molecules and elements from the gas. And the existence of a surface opens completely new channels for chemical reactions to take place.

The research concerned with the formation of solid particles in astrophysical situations consequently has to proceed on two different levels: A) The clarification of the basic microprocesses and B) the study of the complex interaction in a system consisting of gas, dust, and radiation. Only if both aspects of the problem are studied in equal detail, a realistic description can be obtained.

Therefore in this work both aspects are addressed.

An important problem which until now is not clearly understood is the formation of composite grains. In astrophysical situations (unlike laboratory experiments) the composition of a supersaturated vapor is not prepared in a way that only a specific compound will form. Instead, the object (say the late type giant producing dust in its envelope) determines the conditions for dust formation and one has to work out what kind of dust (and composition) actually are produced. In almost all relevant situations a large variety of chemical elements is present and can be incorporated in dust grains. Most probably a dirty kind of dust will form. These grains will neither have the composition of a typical laboratory compound nor will the composition be constant throughout the dust grain. Instead, kinetic factors determine the composition. Since the kinetic factors vary with the conditions in the vapor, the dust grains should grow in layers of more or less smoothly changing composition.

Until now all theoretical approaches to dust formation considered either the formation of pure grains (consisting only of one element) or the formation of composite grains of a fixed stoichiometric composition. In Chap. 1 following the derivation of the moment equations for homogeneous dust [Gail and Sedlmayr 1988] an attempt is made to reformulate this system in order to allow for variable and non-stoichiometric compositions. Furthermore the description of the process is extended by a formalism allowing for the determination of the grain size distribution function.

In Chap. 2 a model of the dust forming envelope of a red giant is outlined. These objects are particularly interesting for two reasons: 1. The bulk of dust in the galaxy originates from these objects. And 2. The winds of red giants are the prototype of a system with a complex interaction between dust formation and hydrodynamics. Chapter 2 summarizes the basic equations describing such a system. Thereby the interaction between the dust particles and the gas which is responsible of the momentum transfer from the dust to the gas is discussed in detail.

In Chap. 3 a solution of the system of Chap. 2 is discussed. Using the derived size distribution the evolution of the process of dust formation is followed in detail.
An important feature of the model is, that it allows for a self-consistent determination of the dust induced mass loss rate of the red giant. This is discussed in the second part of Chap. 3.

Chapter 4 returns to the *Ansatz* of Chap. 1 and presents a test-calculation for the formation of composite grains in the wind of an M-star. Thereby the changing composition of the condensing material and the internal structure of the produced dust grains are discussed.

# Chapter 1

# Theory of Dust Formation

## 1.1 Introduction

The formation of solid particles from the gas phase generally has to be treated as a chain of chemical reactions leading from a single atom via small and subsequently larger molecules to macroscopic solid particles. Therefore the most general description of the dust formation process is a chemical reaction network including all relevant reactions. This problem is, of course, intractable since especially at larger molecule sizes the number of different pathways increases dramatically. It is, therefore, necessary to deal with several approximations which lead to a reduction of the required effort but still describe the main properties of the process.

Though the formation of solids in principle is a continuous process from atoms to macroscopic particles, it usually falls into two parts where different types of approximations can be used.

1. **The Molecular Domain**
   The rate at which new solid particles are formed is laid down in this range. It is characterized by many competing reactions leading to the formation and dissociation of molecules. The properties of every molecule may be quite different and, for a correct description, the rate coefficients for every relevant reaction must be known. If the conditions are favorable there will be some reaction pathways leading to larger and larger molecules that are candidates for a subsequent growth to macroscopic particles. The rate of formation for these large molecules that serve as condensation nuclei will be called the *nucleation rate* $J_*$.

2. **The Macroscopic Particle Domain**
   The description of reactions between macroscopic dust particles and molecules from the gas phase is (to some extent) more simple than the treatment of molecule – molecule reactions since the properties of macroscopic particles become less dependent on the size of the particle. Very often the relaxation time towards internal equilibrium within the grain is much smaller than the time in which the conditions in the gas phase change considerably or in which composition and size of the grain are altered. Therefore the grain has an internal temperature governing the distribution of energy in the different degrees of freedom and the frequency of spontaneous processes like the emission of a photon or a molecule become independent of the properties of the environment. Furthermore, given a chemical composition of the grain, the properties of the surface do not vary with the grain size. Every unit area on the grain has some reactive sites where reactions with gas molecules can occur. The rates of most processes simply become proportional to the surface area of the grain. It is this simplification that allows the treatment of processes concerning dust by appropriate moment methods as they have been used in the literature (e.g. [Fix 1969, Draine and Salpeter 1977, Gail and Sedlmayr 1988]).

The process of grain formation can be seen as a diffusion process in the space of the grain size. Consequently, the classical equation describing this process is a diffusion equation, a partial differential equation second order in the cluster size and first order in time for the size distribution function (for a review of the classical nucleation theory see [Feder *et al.* 1966]). To describe dust formation by this equation (the so called *Zeldovich* equation), it is necessary to approximate the discrete size distribution function $f(N, t)$ which gives the number density of grains of size $N$ at time $t$ by a continuous function of the grain size.

The calculation of grain formation by this equation meets two problems:

1. The continuous approximation becomes critical at very small grain sizes.

2. The solution of the partial differential equation requires a large computing effort especially if $f(N, t)$ is nonvanishing over an extremely broad size interval. This case can occur in astrophysical situations as will be shown later in this thesis (see Sect. 3.1.3): In a dust driven wind the relevant size interval can span from 100 to $10^{18}$ monomers.

Due to this situation, other methods have to be developed in order to allow for realistic studies of the formation of solids in space. The problems with very small grain sizes can be avoided by a discrete theory in this range which will be discussed in some more detail in the next section. Approximate analytical solutions of the Zeldovich equation have been given by several authors (e.g. [Draine and Salpeter 1977, Yamamoto and Hasegawa 1977]). Both authors use Taylor expansions for some important quantities like the temperature as a function of time and the Gibbs energy of formation of small clusters as a function of their size. These approximations can be used in situations in which the external conditions change independently of the produced dust in the temperature range of the most effective nucleation. However, these approximations might lead to large errors in environments, where the dust itself controls the timescales of the changes of temperature and other hydrodynamic and chemical parameters. Furthermore, these methods fail under conditions where parts of the dust are destroyed. Such conditions can be found for example in variable stars.

A numerical treatment of dust formation was carried out by [Fix 1969] and later by [Draine and Salpeter 1977]. They used a kind of moment method where the evolution of the size distribution means (weighted by powers of the particle size) was followed by the solution of a coupled set of ordinary differential equations. The moment equations given by [Draine and Salpeter 1977] are formally the same as given later by [Gail and Sedlmayr 1988].

[Gail and Sedlmayr 1988] gave a complete discrete theory of the formation of homogeneous dust grains (i.e grains built up by a single monomer). They derived the master equation by balancing the gains and losses of every particle size $N$. This discrete theory gives the exact expressions for the nucleation rate as it has to be calculated for very small grain sizes. Neglecting the destructive processes (the evaporation of monomers from the grains by thermal evaporation or sputtering processes) they derived a set of moment equations from the master equation. These moment equations, now rigorously derived, allow for a detailed numerical treatment of dust formation with a minimum of computational effort.

It has been shown by [Gauger *et al.* 1990b] that thermal evaporation as a destructive process can quite easily be incorporated into the moment equations if the size distribution is known. Therefore dust formation in variable situations can be treated in the same very fast way.

One aim of this thesis is to further extend the applicability of the moment equations in order to allow for the condensation of *composite* grains. Composite dust grains are considered to be a mixture of different molecules and crystal structures as well as the so-called *core-mantle* grains where different materials condense in layers on the grains.

For this kind of dust grains the *monomer* no longer has a definite meaning and the moment method in its classical form cannot be used. It will be shown that a moment system can be obtained in a natural way if the size distribution moments are defined with respect to the *volume* of the grains rather than the *number of contained monomers*. The resulting equations have the same form as the equations given by [Gail and Sedlmayr 1988]. However, since the quantities have to be defined differently, different symbols will be used in order to avoid confusion.

It has already been mentioned that the process of dust formation usually falls into two regions; namely the region of small molecules and the region of macroscopic grains. Though the border between the two regions is naturally diffuse, one can define a limiting volume $V_\ell$. Above this critical volume the dust grains shall have macroscopic properties like an internal temperature, broad absorption and emission features (rather than absorption and emission lines) etc.. All particles smaller than this limiting volume will be called molecules and not grains. In the following section it will briefly be sketched how the nucleation rate can be calculated from the processes between the molecules. The derivation of the moment equations will be given in Sect. 1.3.

## 1.2 The Nucleation Rate

One of the most crucial problems in calculations of dust formation is the determination of the nucleation rate, because the nucleation rate is an extremely steep function of the conditions in the gas (especially the gas temperature). The usual *Ansatz* is to consider the most effective chain of reactions leading from atoms to macroscopic particles and to calculate the flux of clusters through the chain. The rate at which particles of a certain minimum size will be produced by the chain is the so called *nucleation rate* $\mathcal{J}_*$.

If the chemical pathway for the nucleation process is known, the nucleation rate can be directly calculated from the solution of the corresponding rate equations. This has been discussed in full generality by [Gail and Sedlmayr 1988].

The application of this theory depends on the data required for the molecules that are part of the reaction chain, and on the knowledge of the reaction rates. The evaluation of the relevant nucleation path ways is presently being worked on (e.g. [Goeres 1991, Köhler 1992]) but unfortunately has not been ready to be used here. Therefore, the classical approaches for the nucleation process have to be used here. This approach has been discussed in detail by [Feder *et al.* 1966].

Since the reaction coefficients of the molecules often vary by orders of magnitude there usually is one clearly defined slowest reaction in the chain which is called the *bottle neck* reaction. The cluster that is produced by this reaction is called the *critical cluster*. Since this formation reaction is the slowest in the chain the next growth step is much faster and, consequently, the concentration of the critical cluster is always very small.

If the chemical timescales in the system below the critical cluster are sufficiently small compared to the timescales in which the thermodynamic conditions in the gas change, one may assume that the chain molecules smaller than the critical cluster achieve their chemical equilibrium concentrations. The nucleation rate is then given essentially by the equilibrium value of the number density of the critical cluster multiplied by the net growth reaction rate for this cluster.

In order to calculate the nucleation rate this way it is necessary to know the equilibrium concentrations of the chain molecules, which can be obtained from the thermodynamic functions of the molecules. Since it is usually impossible to get all the data required, in classical nucleation theory these thermodynamic functions are often calculated from the surface energy of the bulk material. This concept is able to describe the evaporation energy of molecules from large clusters, but for very small clusters it may be far from the truth. Nevertheless, it must be used until detailed data are available. But it should always be kept in mind that the surface energy concept is used only because of the lack of correct data for the thermodynamical properties of the clusters. In this sense the results obtained have a preliminary character.

Adopting the surface energy concept, the Gibbs energy of the different clusters can be calculated analytically. Then the nucleation rate is also an analytical expression. The corresponding equations in the form they are used here are given by [Gail *et al.* 1984].

## 1.3 Growth and Destruction of Dust Grains

### 1.3.1 The Reaction Rates

The dust particles in astrophysical environments are subject to various processes that alter their size and composition. The following different processes will be allowed in the description of dust formation (see Fig. 1.1). $\mathcal{G}$ denotes a dust grain, $\mathcal{G}'$ is the same grain after a reaction. $\mathcal{M}_1$, $\mathcal{M}_2$ and $\mathcal{M}_3$ denote reaction partners (molecules, atoms, ions etc.. For simplicity, all these species will be called *molecules* from now.).

*I* Reaction of the dust grain with a molecule. The molecule sticks completely onto the grain. The chemical equation is

$$\mathcal{G} + \mathcal{M}_1 \rightarrow \mathcal{G}'. \tag{1.1}$$

Henceforth this will be referred to as a reaction of type *I*.

*I'* The reverse process is a spontaneous evaporation of the molecule $\mathcal{M}_1$ from the grain which will be referred to as reaction type *I'* and the chemical equation is

$$\mathcal{G}' \rightarrow \mathcal{G} + \mathcal{M}_1. \tag{1.2}$$

*II* Reaction of a grain with a molecule $\mathcal{M}_2$. One part of $\mathcal{M}_2$ sticks onto the grain while another part of it and, eventually, a molecule that belonged to the grain before the reaction, is returned to the gas phase. This process will be referred to as type *II*. Of course this comprises also chemical and mechanical sputtering processes where more material is returned into the gas phase than has got stuck onto the grain. It also includes the possibility that the molecule $\mathcal{M}_2$ is completely returned. Should $\mathcal{M}_2$ be an UV photon, photo sputtering effects can be treated as well. The chemical equation is

$$\mathcal{G} + \mathcal{M}_2 \rightarrow \mathcal{G}' + \mathcal{M}'_2 \ [+\mathcal{M}'_3 + \ldots]. \tag{1.3}$$

*II'* The reverse process in this case (type *II'*) is

$$\mathcal{G}' + \mathcal{M}'_2 \ [+\mathcal{M}'_3 + \ldots] \rightarrow \mathcal{G} + \mathcal{M}_2. \tag{1.4}$$

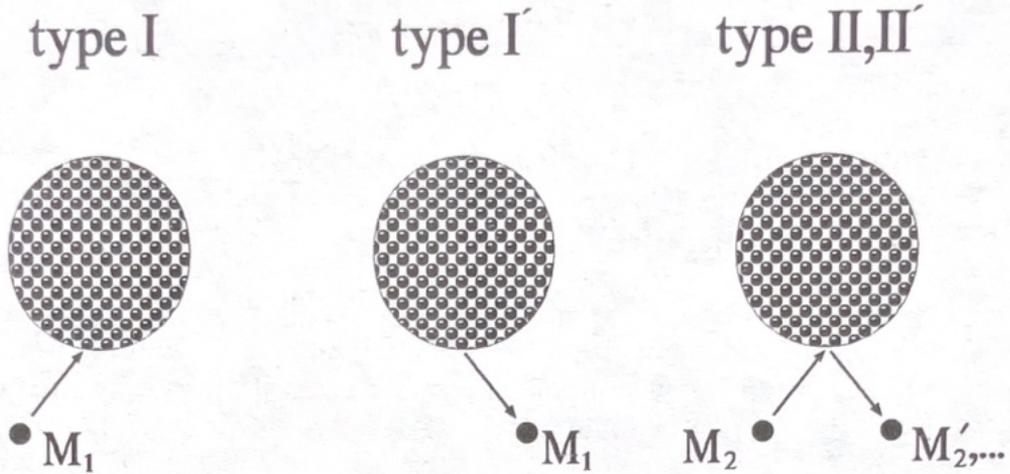

Figure 1.1:
Different types of chemical reactions between gas particles (electrons, atoms, molecules, ions and photons) and the dust grains.

---

Let $f(V,t)$ be the *size distribution of dust particles* such that $f(V,t)dV$ equals the number of particles per unit space volume in the grain volume interval between $V$ and $V+dV$. Let us consider a special reaction incorporating these grains and denote all quantities concerning this reaction by an index $i$. Then the frequency of reactions per unit volume are of the general form

$$R_i = f(V,t)dV \cdot \omega_i A(V) \tag{1.5}$$

where $A(V)$ is the surface of a grain with volume $V$ and $\omega_i$ denotes the *reaction frequency* per unit surface area.

Depending on the special type of the reaction, $\omega_i$ takes different forms:

In the case of reactions of type $I'$ (i.e spontaneous dissociation reactions) $\omega_i$ is a function of the dust properties alone and independent of the conditions within the gas phase.

In the case of reactions with other particles $\omega_i$ may be written as

$$\omega_i = n_i v_i \alpha_i \tag{1.6}$$

where $n_i$ is the number density of the molecule within the gas phase, and $v_i$ is the relative velocity between the grain and the molecule projected onto the normal vector of the grain surface and averaged both over all molecules having a normal component of the relative velocity pointing towards the grain and over the total grain surface. In other words, $v_i$ is defined in order to certify that $A(V)n_i v_i$ equals the total collision rate between a grain with the surface $A$ and a molecule with density $n_i$.[1] $\alpha_i$ is the number of reactions divided by the number of collisions and therefore is called the reaction probability or, in the case of reactions leading to a sticking of the molecule (type $I$), the sticking probability. $\alpha_i$ may depend on the properties of the molecule and the interaction energy of the collision as well as on the grain temperature and the chemical composition of the surface.

### 1.3.2 The Milne Relations

The reaction frequencies $\omega_i$ are not independent from each other. All processes are related to their inverse processes by time reversal. Assuming a state of chemical equilibrium between dust and gas, time reversal requires a detailed balancing between every process and its reverse process. This leads to a condition for the reaction rates which is called the *Milne relation*.

Defining a discrete function $r(i)$ by the requirement that the reaction with the index $r(i)$ is the reverse to the reaction $i$, the Milne relations may simply be written as

$$\overset{\circ}{\omega}_{r(i)} = \overset{\circ}{\omega}_i \qquad (1.7)$$

where $\circ$ denotes the frequencies in thermodynamical equilibrium. Since it is possible that dust and gas have different temperatures, it is important to define the temperature at the equilibrium state (c.f. [Gauger *et al.* 1990b]). From now all equilibrium quantities will refer to the dust temperature denoted by $T_d$.

The Milne relations can be used in order to express one half of the reaction frequencies by the other half and some thermodynamical quantities. To proceed it is advantageous to distinguish forward and backward reactions. From a pair of reverse reactions the *forward reaction* will be defined to be the one leading to an *increase* of the volume of the dust grain. These reactions will be labeled by an index $\iota$ (different from $i$) in order to indicate that the sum over $\iota$ adds only half as many terms as the sum over $i$. The corresponding *backward reaction* is of course labeled by the index $r(\iota)$. All reactions leading to an increase of the volume of the grain are connected with a collision of the grain with its reaction partner and thus are of the form (1.6). Therefore $\overset{\circ}{\omega}_\iota$ may be specified to

$$\overset{\circ}{\omega}_\iota = \overset{\circ}{n}_\iota \overset{\circ}{v}_\iota \overset{\circ}{\alpha}_\iota . \qquad (1.8)$$

---

[1] For a spherical dust grain at rest in a gas with kinetic temperature T we simply have $v_i^2 = kT/(2\pi m_i)$ where $k$ is Boltzmann's constant and $m_i$ is the mass of the molecules. For a grain moving at a drift velocity $v_D$ much larger than the mean thermal velocity ($v_{th,i}^2 = 2kT/m_i$) of the molecule $v_i$ is given by $v_i^2 = v_D/4$. For more details see [Schaaf 1963] and Sect. 2.3.

Since the reactions may occur in both directions one may define a net reaction frequency by

$$\tilde{\omega}_\iota = \omega_\iota - \omega_{r(\iota)} = \omega_\iota(1 - S_\iota^{-1}). \tag{1.9}$$

In Eq. (1.9), the *generalized supersaturation ratio* of the reaction $\iota$ has been introduced. It is defined by

$$S_\iota = \frac{\omega_\iota}{\omega_{r(\iota)}}. \tag{1.10}$$

The generalized supersaturation ratio shows the direction of the net effect of a forward-backward pair of reactions. If $S_\iota$ is larger than unity, the reaction pair net leads to an increase in volume of the grain. If $S_\iota$ is less than unity, the net effect points to the destruction of dust grains. And if $S_\iota$ equals unity, the reaction $\iota$ is completely balanced by its inverse reaction.

Since the quotient $\overset{\circ}{\omega}_\iota / \overset{\circ}{\omega}_{r(\iota)}$ always equals unity one may multiply $S_\iota$ by this quotient:

$$S_\iota = \frac{\omega_\iota}{\overset{\circ}{\omega}_\iota} \frac{\overset{\circ}{\omega}_{r(\iota)}}{\omega_{r(\iota)}}. \tag{1.11}$$

In the case of reactions of type $I$, the backward reaction is a spontaneous evaporation process of a molecule from the dust grain. Then the backward reaction is independent of the conditions in the gas and always equals the equilibrium rate:

$$\omega_{r(\iota)} = \overset{\circ}{\omega}_{r(\iota)} \qquad \text{type } I, \tag{1.12}$$

Depending on the reaction type, the generalized supersaturation ratio therefore may be written as

$$\begin{aligned} S_\iota &= \frac{n_\iota v_\iota \alpha_\iota}{\overset{\circ}{n}_\iota \overset{\circ}{v}_\iota \overset{\circ}{\alpha}_\iota} & \text{type } I \\ S_\iota &= \frac{n_\iota\, \overset{\circ}{n}_{r(\iota)}\, v_\iota\, \overset{\circ}{v}_{r(\iota)}\, \alpha_\iota\, \overset{\circ}{\alpha}_{r(\iota)}}{\overset{\circ}{n}_\iota\, n_{r(\iota)}\, \overset{\circ}{v}_\iota\, v_{r(\iota)}\, \overset{\circ}{\alpha}_\iota\, \alpha_{r(\iota)}} & \text{type } II. \end{aligned} \tag{1.13}$$

In order to determine both forward and backward rates it suffices to know the forward rate and the equilibrium values of $n$, $v$, and $\alpha$.

Since in thermodynamic equilibrium the velocities are always given by a Maxwell distribution, the relative velocities (in the case of a spherical dust grain) are given by

$$\overset{\circ}{v}_i = \sqrt{\frac{kT_d}{2\pi m_i}}. \tag{1.14}$$

The terms in Eq. (1.13) including $\alpha$ and $\overset{\circ}{\alpha}$ describe the deviations of the population of state from the TE distribution in the cluster and in the molecules. Large deviations from unity have only to be expected if the considered reactions require activation energies (for a discussion see [Gail and Sedlmayr 1988]).

The remaining problem is the determination of the equilibrium densities of the molecules involved in the gas phase. At least in the case of compounds with a definite stoichiometric composition these densities may be calculated from the following consideration:

Consider a solid particle consisting of a number $J$ of chemical elements denoted by $E^{(j)}$. Its composition is given by the notation

$$E^{(1)}_{s_1} E^{(2)}_{s_2} \ldots E^{(J)}_{s_J}. \tag{1.15}$$

The $s_j$, $j = 1, \ldots, J$ are the stoichiometric coefficients of the solid.

In an experiment one may put this solid into a vacuum box and wait until equilibrium between the solid and its vapor has been established. From thermodynamics one knows that the equilibrium between atoms in the gas and the solid may be expressed by

$$\prod_{j=1}^{J} \overset{\circ}{p}{}_{E^{(j)}}^{s_j} = \exp\left\{ \frac{\Delta G_s - \sum_{j=1}^{J} s_j \Delta G(E^{(j)})}{kT} \right\} \tag{1.16}$$

where $\Delta G_s$ is the Gibbs energy of the solid state per monomer and $\Delta G(E^{(j)})$ is the Gibbs energy for the atoms in the gas phase, both Gibbs energies taken relative to their reference states. $\overset{\circ}{p}_{E^{(j)}}$, $j = 1, \ldots, J$ are the equilibrium pressures of the atoms in the gas phase.

Furthermore there holds chemical equilibrium in the vapor. Since the vapor is the product of the evaporation process of the solid, the element abundances in the vapor are given by the abundance ratios in the solid.

To write that as a formula let us define the *total density* $n_{<E^{(j)}>}$ of an element $E^{(j)}$ by

$$n_{<E^{(j)}>} = \sum_{k=1}^{K_j} \nu_{j,k} n_{X(j,k) E^{(j)}_{\nu_{j,k}}} \tag{1.17}$$

where the sum runs over all molecules containing the element $E^{(j)}$. $\nu_{j,k}$ is the stoichiometric coefficient of the element $E^{(j)}$ in the molecule $X(j,k) E^{(j)}_{\nu_{j,k}}$. $X(j,k)$ denotes the rest of the molecule containing all elements different from $E^{(j)}$.

Let the *total partial pressure* of the element $E^{(j)}$ be defined by

$$p_{<E^{(j)}>} = n_{<E^{(j)}>} kT. \tag{1.18}$$

The chemical abundance relation of the vapor may then be written as

$$s_1 n_{<E^{(1)}>} = s_2 n_{<E^{(2)}>} = \ldots = s_J n_{<E^{(J)}>}. \tag{1.19}$$

If this equation is multiplied by $kT$, the same expression holds for the total partial pressures $p_{E^{(j)}}$.

Together with the equations for chemical equilibrium (these equations are described in detail in Sect. 2.5) between the atoms and molecules consisting of the $J$ elements, Eqs. (1.16) and (1.19) form a closed system of $2J$ equations for the $2J$ unknowns: $p_{E^{(1)}}, \ldots, p_{E^{(J)}}, \overset{\circ}{p}_{<E^{(1)}>}, \ldots, \overset{\circ}{p}_{<E^{(J)}>}$. The solution of the system provides the equilibrium densities in the gas of all elements contained in the solid and, furthermore, of all molecules consisting of these elements only.

Sometimes the reactions leading to the growth of grains are due to molecules that contain an element that will not be incorporated in the solid. In astrophysical environments, especially the most abundant element, hydrogen, forms such molecules (for example in carbon stars the most important molecule for grain growth is $C_2H_2$). For these molecules, the reaction coefficients must be known explicitly for both forward and backward reactions.

## 1.4  The Moment Equations

### 1.4.1  Definition of the Moments of the Distribution Function

The moments of the size distribution function $f(V, t)$ of dust grains are defined by

$$\mathcal{L}_j(t) = \int_{V_\ell}^{\infty} f(V, t) V^{\frac{j}{d}} dV \tag{1.20}$$

where $V_\ell$ is the lower boundary of the region where the particles are considered to be macroscopic dust particles (usually about 100 atoms). The moments have units $[\text{cm}^j \text{cm}^{-3}]$ and are closely related to the averages of the powers of the particle radius with respect to the size distribution. For spherical grains ($d = 3$) one has the (see [Gail *et al.* 1984, Gail and Sedlmayr 1988]):

total number of grains per cm$^3$

$$n_d = \mathcal{L}_0 \qquad (1.21)$$

mean grain radius

$$<a> = \sqrt[3]{\frac{4\pi}{3} \cdot \frac{\mathcal{L}_1}{\mathcal{L}_0}} \qquad (1.22)$$

mean grain surface

$$<A> = \sqrt[3]{\frac{1}{36\pi} \cdot \frac{\mathcal{L}_2}{\mathcal{L}_0}} \qquad (1.23)$$

mean grain volume

$$<V> = \frac{\mathcal{L}_3}{\mathcal{L}_0}. \qquad (1.24)$$

### 1.4.2 Derivation of the Moment Equations

In this section the equations describing the growth and destruction of dust grains will be derived. Like in the previous section all possible reactions will be labeled by an index $i$. Every reaction is assigned by a *change in volume* of the dust grain, denoted by $V_i$. In the case of three dimensional dust grains, $V_i$ is the *volume* that is occupied by the involved molecule in the actual crystal structure at the grain surface. In the case of two dimensional grains, $V_i$ denotes the *area* occupied by the involved molecule in the grain. If two or more molecules are involved in the process, $V_i$ denotes the net change in volume (area) due to reaction $i$.

Depending on the reaction, $V_i$ may be positive (if the size of the dust grain is increasing due to reaction $i$) or negative (if the size of the grain is decreasing). Furthermore $V_i$ may depend on the composition of the actual grain surface and the temperature of gas and dust, since the special structure which is formed in the solid particle may well depend on these conditions. It will be assumed that all grains of the same size have identical properties concerning all relevant parameters like grain temperature and chemical composition of the surface.

The number density of grains in the volume interval between $V$ and $V + dV$ is

- increased due to reactions of dust particles with a volume in the range between $V - V_i$ and $V - V_i + dV$ where the volume is changed by $V_i$,

- decreased due to reactions of dust particles with a volume in the range between $V$ and $V + dV$ where the volume is changed by $V_i$,

and the total change in a co-moving frame is given by

$$\frac{d}{dt}f(V,t)dV = \sum_{i=1}^{I} f(V-V_i,t)dV\, A(V-V_i)\omega_i$$
$$- \sum_{i=1}^{I} f(V,t)dV\, A(V)\omega_i \qquad (1.25)$$

where it has been assumed additionally that the reaction frequencies are the same for grains in the volume range $V \pm \max_{i=1,\ldots,I}|V_i|$. The volume interval $dV$ is not changed due to this translation in the volume space and, therefore, can be omitted in the following.

To proceed further the reactive surface of the grains has to be expressed as a function of the volume. Here the possibility of fractal grain surfaces (e.g. [Wright 1989]) is neglected since this complication cannot be treated in the frame of this description. Then the surface of the grain is always proportional to $V^{(d-1)/d}$ where $d$ denotes the dimension of the grain. The surface may then be written as

$$A(V) = \xi_A^{(d)} V^{\frac{d-1}{d}} \qquad (1.26)$$

where $\xi_A^{(d)}$ is a constant depending on the geometry of the grains. For example in the case of spherical grains (i.e. spheres and circles for $d=3$ and $d=2$, respectively)

$$\xi_A^{(d)} = \begin{cases} \sqrt{4\pi} \cdot 2a_{\text{reactive}} & d=2 \\ \sqrt[3]{36\pi} & d=3 \end{cases} \qquad (1.27)$$

where $a_{\text{reactive}}$ is the width of the reactive ring at the edge of the circular grain. For later use the quantity $\xi_a^{(d)}$ is also introduced which allows for the calculation of the particle radius $a$ by

$$a(V) = \xi_a^{(d)} V^{\frac{1}{d}}. \qquad (1.28)$$

For spherical particles $\xi_a^{(d)}$ is given by

$$\xi_a^{(d)} = \begin{cases} \sqrt{1/\pi} & d=2 \\ \sqrt[3]{3/(4\pi)} & d=3 \end{cases}. \qquad (1.29)$$

Inserting Eq. (1.26) into Eq. (1.25) one finds

$$\frac{d}{dt}f(V,t)dV = \sum_{i=1}^{I} \omega_i \xi_A^{(d)} \left\{ (V-V_i)^{\frac{d-1}{d}} f(V-V_i,t) - V^{\frac{d-1}{d}} f(V,t) \right\}. \qquad (1.30)$$

It should be emphasized that until here the reaction frequencies $\omega_i$ may depend on the size of the particles. It has only been assumed that particles of a given size should have the same chemical composition and temperature because they have been processed in the same way since their nuclei had been formed. If this is in fact the case, every grains size has to be treated separately. Fortunately, there are several reasons why the situation is not that bad and why most processes act on different sized grains in the same way.

In general a size dependence of the reaction frequencies $\omega$ might be due to a size dependence of the reaction probabilities $\alpha$ or a size dependence of the relative velocity between grains and molecules, or both. Possible reasons are:

1. *Different Temperatures of Different Grain Sizes*
   Usually, the grain temperature is governed by the interactions of the grains with the radiation field. If the grains are sufficiently small in order to be treated by the small particle limit of the Mie-theory, the temperature becomes independent of the size of the grain (c.f. [Gauger *et al.* 1990b]) if the grains are of uniform material. The latter condition may be violated in situations where the conditions in the gas phase have changed considerably in the time between the formation of the larger and the smaller particles and then, the grain temperature may well be a function of the size. However, dust formation usually occurs at temperatures much lower than the boiling point of the material. In this range the temperature of the grain is only of minor importance. This is different, however, in the envelope of variable stars where the temperature may be increased later to higher values and strong evaporation processes may take place. Then, the size dependence of the temperature may be crucial.

2. *Size Dependence of the Relative Velocity $v_i$ Between Grains and Molecules*
   However, all grains are assumed to be macroscopic. Then, their mass is much larger than that of the molecules and the mean relative velocity between grains and molecules becomes independent of the grain size if their movement is governed by thermal motion. This of course is not valid in the case of drift velocities where the relative velocities may depend strongly on the grain size. However, if the grains are small the momentum coupling to the gas is strong. Under red giant like conditions the drift velocities are then limited at approximately the speed of sound. Then, the effects of different drift velocities are small (see Sect. 2.3.1).

3. *Different Chemical Composition of the Surface on Grains of Different Size*
   The chemical grain properties which are important for the reaction probabilities $\omega$ are the composition of the grain *surface*. Since the conditions in the gas are time dependent, the composition of the surface can also be a function of time. However, since at any time $t_0$ the surface of all grains was formed in the conditions just before $t_0$, all grains have the same chemical surface composition at $t_0$ (c.f. Fig. 1.3).

4. *Size Dependence of the Grain Charge*
If the dust grains are charged it is clear that the charge will considerably depend on the size [Augustin 1990, Taylor *et al.* 1991]. A negative charge on the grains will increase the reaction frequencies with positive ions in the gas phase. A positive charge might totally suppress any reaction with positive ions. The reactions with neutral molecules will remain essentially unaffected [Augustin 1990]. However, it has been shown by several authors that charge on dust grains is important only if a sufficiently strong UV radiation field is present in order to ionize both grains and molecules. In very cold giants and supergiants studied here, no active chromospheres are present and thus one may neglect the possibility of charged grains

From now on it will be assumed that the reaction frequencies are independent of the grain size.

The next step is multiplication of Eq. (1.30) by $V^{j/d}$ and integration from the lower boundary $V_\ell$ to infinity. For the left hand side of the equation one finds

$$l.h.s. = \int_{V_\ell}^{\infty} dV V^{\frac{j}{d}} \frac{d}{dt} f(V,t) = \frac{d}{dt} \int_{V_\ell}^{\infty} dV V^{\frac{j}{d}} f(V,t) = \frac{d}{dt} \mathcal{L}_j. \qquad (1.31)$$

For the evaluation of the right hand side one has to note that the integral is taken from a lower boundary $V_\ell$. Due to various processes it is possible that particles diffuse in the cluster size space through this boundary. The flux $\mathcal{J}_\ell$ through this boundary is given by

$$\begin{aligned} \mathcal{J}_\ell &= \sum_{\substack{i=1 \\ V_i > 0}}^{I} \omega_i \xi_A^{(d)} \int_{V_\ell}^{V_\ell + V_i} f(V - V_i, t)(V - V_i)^{(\frac{d-1}{d})} dV \\ &- \sum_{\substack{i=1 \\ V_i < 0}}^{I} \omega_i \xi_A^{(d)} \int_{V_\ell}^{V_\ell + V_i} f(V,t) V^{\frac{d-1}{d}} dV \end{aligned} \qquad (1.32)$$

which may also be written as

$$\mathcal{J}_\ell = \sum_{i=1}^{I} \omega_i \xi_A^{(d)} \int_{V_\ell - V_i}^{V_\ell} f(V,t) V^{\frac{d-1}{d}} dV. \qquad (1.33)$$

If the processes associated with a positive change in Volume $V_i$ are dominant in the region around $V_\ell$, $\mathcal{J}_\ell$ may be replaced by $\mathcal{J}_*$ as argued in Sect. 1.2. Thus $\mathcal{J}_\ell$ can be calculated by means of a microscopic theory. If on the other hand the processes associated with a negative change in Volume are dominant, one needs to know the values of $f(V,t)$ in the interval $(V_\ell, V_\ell + \max_{i=1,\ldots,I}|V_i|)$. Due to the condition $V_\ell \gg \max_{i=1,\ldots,I}|V_i|$ one may assume that $f(V,t)$ is only varying slowly in this small interval and may approximate $f(V_\ell + V_i, t)$ by $f(V_\ell, t)$. Then

$$\mathcal{J}_\ell \simeq \sum_{i=1}^{I} \omega_i \xi_A^{(d)} f(V_\ell, t) V_i V_\ell^{\frac{d-1}{d}}. \qquad (1.34)$$

Hence the only information one needs to know for the calculation of $\mathcal{J}_\ell$ in this case is the value of the size distribution at the lower boundary (see also [Gauger et al. 1990b]). The integral of Eq. (1.25) multiplied by $V^{j/d}$ yields

$$\frac{d}{dt}\mathcal{L}_j = \sum_{i=1}^{I} \omega_i \xi_A^{(d)} \int_{V_\ell}^{\infty} V^{\frac{j}{d}} \left\{ (V-V_i)^{\frac{d-1}{d}} f(V-V_i, t) - V^{\frac{d-1}{d}} f(V,t) \right\} dV. \qquad (1.35)$$

In the first term in the integral $V$ is substituted by $V - V_i$ and one finds

$$\frac{d}{dt}\mathcal{L}_j = \sum_{i=1}^{I} \omega_i \xi_A^{(d)} \left\{ \int_{V_\ell - V_i}^{\infty} (V+V_i)^{\frac{j}{d}} V^{\frac{j}{d}} f(V,t) dV - \int_{V_\ell}^{\infty} V^{\frac{j}{d}} V^{\frac{d-1}{d}} f(V,t) dV \right\}. \qquad (1.36)$$

By definition of the current through the boundary $V_\ell$ (Eq. (1.33)) one has

$$\frac{d}{dt}\mathcal{L}_j = \mathcal{J}_\ell V_\ell^{\frac{j}{d}} + \sum_{i=1}^{I} \omega_i \xi_A^{(d)} \int_{V_\ell - V_i}^{\infty} \left\{ (V+V_i^{\frac{j}{d}} - V^{\frac{j}{d}}) \right\} V^{\frac{d-1}{d}} f(V,t) dV. \qquad (1.37)$$

Since $\max_{i=1,\ldots,I}|V_i| \ll V$, one may approximate the bracket by

$$(V-V_i)^{\frac{j}{d}} - V^{\frac{j}{d}} \simeq \frac{j}{d} V^{\frac{j}{d}-1} V_i. \qquad (1.38)$$

Introducing this result one finally finds

$$\begin{aligned}\frac{d}{dt}\mathcal{L}_j &= \mathcal{J}_\ell V_\ell^{\frac{j}{d}} + \sum_{i=1}^{I} \omega_i \xi_A^{(d)} V_i \frac{j}{d} \int_{V_\ell}^{\infty} f(V,t) V^{\frac{j-1}{d}} dV \\ &= \mathcal{J}_\ell V_\ell^{\frac{j}{d}} + \sum_{i=1}^{I} \omega_i \xi_A^{(d)} V_i \frac{j}{d} \mathcal{L}_{j-1} \\ &= \mathcal{J}_\ell V_\ell^{\frac{j}{d}} + \frac{j}{d} \chi \mathcal{L}_{j-1} \end{aligned} \qquad (1.39)$$

with the definition

$$\chi = \xi_A^{(d)} \sum_{i=1}^{I} \omega_i V_i. \tag{1.40}$$

These are the equations describing the evolution of the dust component in the macroscopic limit.

### 1.4.3 The Meaning of the Quantity $\chi$

Applying Eq. (1.20) to a single spherical particle with a volume $V_0 > V_\ell$, the moments of this particle are

$$\mathcal{L}_j = \int_{V_\ell}^{\infty} V^{\frac{j}{d}} \delta(V - V_0) dV = V_0^{\frac{j}{d}} \tag{1.41}$$

and the moment equations read

$$\frac{d}{dt} V_0^{\frac{j}{d}} = \frac{j}{d} \chi V_0^{\frac{j-1}{d}}. \tag{1.42}$$

From the first moment equation ($j = 1$) and Eq. (1.28) one can easily see that the growth velocity $da/dt$ of the particle radius $a$ in units [cm/s] is simply given by

$$\frac{d}{dt} a = \frac{\xi_a^{(d)}}{d} \chi. \tag{1.43}$$

One also sees immediately that this is independent of the particle size and, therefore, all particle radii change at the same rate. This is of course a consequence of the assumption that all processes act proportional to the grain surface and are independent of the grain size.

### 1.4.4 Growth or Destruction?

Further insight into the total effect of the reactions between grains and molecules can be obtained by incorporation the Milne relations (1.7) into the definition of $\chi$. One may rewrite Eq. (1.40) by combining the terms for the forward and reverse reaction as follows:

$$\begin{aligned} \chi &= \xi_A^{(d)} \sum_{i=1}^{I} \omega_i V_i \\ &= \xi_A^{(d)} \sum_\iota \omega_\iota \left(1 - S_\iota^{-1}\right) V_\iota. \end{aligned} \tag{1.44}$$

The generalized supersaturation ratio $S_\iota$ expresses for a specific reaction, whether the reaction acts in the direction of grain growth ($S_\iota > 1$) or grain destruction ($S_\iota < 1$). In the case of thermodynamic equilibrium all reactions are completely balanced with the reverse reactions. This fact is expressed by

$$S_\iota = 1 \quad \text{for all } \iota. \tag{1.45}$$

In this case, $\chi$ is of course zero and the dust grains do not change their size. Obviously a sufficient condition for the growth of dust grains is

$$S_\iota > 1 \quad \text{for all } \iota \tag{1.46}$$

since here all reactions with a positive change in volume are more frequent than their reverse reactions. A sufficient destruction condition for dust grains is

$$S_\iota < 1 \quad \text{for all } \iota. \tag{1.47}$$

It is possible that not all $S_\iota$ are either greater or less than unity. In this case, different competing processes take place on the surface of the dust grains and the net effect (determined by the value of $\chi$) cannot be seen from a single supersaturation ratio alone [Gauger et al. 1990b].

### 1.4.5 The Normalized Moment Equations

In an environment with a nonvanishing hydrodynamic velocity field $\mathbf{v}$ the time derivative $d/dt$ in the Eqs. (1.39) has to be the substantial derivative

$$\frac{df}{dt} = \frac{\partial f}{\partial t} + \text{div}(\mathbf{v}f). \tag{1.48}$$

The divergence term can be eliminated if all densities connected with the size distribution (like $f, \mathcal{J}_*, \mathcal{L}_j$) are normalized by the total density of hydrogen $n_{<H>}$ (c.f. Eq. (1.17)). From here on all quantities normalized with $n_{<H>}$ will be designated by a hat: "$\hat{\phantom{x}}$". It is then simple to show that in a co-moving frame the following moment equations are valid ([Gauger 1991]):

$$\hat{\mathcal{L}}_j = \int_{V_\ell}^\infty \hat{f}(V,t) V^{\frac{j}{d}} dV \tag{1.49}$$

$$\frac{d}{dt}\hat{\mathcal{L}}_j = \hat{\mathcal{J}}_\ell V_\ell^{\frac{j}{d}} + \frac{j}{d}\chi \hat{\mathcal{L}}_{j-1}. \tag{1.50}$$

### 1.4.6 The Limit of Homogeneous Dust Particles

Not in all situations is it necessary to deal with the possibility of heterogeneous dust formation. For example in the case of carbon stars the abundance of the dust forming species $C$ is much larger than the abundance of any other species that may form dust. Therefore, apart from small impurities, these grains may be considered as consisting of a single monomer species, the carbon atom. Then all significant processes are the addition or sublimation of one or more monomers. One can factorize the monomer volume from the moment equations and describe the dust component by a size distribution $f(N,t)$ over the number $N$ of monomers contained in the dust grain. Properly defined moments in this case are

$$\mathcal{K}_j = \int_{N_\ell}^{\infty} f(N,t) N^{\frac{j}{d}} dN. \tag{1.51}$$

With the transformation

$$N(V) = \frac{V}{V_{mono}} \tag{1.52}$$

the moment equations transform to the well known equations for homogeneous dust formation and growth as given by [Gail and Sedlmayr 1988, Gauger *et al.* 1990b].

$$\frac{d}{dt}\mathcal{K}_j = \mathcal{J}_\ell N_\ell^{\frac{j}{d}} + \frac{j}{d}\frac{1}{\tau}\mathcal{K}_{j-1} \tag{1.53}$$

where the growth time $\tau$ is defined by

$$\frac{1}{\tau} = A_{\mathrm{mono}} \sum_{i=1}^{I} \omega_i N_i. \tag{1.54}$$

$A_{mono}$ is the hypothetical surface of the monomer defined by

$$A_{\mathrm{mono}} = \xi_A^{(d)} V_{\mathrm{mono}}^{\frac{d-1}{d}}. \tag{1.55}$$

$N_i$ is the number of monomers added to the grain in the reaction i. (If the reaction is destructive, $N_i$ is of course negative.)

## 1.5 The Size Distribution

The calculation of the time variation of the dust component via the moment equations has the striking advantage that the important averages of the powers of the particle volume can be calculated by a simple set of coupled first order differential equations instead of either an almost infinite set of rate equations or the second order Zeldovich equation. However, this advantage is gained at the cost of loosing all the information on the full size distribution as well as (in the case of composite grain formation) the information on the detailed chemical structure of the grains. This information cannot be expressed in terms of a finite number of moments of the distribution function. However, if the full time evolution of the dust component is followed by the moment equations, all necessary information is produced. The only problem is finding a way that allows for a simple reconstruction of this information but keeps the advantages of fast calculation. This way has already been published in [Dominik et al. 1989] and some modifications are due to [Gauger et al. 1990b]. In [Dominik et al. 1989] a formal derivation was given. Here a different and more intuitive method will be used to show how the size distribution is obtained.

The following derivation is especially simple if the size of a particle is expressed by its radius $a$ rather than its volume $V$. The transformation from $V$ to $a$ is given by Eq. (1.28) and the relation between the size distribution in $f(V,t)$ and the size distribution $f(a,t)$ is given by

$$f(a,t) = f(V,t)\frac{dV}{da} = \frac{d}{\xi_a^{(d)}} V^{\frac{d-1}{d}} f(V,t). \qquad (1.56)$$

In order to construct the size distribution at the instant $t_1$, one has to follow in Fig. 1.2 the time evolution of the size of two dust particles $A$ and $B$. The lower limiting size $a_\ell = \xi_a^{(d)} V^{(1/d)}$ is exceeded by $A$ at time $t_0 - dt_0$ and and by $B$ at time $t_0$, respectively. At time $t_0$ the particle $A$ has already grown somewhat and the difference in size is given by

$$da(t_0) = a_A(t_0) - a_B(t_0) = a_A(t_0) - a_\ell. \qquad (1.57)$$

Subsequently, both particles change in size. During this time the curves $A$ and $B$ cannot intersect: Were they to intersect at any time $t_{is}$, $A$ and $B$ would reach identical size at $t_{is}$. Hence from then on, $A$ and $B$ would be subject to the same processes and, therefore, have identical properties for all following times[2].

---

[2]One can think of situations where this is not true. As an example consider a particle that is injected at high velocities into a medium where other dust particles are present. Then, there can be two particles of equal size that are suffering different processes at the same time. This possibility will excluded be here.

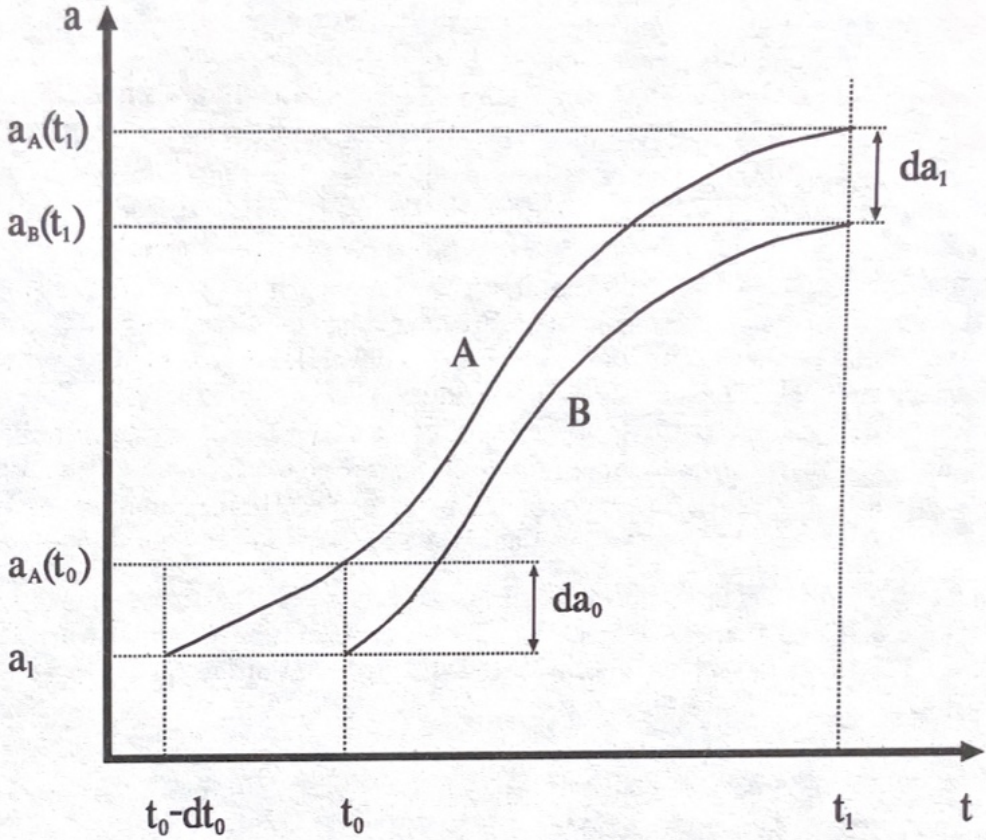

Figure 1.2:

The time evolution of the particle radius a of two dust grains of different size.

At time $t_1$ the difference in size is given by

$$da(t_1) = a_A(t_1) - a_B(t_1). \tag{1.58}$$

How many particles do exist in the size interval $[a_A(t_1), a_B(t_1)]$? Since the curves cannot intersect, all particles in this *size* interval must have exceeded the lower boundary $a_\ell$ in the *time* interval $[t_0 - dt_0, t_0]$ and vice versa, thus

$$\int_{a_A(t_1)}^{a_B(t_1)} \hat{f}(a, t_1) da = \int_{t_0 - dt_0}^{t_0} \hat{\mathcal{J}}_\ell(t) dt. \tag{1.59}$$

Since $a(t)$ is an almost continuous function of $t$, for small values of $dt_0$ and $da_1$ both integrals may be approximated by the assumption of a constant integrand and one finds

$$\hat{f}(a,t_1)da_1 = \hat{J}_\ell dt_0 \quad \text{or}$$
$$\hat{f}(a,t_1) = \hat{J}_\ell(t_0)\frac{dt_0}{da_1}. \quad (1.60)$$

This equation is generally valid if the growth curves do not intersect. One has to note, however, that the quantities $da$ and $dt$ have to be taken at different times. In order to evaluate the full size distribution it is necessary to follow the growth of a number of particles (according to the desired resolution in $a$) individually by use of Eq. (1.43).

A great simplification comes from our assumption that the growth of the particle radius is independent of the radius itself. In this case the difference in size $da(t)$ is independent of time and, especially, equal to $da_0$. Inserting this and Eq. (1.43) into Eq. (1.60) one has

$$\hat{f}(a,t) = \hat{J}_\ell(t_0)\frac{dt_0}{da_0}$$
$$= \frac{d}{\xi_a^{(d)}}\left(\frac{\hat{J}_\ell}{\chi}\right)_{t_0}. \quad (1.61)$$

Except for a geometric factor the size distribution is given by the quotient of the nucleation rate and the growth velocity of the particle radius at the instant when the particle size exceeded the lower boundary.

Due to the parallelism of the curves $a(t)$ it is not necessary to follow the curves of many particles individually. Instead it suffices to follow $a(t)$ for one special particle, most suitably the first particle formed which also is the largest particle.

Starting at this point, [Gauger et al. 1990b] have developed a representation for the size distribution which allows for an easy access to $f(a,t)$ at every instant of the calculation. In order to do that they introduce a function $g$ by

$$g(a_{\max}) = \frac{d}{\xi_a^{(d)}}\left(\frac{\hat{J}_\ell}{\chi}\right)_t. \quad (1.62)$$

$a_{\max}$ is the radius of the largest dust grain at time $t$ given according to Eq. (1.43) by

$$a_{\max} = a_\ell + \frac{\xi_a^{(d)}}{d}\int_{t_i}^{t}\chi(t')dt' \quad (1.63)$$

where $t_i$ is the time of the onset of dust formation when the first dust particle exceeds the size $a_\ell$. It can then be shown that the size distribution is given by

$$\hat{f}(a,t) = g(a_{\max} - a + a_\ell) \qquad a_{\max} \geq a \geq a_\ell. \qquad (1.64)$$

In order to obtain this representation one has to integrate Eq. (1.63) simultaneously with the moment equations and to store the function $g$.

## 1.6 Chemical Composition of the Grains

### 1.6.1 Chemical Depletion in the Gas

The chemical depletion in the gas phase can be treated in two ways:

- **Depletion of chemical species.** This form of balance is necessary if the chemistry in the gas phase is treated as a non equilibrium process. Then a rate equation is solved for every chemical species (i.e. every atom, molecule, ion).

  Let $M_j$, $j = 1,\ldots,J$ be the symbols for the different chemical species and let $\gamma_{j,i}$ be the net number of molecules of kind $M_j$ extracted from the gas by destruction or condensation in the reaction $i$. ($\gamma_{j,i}$ is negative if a molecule $M_j$ is injected into the gas phase). Since the number of reactions of kind $i$ per second is given by $\xi_A^{(d)} \omega_i \mathcal{L}_{d-1}$, the rate of formation of the molecule $M_j$ due to grain-gas reactions is given by

  $$\left[\dot{n}_{M_j}\right]_{\text{dust}} = -\sum_{i=1}^{I} \gamma_{j,i} \xi_A^{(d)} \omega_i \mathcal{L}_{d-1}. \qquad (1.65)$$

- **Depletion of elements.** If the chemistry in the gas phase is assumed to be in local equilibrium it is sufficient to know the extraction of chemical elements from the gas phase.

  Let $E_j$, $j = 1,\ldots,J$ be the symbols for the chemical elements and let $\nu_{j,i}$ be the net number of atoms of kind $E^j$ extracted from the gas during the reaction $i$. The change in the *abundance* $\varepsilon_j = n_{<E^j>}/n_{<H>}$ is given by

  $$\dot{\varepsilon}_j = -\frac{1}{n_{<H>}} \sum_{i=1}^{I} \nu_{i,j} \xi_A^{(d)} \omega_i \mathcal{L}_{d-1}. \qquad (1.66)$$

### 1.6.2  Chemical Composition of the Dust Grains

Now it is quite simple to calculate the chemical composition of the dust grains. If again one stresses the assumption that all processes act on all grains in the same way it is clear that all grains have the same chemical composition at their surface. This composition is determined by the composition of the actually condensing material. The number of atoms of kind $E_j$ condensing per unit time on a unit surface of the grains is given by $\sum_{i=1}^{I} \nu_{i,j} \omega_i$. The part of this contribution to the total growth is given by

$$s_j = \frac{\sum_{i=1}^{I} \nu_{i,j} \omega_i}{\sum_{j=1}^{J} \sum_{i=1}^{I} \nu_{i,j} \omega_i} \quad (1.67)$$

and the *relative chemical composition* of the surface layer may be written as

$$E_{s_1}^{(1)} E_{s_2}^{(2)} \ldots E_{s_J}^{(J)}. \quad (1.68)$$

As an example consider the formation of pure $MgSiO_4$. Here the relative chemical composition is

$$Mg_{1/6} Si_{1/6} O_{2/3}. \quad (1.69)$$

Since the growth of all particles occurs in the same way at all times this assumption furthermore implies that all grains have the same composition as a function of depth below the surface (compare Fig. 1.3). If one knows the chemical composition of the largest grain, one has simultaneously evaluated the composition of all smaller grains.

In order to obtain the chemical composition one only has to store the functions $s_j(a_{\max})$ during the calculation of dust formation. The chemical composition of the grain surface at time $t$ is then given by

$$E_{s_1(a_{\max}(t)-a_{atom})}^{(1)} E_{s_2(a_{\max}(t)-a_{atom})}^{(2)} \ldots E_{s_J(a_{\max}(t)-a_{atom})}^{(J)} \quad (1.70)$$

where $a_{atom}$ is the typical extension of a one atom surface layer.

The composition of a grain of radius $a(t)$ as a function of the distance $\ell$ from the center of the grain is given by

$$E_{s_1(a_{\max}(t)-a(t)+\ell)}^{(1)} E_{s_2(a_{\max}(t)-a(t)+\ell)}^{(2)} \ldots E_{s_J(a_{\max}(t)-a(t)+\ell)}^{(J)}$$
$$a_\ell < \ell \leq a(t). \quad (1.71)$$

This formula is not valid for values $\ell < a_\ell$ since it takes into account only the composition of the growth material. In this way no information about the composition of the grain nuclei is retained.

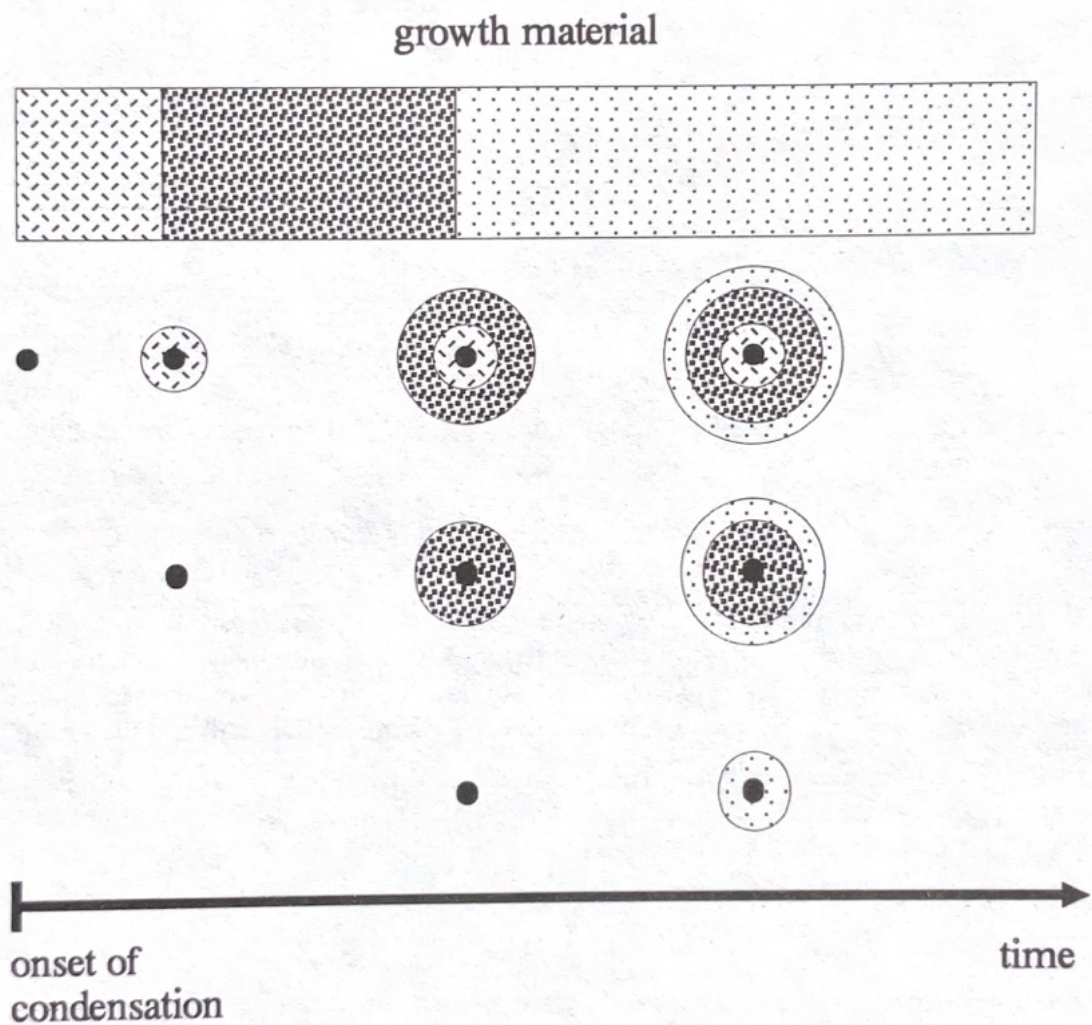

Figure 1.3:
Sketch of the time evolution of the internal chemical composition of three dust grains originating at different times. The bar at the top shows the composition of the condensing material which later is found as a mantle on the grains. Note that, at every instant, all grains have the same composition as a function of depth below the surface. The black cores are the grain nuclei of unknown composition.

# Chapter 2

# Construction of a Model of Dust Forming Stationary Winds Around Late Type Stars

## 2.1 Overview

There is an unambiguous answer to the question of where the solid particles in the interstellar medium originate from: They are formed by stellar matter returning to the interstellar medium. This returning occurs continuously in the stellar wind of every star or discontinuously in explosive ejections like novae and supernovae. As far as the production of dust grains is concerned, the most important sites are the late type giants and supergiants [Weinzierl 1991, Gehrz 1989]. All other objects play only a minor role. Essentially, there are two reasons for this:

1. The winds of red giants are the source of about 85% of the total returning mass from stars to the galaxy [Weinzierl 1991]. Therefore, even if other objects would produce dust at the same efficiency as red giants do, their dust production would still not dominate.

2. Furthermore the winds of red giants provide rather high gas densities in the cooling outflow accompanied by low expansion velocities (i.e. sufficient time for dust formation and growth in the relevant pressure-temperature range) and small UV radiation fields (necessary for a rather undisturbed evolution of the chemistry). All other objects where dust formation is possible lack at least one of these important conditions.

There is a further reason why these objects are especially interesting in the context of this work: In the winds of late type giants the coupling between the dust and the hydrodynamic and thermodynamic properties of the wind is much stronger than in other dust forming objects. The hydrodynamics of the ejecta of novae and supernovae cannot be significantly influenced by the presence of dust since the energy contained in the expansion of the material is much larger than that accessible to dust grains by the absorption of light.

This is completely different in the winds of cool giants. [Jura 1986] has estimated that at least in carbon stars the momentum flux carried by the mass loss is about the same as that in the radiation field. Furthermore [Jura 1986] showed that in almost all carbon stars with massive mass loss the opacity due to dust grains is enough for driving a wind. [Knapp 1985] demonstrated that the momentum flux in the wind is approximately proportional to the dust opacity in the shell. The natural suggestion would be that at least the final acceleration of the wind to its terminal velocity is due to the absorption of dust grains.

Late type giants with dust forming winds first attracted attention because of their large infrared excess [Gehrz and Woolf 1971]. Some of the objects essentially show the spectrum of a black body of about $2400K$ superposed by that of a $\approx 600K$ black body due to thermal emission of an extended dust shell [Merrill and Stein 1976]. Others are completely obscured by the dust and their spectrum essentially is that of a cold black body. The observation of the profiles of molecular lines showed that the late type stars with massive infrared excesses also have massive winds. Since the central stars are too hot for solid particles to exist, the dust has to be formed in the cooling outflow from the star.

Further information has been obtained during the last few years mainly from infrared and radio observations. In the radio band the rotational transitions of several important molecules present in these objects can be observed, among them the $CO$ molecule. From $CO$ observations (e.g. [Knapp et al. 1982, Knapp and Morris 1985]) the terminal wind velocities and the densities of the $CO$ molecule can be estimated. Terminal outflow velocities were found in the range between $5 km s^{-1}$ and $40 km s^{-1}$. From the $CO$ densities (combined with a reasonable assumption about the $CO$-to-$H$ ratio) mass loss rates in the order of $10^{-4} M_\odot yr^{-1} - 10^{-8} M_\odot yr^{-1}$ have been obtained. These are the largest mass loss rates known in stars[1].

A further possibility for the determination of mass loss rates is the analysis of the infrared spectra. Especially the far infrared spectra are completely determined by the thermal emission of dust grains at larger distances from the star. Appropriate models for the radiative transfer in dust envelopes allow the evaluation of the rate at which *dust* is leaving the star [Veen and Olofsson 1989]. Here one needs a reasonable assumption about the dust-to-gas ratio in order to estimate the *gas* mass loss from the star. The range of derived mass loss rates here is consistent with that derived by $CO$ observations though for single objects rather large differences may occur [Veen and Rogers 1989].

The dust-to-gas ratios are apparently rather high in late type stars with massive mass loss. Estimates for this ratio have been given by [Knapp 1985, Jura 1986]. The results are values of some $10^{-3}$ by mass. According to the chemical abundances in the stars this shows that the condensation efficiency must be of the order of unity.

Chemically the stars can be divided into three classes, the M-, S-, and C-stars. In M-stars, oxygen is more abundant than carbon and blocks all the carbon by binding it in the $CO$ molecule. The prominent absorption bands in the spectrum are therefore due to oxygen bearing molecules. The opposite is true for C-stars. Here oxygen is blocked in $CO$, and the spectrum is characterized by the absorption bands of carbon bearing molecules. S-stars are the intermediate stars where oxygen and carbon almost have the same abundance. The three classes can also be distinguished by the dust present in their envelopes: in carbon stars this dust maily is carbon of graphite or amorphous structure or possibly polyaromatic hydrocarbons (PAH's). In M-stars the infrared observations show silicate like features. Some stars show the properties of carbon stars but seem to have also silicate like dust at larger distances from the star [Willems and deJong 1988]. This is believed to be an evolutionary effect: Due to the dredge up processes in the evolution of AGB stars carbon is mixed up from the center of the star to the surface. Thereby the star is converted from an M-star to a C-star and the older silicate dust surrounds the fresh carbon dust.

---

[1] Only early supergiants may reach mass loss rates of the same magnitude.

Both the mass and the luminosity of late type giants are very uncertain and can only be estimated. The results of theoretical studies of the evolution of low mass stars (e.g. [Boothroyd and Sackmann 1988a, Boothroyd and Sackmann 1988b]) show that top AGB stars should have masses between $1M_\odot$ and $6M_\odot$ and luminosities of about $10^4 L_\odot$.

Another important property of dust forming late type stars is that many of them are long period variable stars. In these stars radial pulsations (which have their cause in the interior of the star) are responsible for the changing luminosity and temperature. The most remarkable ones among the long period variables are the Mira variables. Here the radial pulsations are so large that strong shock waves travel periodically through the atmosphere. These shock waves lead to a deposit energy and momentum in the atmosphere and thereby may contribute to the mass loss mechanism of the stars.

Several mass loss mechanisms have been discussed for the winds. Apart from radiation pressure on dust grains these include sound waves ([Pijpers and Hearn 1989]), Alfvèn waves (e.g. [Hartmann and MacGregor 1980]), and in the case of long period variables shock waves (e.g. [Bowen 1988]). It seems to be widely accepted that the mass loss mechanism is strongly influenced by radiation pressure on dust grains. The mechanism is assumed to be a two step process (e.g. [Morris 1987] and references therein):

1. Levitation of the atmosphere by the dissipation of shock waves traveling through the atmosphere.
   In stars without large radial pulsations smaller shock waves originating from acoustic waves leaving the star may have a similar though less prominent effect (e.g. [Cuntz 1987, Gail 1990]).

2. Acceleration of the wind by radiation pressure on dust grains.
   In stars with effective temperatures larger than $3000K$, dust formation only results in an additional acceleration of the material. In very cool and luminous stars, particularly in cool carbon giants, dust formation may occur close enough to the photosphere in order to be capable of driving a massive wind even without the support of dissipating waves.

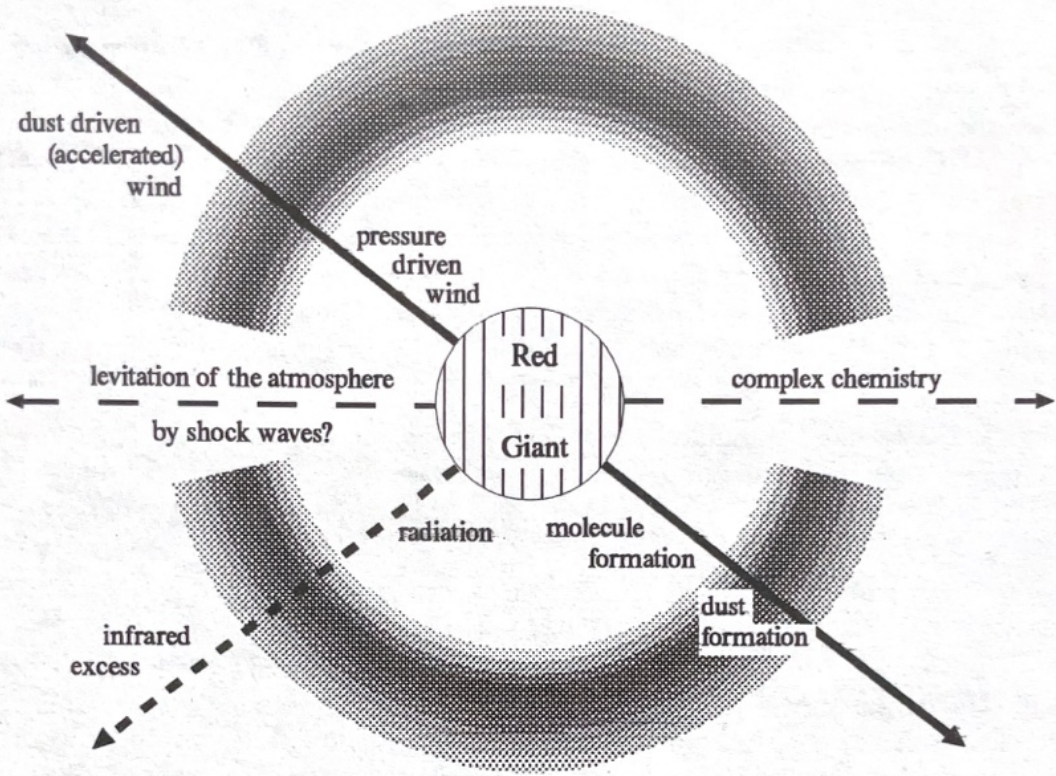

Figure 2.1:
Scenario of a red giant. Adopted with changes from [Sedlmayr 1989a].

A model of the dust forming envelope of a red giant therefore has to simultaneously describe the hydrodynamic structure, the thermodynamic structure, the chemical composition of the gas, and the process of dust formation including the feedback on the hydrodynamic and thermodynamic structure.

Fig. 2.1 shows a sketch of the basic shape and the processes in the envelope of a red giant having a spherical dust forming wind. The red giant is characterized by typical parameters $M_\star \simeq 1 M_\odot$ and $L_\star \simeq 10^4 L_\odot$. The effective temperature of the stars photosphere is in the range of $T_\star \simeq 2000\ldots 3000K$. The radius of the photosphere is of the order of $1000 R_\odot$.

The giant is the source for the outflow of gas. Near the star this outflow may be supported by the dissipation of several types of waves. As the gas cools down, a complex chemistry evolves, starting with atoms and small molecules like $CO$ and $N_2$. Subsequently, more complex molecules are formed. At some distance from the star where the temperature becomes sufficiently low, the formation of solid particles starts. This process has dramatic consequences on the structure of the whole object: Compared to the gas, the dust has an enormous cross section for the absorption of radiation. Therefore, it absorbs a large fraction of the stellar radiation field. The absorbed energy is transformed into internal energy of the dust and then reemitted at infrared wavelengths. Thereby the characteristic infrared excess of these objects is produced. Furthermore the dust absorbs the momentum contained in the radiation field. This momentum is transferred to the gas by grain-gas collisions. The consequence is the acceleration of the material in the envelope, i.e. a massive wind from the star.

In the following the basic equations describing such a physical system will be discussed. The model is restricted to a stationary, spherically symmetric wind. The dissipation of shock waves is neglected and the scenario of a purely dust driven wind is discussed. The discussion will concentrate on some aspects where I have made contributions. The other parts of the model are essentially the same as in [Gail and Sedlmayr 1987b].

## 2.2 The Hydrodynamic Equations

The basic hydrodynamic equations describing the motion of the outflowing gas are the law of mass conservation

$$\frac{\partial \rho}{\partial t} + \mathrm{div}(\rho \mathbf{v}) = 0 \tag{2.1}$$

where $\rho$ is the mass density and $\mathbf{v}$ is the velocity of the gas, and Eulers equation

$$\frac{\partial \mathbf{v}}{\partial t} + (\mathbf{v} \cdot \mathrm{grad})\mathbf{v} = -\frac{1}{\rho}\mathrm{grad}p + \mathbf{g} + \mathbf{g}_{\mathrm{rad}}. \tag{2.2}$$

Here $p$ is the gas pressure, $\mathbf{g}$ is the gravitational acceleration, and $\mathbf{g}_{\mathrm{rad}}$ is the acceleration due to absorption and emission of radiation. The radiative force will be discussed in detail in Sect. 2.3.

Since in this work only stationary winds are considered, the partial time derivative always vanishes. Furthermore it will be assumed that the outflow from the star is spherically symmetric. Then the law of mass conservation reduces to

$$\frac{1}{r^2}\frac{\partial}{\partial r}(r^2 \rho v) = 0 \tag{2.3}$$

where $v$ is the radial component of $\mathbf{v}$. Eq. (2.3) may directly be integrated to

$$r^2 \rho v = \mathrm{const} = \frac{\dot{M}}{4\pi} \tag{2.4}$$

where the integration constant $\dot{M}$ denotes the *mass loss rate*.

Eulers equation may be written as

$$v\frac{\partial v}{\partial r} = -\frac{1}{\rho}\frac{\partial p}{\partial r} - \frac{GM(r)}{r^2} + g_{\rm rad} \qquad (2.5)$$

where $G$ is the gravitational constant and $M(r)$ is the mass enclosed in the sphere of radius r. Since the mass outside the photosphere of the star is negligible (except for the models with very large mass loss rates), $M(r) = M_*$ will be assumed. $g_{\rm rad}$ is the radial component of the radiative acceleration.

The appropriate equation of state in the diluted gas in the stellar envelope is provided by the ideal gas law

$$p = nkT \qquad (2.6)$$

where $n$ is the number density of all particles, $k$ is Boltzman's constant, and T is the gas temperature. If no significant ionization occurs, it is sufficient to consider only the contributions of $H$, $H_2$, and $He$ to $n$. All other species are less abundant by at least a factor of about 1000:

$$n = n_H + n_{H_2} + n_{He}. \qquad (2.7)$$

## 2.3 The Radiative Force

It is well known that the radiation pressure in the envelope of giant stars might be a source of force exceeding the gravitational force. However, in order to obtain high radiative accelerations, it is necessary for the atmosphere to effectively absorb a significant fraction of the radiation. In the envelopes of hot giant stars (spectral type $O$ and $B$) the radiation of the central star is most intensive at UV wavelengths due to the high effective temperatures between $10^4 K$ and $10^5 K$. Therefore, the atoms and ions are able to absorb the radiation and thereby extract energy and momentum from the radiation field. A large fraction of the spectrum can be absorbed due to two different effects: First, the number of contributing lines is very large (up to some $10^5$) and, second, the Doppler effect shifts all lines by the large expansion velocity (some $10^3$ km/s). Due to the continuous acceleration of the matter, the narrow absorption lines of the atoms are able to tap a much broader part of the spectrum.

The situation is quite different in the envelopes of red giant stars. Since the effective temperatures are located around $2000K$ - $3000K$, most radiation is emitted at IR wavelengths. In this case atoms are ineffective absorbers. Another opacity source is needed.
One possibility is the IR absorption of vibrational transitions in molecules like $CO$ and $H_2O$. This absorption may in fact strongly increase the absorption coefficient of the gas, but it is not efficient enough to surmount the gravitation and to drive substantial mass loss [Plez 1991].

The second possibility is the presence of dust grains in the atmosphere. Since dust grains are solid particles, they absorb radiation at almost all wavelengths. Therefore the grains extract energy and momentum from the radiation field. For a single dust grain, the radiation force immediately exceeds the gravitational force and it starts to move outward. Due to this movement the grain suffers collisions with gas particles leading to a momentum transfer from the grain to the gas. In order to create an effective stellar mass loss (i.e. to remove not only dust grains but also gas) it is of essential importance that the grains are strongly coupled to the gas and thus do transmit momentum from the radiation field to the gas most effectively.

Therefore the momentum coupling between gas and dust is discussed in detail in the following section.

### 2.3.1 Mechanical Coupling Between Dust and Gas

**The Collision Rates**

To estimate the rates at which momentum and energy are exchanged between dust and gas, it is important to know the collision rate between a dust grain moving at a *drift velocity* $\mathbf{v}_D$ relative to the gas.

Considering the mean free path $\lambda$ of the gas particles

$$\lambda = \frac{1}{\sigma n} = 10^6 \text{cm} \left[\frac{10^{-16}\text{cm}^2}{\sigma}\right] \left[\frac{10^{10}\text{cm}^{-3}}{n}\right] \tag{2.8}$$

where $\sigma$ is the cross section for collisions between gas particles and $n$ is the total density of gas particles, one clearly sees that the mean free path of the gas particles is always large compared to the typical dimensions of the dust grains ($\lesssim 1\mu m$). In this case the gas is called strongly rarefied. Then, the velocity distribution of the gas particles near the surface of the dust grain may be considered as undisturbed by the existence of the surface. Hence, the velocity distribution is a *Maxwell* distribution $f_i(\mathbf{v})$ for every gas species $i$. These are characterized by the mean thermal velocities $v_{\text{th},i}$

$$v_{\text{th},i} = \sqrt{\frac{2kT}{m_i}}, \tag{2.9}$$

where $m_i$ is the mass of the species $i$ and $T$ is the gas temperature. Since the mass of the dust grains always is large compared to the mass of the molecules, one may neglect the thermal motion of the grains.

Now consider a surface element $d\sigma$ moving at the velocity $\mathbf{v}_D$ and count the number of collisions of gas particles of species $i$ with this surface. If the normal to the surface

points into the negative $x$-direction, the number of collisions per unit time is given by [Schaaf 1963]

$$dZ_i = \int_{-\infty}^{\infty} d(\mathbf{v} - \mathbf{v_D})_z \int_{-\infty}^{\infty} d(\mathbf{v} - \mathbf{v_D})_y \int_{0}^{\infty} d(\mathbf{v} - \mathbf{v_D})_x n_i |(\mathbf{v} - \mathbf{v_D})_x| f_i(\mathbf{v}). \qquad (2.10)$$

The total collision rate is then given by integration of Eq. (2.10) over the surface of the dust grain. For a sphere of radius $a$ the result is [Schaaf 1963]

$$Z_i = \pi a^2 \frac{v_{\text{th},i}}{\sqrt{\pi}} \frac{n_i}{s_i} \left[ \frac{\sqrt{\pi}}{2}(2s_i^2 + 1)\Phi(s_i) + s_i \exp(-s_i^2) \right] \qquad (2.11)$$

where $s_i = v_D/v_{\text{th},i}$ is the ratio of the drift velocity ($v_D = |\mathbf{v_D}|$) to the mean thermal velocity of species i. $\Phi(s)$ is the well known error function

$$\Phi(s) = \frac{2}{\sqrt{\pi}} \int_0^s e^{-t^2} dt. \qquad (2.12)$$

The limiting cases for large and small drift velocities can be written as[2][3]:

$$Z_i = \begin{cases} 4\pi a^2 n_i \frac{v_{\text{th},i}}{2\sqrt{\pi}} & s \ll 1 \\ \pi a^2 n_i |\mathbf{v_D}| & s \gg 1 \end{cases}. \qquad (2.16)$$

---

[2]The error function makes the evaluation of Eq. (2.11) rather time consuming. For normal use, the following approximation may serve as well:

$$\Phi(s) \simeq \sqrt{1 - \exp\left(-\frac{4s^2}{\pi}\right)}.$$

The maximum error produced with this formula is about 0.7% at s=1.1.

[3]With the exact expression for the collisional rate the *mean relative velocity* $v_i$ between the dust grain surface and the gas particles needed in the calculation of the reaction frequencies (cf. Eq. (1.6)) may now be calculated properly. Due to the definition of $v_i$ one has

$$Z_i = 4\pi a^2 n_i v_i. \qquad (2.13)$$

According to Eq. (2.11) $v_i$ is then given by

$$v_i = \frac{v_{\text{th},i}}{4\sqrt{\pi}} \frac{1}{s_i} \left[ \frac{\sqrt{\pi}}{2}(2s_i^2 + 1)\Phi(s_i) + s_i \exp(-s_i^2) \right] \qquad (2.14)$$

with the limiting cases

$$v_i = \begin{cases} v_{\text{th},i}/\sqrt{4\pi} & s \ll 1 \\ |\mathbf{v_D}|/4 & s \gg 1 \end{cases}. \qquad (2.15)$$

### The Drag Force

For symmetry reasons it is clear, that the mean momentum transfer between a spherically symmetric dust grain and the gas vanishes, if the drift velocity is zero. On the other hand, if there is a drift velocity, the collisions occur predominantly at that side of the surface pointing in the direction of the drift motion and the result will be a decelerating drag force. For the calculation of this force one has to know details about the interaction between dust and gas: The gas particles can be reflected elastically, adsorbed by the grain for some time and then evaporated again[4], or adsorbed and not evaporated again. The transfer of both momentum and energy depends on this question. However, until now there is no physical theory that allows for an accurate description of these processes. If the momentum transfer is calculated for all three cases named above (e.g. [Epstein 1924]), it varies by about a factor of two. It will be shown later that this factor is of no importance in this context. Therefore, only elastic reflections of the particles will be considered. Then, the drag force on a spherical dust grain due to the gas particle species $i$ is given by [Schaaf 1963]

$$\begin{aligned}\mathbf{F}_{\text{drag},i} = & -\pi a^2 m_i n_i v_{\text{th},i}^2 \frac{2}{\sqrt{\pi}} \frac{\mathbf{v}_D}{|\mathbf{v}_D|} \\ & \left\{ \frac{(s_i^2+1)\exp(-s_i^2) + \sqrt{\pi} s_i (s_i^2 + \frac{3}{2})\Phi(s_i)}{2s_i} \right. \\ & \left. + \frac{s_i \exp(-s_i^2) + \sqrt{\pi}(s_i^2 + \frac{1}{2})\Phi(s)}{4s_i^2} \right\}.\end{aligned} \qquad (2.17)$$

The drag force is always directed against the drift velocity. If $F_{\text{drag}}$ denotes the rate of the drag force, the limiting cases for this formula can be written as[5]

$$F_{\text{drag},i} = \begin{cases} -\pi a^2 m_i n_i \dfrac{8}{3}\dfrac{v_{\text{th},i}}{\sqrt{\pi}} v_D & s \ll 1 \\ -\pi a^2 m_i n_i v_D^2 & s \gg 1 \end{cases}. \qquad (2.18)$$

The total drag force is given by the sum over all contributing gas species $i$. The drag force is always proportional to the product $m_i n_i$ of each species which can be interpreted as the mass density of the species. Thus in astrophysical environments

---

[4]then the gas particles leave the grain with a mean velocity according to the grain temperature and independent of their velocity before the collision

[5]The following approximation for the drag force reproduces both limiting cases and is always better than 1.2% :

$$F_{\text{drag},i} = -\pi a^2 m_i n_i v_{\text{th},i}^2 \sqrt{\frac{64}{9\pi} s_i^2 + s_i^4}.$$

with normal chemical abundances it is usually sufficient to consider the contributions of hydrogen and helium to the drag force. The only other dependence on the particle mass comes from the mean thermal velocity. It contains the square root of the mass and is important only in the case of small drift velocities. Neglecting this dependence, a simple expression for the total drag force $F_{\text{drag}}$ may be obtained by replacing $m_i n_i$ in Eq. (2.18) by the total mass density $\rho$. This expression is correct for $s > 1$ and fails by 20% for small values of $s$. From now this expression will be used and $v_{\text{th}}$ will denote the thermal velocity of the $H_2$ molecule.

### Equilibrium Drift Velocity and Acceleration Timescales for Dust Grains

A dust particle which emerges in the envelope around a red giant initially is at rest relative to the gas phase. But when the size exceeds a certain limiting value, the grain starts to absorb radiation like a macroscopic particle which is very effective. Thus it is accelerated by the radiative force $F_{\text{rad}}$ and begins to drift relative to the gas component. Collisions with the gas lead to a drag force in the opposite direction. Since the drag force increases with increasing drift velocity, there exists an equilibrium drift velocity $\overset{\circ}{v}_D$. Neglecting for the time being the change in mass of the dust grain due to growth or evaporation, the equation of motion for the grain in the frame of the gas is given by

$$\frac{dv_D}{dt} = \frac{1}{m_a} \{F_{\text{rad}} - F_{\text{drag}}\} \qquad (2.19)$$

where $m_a$ is the mass of a dust grain of radius $a$. According to Eq. (2.18) the drag force is proportional to the drift velocity in the case of small drift velocities and proportional to the square of the drift velocity in the case of large drift velocities:

$$F_{\text{drag}} = \begin{cases} bv_D & s \ll 1 \\ Bv_D^2 & s \gg 1 \end{cases} \qquad (2.20)$$

with the abbreviations

$$b = \pi a^2 \rho \frac{8}{3} \frac{v_{\text{th}}}{\sqrt{\pi}} \qquad (2.21)$$

$$B = \pi a^2 \rho. \qquad (2.22)$$

The equilibrium drift velocities $\overset{\circ}{v}_D$ are defined by $F_{rad} = F_{drag}$ and the results for the two limiting cases are[6]

$$\overset{\circ}{v}_D = \begin{cases} F_{rad}/b & s \ll 1 \\ \sqrt{F_{rad}/B} & s \gg 1. \end{cases} \quad (2.23)$$

The radiation force on a dust grain of radius $a$ at a distance $r$ from the star is given according to Mie theory[7] by

$$F_{rad}(r) = \frac{4\pi}{c}\pi a^2 \int_0^\infty Q_{ext}(\nu, a) H_\nu(r) d\nu \quad (2.24)$$

where $H_\nu$ is the Eddington flux and $Q_{ext}(\nu, a)$ is the extinction efficiency of the dust grain. $Q_{ext}$ goes to zero as the quotient $(2\pi a/\lambda)$ does and approaches 2 if $(2\pi a/\lambda)$ is much greater than unity. One may define a mean extinction efficiency $\overline{Q}$ by

$$\overline{Q} = \frac{\int_0^\infty Q_{ext}(\nu, a) H_\nu d\nu}{\int_0^\infty H_\nu d\nu}. \quad (2.25)$$

With Eq. 2.4 and with values typical for the envelopes of red giants $(L_*, \dot{M}, v)$ the equilibrium drift velocities are then given by

$$\overset{\circ}{v}_D \simeq 13\,\text{km/s}\,\frac{\overline{Q}}{1}\,\frac{L_*}{10^4 L_\odot}\,\frac{1\,\text{km/s}}{v_{th}}\,\frac{v}{1\,\text{km/s}}\,\frac{10^{-5} M_\odot/\text{yr}}{\dot{M}} \quad (2.26)$$

for subsonic drift and by

$$\overset{\circ}{v}_D \simeq 4.5\,\text{km/s}\left[\frac{\overline{Q}}{1}\,\frac{L_*}{10^4 L_\odot}\,\frac{v}{1\,\text{km/s}}\,\frac{10^{-5} M_\odot/\text{yr}}{\dot{M}}\right]^{1/2} \quad (2.27)$$

for supersonic drift. It can easily be seen that for particles considerably smaller than all relevant wavelengths $(\overline{Q} \ll 1)$ the drift velocities are limited to approximately the speed of sound.

---

[6]With the approximate expression for the drag force (see footnote on page 37) a simple expression for the equilibrium drift velocity, which is valid for all values of $s_i$ may also be obtained:

$$\frac{\overset{\circ}{v}_D^2}{v_{th}^2} = -\frac{32}{9\pi} + \sqrt{\left(\frac{32}{9\pi}\right)^2 + \left(\frac{F_{rad}}{\pi a^2 \rho v_{th}^2}\right)^2}.$$

[7]isotropic scattering is assumed

The typical timescales for the acceleration of the dust grain from the initial drift velocity zero to its equilibrium drift velocity $\overset{\circ}{v}_D$ can be obtained from the solution of Eq. (2.19). For equilibrium drift velocities smaller than the thermal velocity one has to solve Eq. (2.19) with the force term proportional to $b$. One obtains the acceleration timescale $\tau_b$. For equilibrium drift velocities larger than the thermal velocity Eq. (2.19) is solved with the drag force proportional to $B$. Here one obtains another acceleration timescale $\tau_B$. The correct timescale for the acceleration of dust grains to drift velocities above the thermal velocity is the maximum of $\tau_b$ and $\tau_B$.

The two different solutions of Eq. (2.19) and the relevant timescales are:

$$s \ll 1: \quad v_D(t) = \frac{F_{\rm rad}}{m_a}\left(1 - \exp\left\{-\frac{b}{m_a}t\right\}\right) \tag{2.28}$$

$$\tau_b = \frac{m_a}{b} \tag{2.29}$$

$$s \gg 1: \quad v_D(t) = \sqrt{\frac{F_{\rm rad}}{B}}\left(1 - \frac{1}{1 + \frac{\sqrt{F_{\rm rad}B}}{m_a}t}\right) \tag{2.30}$$

$$\tau_B = \frac{m_a}{\sqrt{F_{\rm rad}B}}. \tag{2.31}$$

With the specific density of the dust material

$$\rho_d = \frac{m_a}{(4/3)\pi a^3} \tag{2.32}$$

and with regard to values typical for the envelopes of red giants $(L_*, r, \rho)$ the acceleration timescales are given by

$$\tau_b[s] \simeq 2.6 \cdot 10^4 \frac{a}{0.1\mu} \frac{\rho_d}{3{\rm gcm}^{-3}} \frac{10^{-14}{\rm gcm}^{-3}}{\rho} \frac{1{\rm kms}^{-1}}{v_{\rm th}} \tag{2.33}$$

$$\tau_B[s] \simeq 4 \cdot 10^3 \frac{a}{0.1\mu} \frac{\rho_d}{3{\rm gcm}^{-3}} \left[\frac{1}{\overline{Q}}\right]^{1/2} \left[\frac{10^4 L_\odot}{L_*}\right]^{1/2} \frac{r}{10^{14}{\rm cm}} \frac{10^{-14}{\rm gcm}^{-3}}{\rho}. \tag{2.34}$$

These acceleration timescales have to be compared to the typical hydrodynamic timescales which in a stationary wind situation are given by

$$\tau_{\rm hyd}[s] \simeq \frac{r}{v} = 10^8 \frac{10{\rm kms}^{-1}}{v} \frac{r}{10^{14}{\rm cm}}. \tag{2.35}$$

Therefore the acceleration timescale for the dust grains to reach the equilibrium drift velocity is always small compared to the typical hydrodynamic timescale. One may assume that the grains drift always at their equilibrium speed.

### The Growth Timescale of the Dust Grain

For the estimate of the acceleration timescales in the previous section it has been ignored that the dust grains are growing while they are accelerated outward. Therefore the estimated acceleration timescales are valid only if they are smaller than the typical timescales for changes in the mass or the surface area of the grain. Since the latter is always smaller than the former, the timescale in which the *mass* of the dust grain changes will be used. The definition of this timescale is somewhat arbitraryly. Here the time in which the *volume* of the grain is doubled, will be estimated. The timescale will be called $\tau_{2V}$. The radius increase of a spherical dust particle ($d=3$) is given according to Eq. (1.43) by

$$\frac{da}{dt} = \frac{\xi_a^{(3)}}{3}\chi. \tag{2.36}$$

A change of the volume from $V$ to $2V$ is (for a spherical particle) equivalent to the growth of the radius by a factor $\sqrt[3]{2}$. Thus one finds

$$\tau_{2V}(a) = \frac{3a(\sqrt[3]{2}-1)}{\xi_a^{(3)}\chi}. \tag{2.37}$$

Furthermore it will now be assumed that the growth of the dust grain is essentially due to the condensation of only one kind of molecules $i = m$ which is characterized by the following quantities: volume $V_m$, particle density in the gas $n_m =: \varepsilon_m n_{<H>}$, mass $m_m$, reaction probability $\alpha_m$ and thermal velocity $v_{th,m}$. Now again the two cases $s \ll 1$ ($\tau_{2V} =: \tau_{2V,b}$) and $s \gg 1$ ($\tau_{2V} =: \tau_{2V,B}$) have to be considered. By means of the Eqs. (1.40), (1.6), and (2.15) one finds

$$\tau_{2V,b} = \frac{3a(\sqrt[3]{2}-1)}{\xi_a^{(3)}\xi_A^{(3)} V_m \varepsilon_m n_{<H>} \alpha_m}\sqrt{\frac{4\pi}{v_{th,m}}} \tag{2.38}$$

$$\tau_{2V,B} = \frac{3a(\sqrt[3]{2}-1)}{\xi_a^{(3)}\xi_A^{(3)} V_m \varepsilon_m n_{<H>} \alpha_m}\frac{4}{v_D}. \tag{2.39}$$

In order to compare the growth timescale, and the acceleration timescale the expressions $\tau_b/\tau_{2V,b}$ and $\tau_B/\tau_{2V,B}$, respectively, have to be computed.

$$\frac{\tau_b}{\tau_{2V,b}} = \frac{\xi_a^{(3)}\xi_A^{(3)} V_m \varepsilon_m n_{<H>} \alpha_m v_{th,m}}{6a(\sqrt[3]{2}-1)\sqrt{\pi}}\frac{m_a}{(8/3)\pi a^2 \rho v_{th}} \tag{2.40}$$

$$= \frac{\alpha_m}{4(\sqrt[3]{2}-1)}\frac{v_{th,m}}{v_{th}}\frac{m_a V_m}{(4/3)\pi a^3}\frac{\varepsilon_m n_{<H>}}{\rho}. \tag{2.41}$$

By means of the relation

$$\frac{m_a}{m_m} = \frac{(4/3)\pi a^3}{V_m} \qquad (2.42)$$

and by the definition of the mass density of the molecule $\rho_m = n_m m_m$ one obtains

$$\tau_{2V,b}^{-1}\tau_b = \frac{\alpha_m}{4(\sqrt[3]{2}-1)} \frac{v_{th,m}}{v_{th}} \frac{\rho_m}{\rho} \qquad (2.43)$$

$$\simeq 1.03 \underbrace{\frac{\alpha_m}{1}}_{\leq 1} \underbrace{\sqrt{\frac{\mu}{m_m}}}_{\leq 1} \underbrace{\frac{\rho_m}{\rho}}_{\leq 0.01} \ll 1. \qquad (2.44)$$

Likewise one finds

$$\frac{\tau_B}{\tau_{2V,B}} = \frac{\xi_a^{(3)}\xi_A^{(3)} V_m \varepsilon_m n_{<H>} \alpha_m}{3a(\sqrt[3]{-1})} \cdot \frac{4\sqrt{F_{\text{rad}}}}{\sqrt{B}} \frac{m_a}{\sqrt{BF_{\text{rad}}}} \qquad (2.45)$$

$$\simeq 5.1 \underbrace{\frac{\alpha_m}{1}}_{\leq 1} \underbrace{\frac{\rho_m}{\rho}}_{\leq 0.01} \ll 1. \qquad (2.46)$$

Therefore in both cases, subsonic and supersonic drift motion, the acceleration timescale is always less than the growth timescale of the dust grains. The dust grains are able to reach their equilibrium drift velocity during all stages of their growth and may be assumed to be always drifting at this equilibrium speed. The only crucial point is the ratio of the gas density to the density of the material responsible for the growth of the particles. As indicated above, in C- and M- type giant abundances this quotient is always much smaller than unity. It may be different in chemically peculiar giants like R Coronae Borealis stars. There the ratio of the two densities is approximately unity and, especially in the case of supersonic drift, one cannot simply exclude the possibility of sub-equilibrium drift. However, if the reaction probability $\alpha_m$ is much smaller than unity, one may again draw the same conclusion.

## 2.4  Energy Conservation in the Wind

Apart from the conservation equations for mass and momentum in the wind, the third important conservation law is the conservation of energy which determines the temperature structure of the model. To calculate the temperatures of gas and dust, it is assumed that absorption and emission of radiation are the dominant heating and cooling terms. Therefore radiative equilibrium is assumed in the wind. Then, the radiation field can be calculated from the solution of the radiative transfer problem and the temperature at every site is given from the condition that energy gains and losses due to radiative processes are equal.

### 2.4.1 Lucy's Treatment of the Radiative Transfer Problem

An approximate treatment for the radiative transfer problem in an extended stellar atmosphere has been given by Lucy [Lucy 1971, Lucy 1976]. He gave a new closure procedure for the system of moment equations for the radiation field. The central idea is to split the total solid angle into two regions. One part is the region where the photosphere of the central star is seen, the other part is the rest of the solid angle where only radiation emitted or scattered by the extended atmosphere outside the photosphere is seen. Lucy assumes that the intensity seen from any point in the envelope in each angle part is a function of the distance from the star $r$ only and independent of the angle. This approximation of the intensity leads to a closing condition for the moment equations. If radiative equilibrium is assumed, the radiative transfer problem may be solved analytically and the equilibrium temperature is given by

$$T^4(r) = \frac{T_*^4}{2} \frac{\kappa_J}{\kappa_B} \left\{ 1 - \sqrt{1 - \frac{R_*^2}{r^2}} + \frac{3}{2}\tau_L \right\} \qquad (2.47)$$

$$\tau_L(r) = -\int_\infty^r \kappa_H \frac{R_*^2}{r^2} \qquad (2.48)$$

where $\kappa_J$, $\kappa_B$, and $\kappa_H$ are the frequency means of the intensity, weighted by the mean intensity $J_\nu$, the Planck function $B_\nu$, and the Eddington flux $H_\nu$, respectively. $R_*$ is the radius of the photosphere and $\tau_L$ is the diluted optical depth. This treatment of the radiative transfer is consistent only if

$$\tau_L(R_*) = \frac{2}{3}\frac{\kappa_B}{\kappa_J} \qquad (2.49)$$

which is a second condition to the first order ordinary differential Eq. (2.48) and therefore defines an eigenvalue problem for the complete set of equations which in this way allows to determine the effective temperature $T_* = T(R_*)$.

With the additional assumption of grey absorption of both gas and dust (this also implies equal dust and gas temperatures), i.e. $\kappa_J = \kappa_B$, Eq. (2.49) reduces to

$$\tau_L(R_*) = \frac{2}{3}. \qquad (2.50)$$

## 2.5 Gas Phase Chemistry

For the treatment of dust formation, it is of course essential to know the chemical composition of the gas phase, i.e. the concentrations of all relevant atoms and molecules. The general way of calculating the chemical composition is the solution of the system of rate equations for all chemically relevant species (atoms, molecules, ions, electrons). This means the solution of a set of ordinary differential equations with appropriate initial conditions. It has been done in an exemplary way for the chemistry of an oxygen rich shell around a late type giant (e.g. [Goeres et al. 1988]). However, the solution of this non linear stiff set of equations requires a considerable amount of computing effort. Furthermore, enormous problems are associated with the input data needed for the calculation of the reaction rates (data are uncertain if at all available). Therefore, these calculations have a more demonstrative character. Extended model calculations that include both complex hydrodynamic effects (like the models of dust forming winds studied in this work) and a non equilibrium chemistry quickly exceed the maximum capacity of nowadays supercomputers. In this case less expensive methods have to be applied.

Chemical equilibrium is one possible assumption. This is a state where every reaction is balanced by its own reverse reaction. The concentrations become independent of time (and, therefore, of previous conditions). All concentrations are functions of temperature $T$, density $n_{<H>}$, and the chemical abundances $\varepsilon_j$ of the elements only.

If the equation of state of the system is the ideal gas law, the assumption of chemical equilibrium leads to the so called mass action law:

Let $E^{(j)}, j = 1, \ldots, J$ denote the chemical elements and let $\overset{\circ}{p}_j$ be the partial pressure of the atomic kind $E^j$ in chemical equilibrium. Then, the partial pressure of a molecule $\mathcal{M}_\ell$ with the chemical formula $E^{(1)}_{\nu_{\ell,1}} E^{(2)}_{\nu_{\ell,2}} \ldots E^{(J)}_{\nu_{\ell,J}}$ is given by

$$\overset{\circ}{p}_{\mathcal{M}_\ell} = k_{\mathcal{M}_\ell} \prod_{j=1}^{J} \left(\overset{\circ}{p}_j\right)^{\nu_{\ell,j}}, \qquad (2.51)$$

where

$$k_{\mathcal{M}_\ell} = \exp\left\{-\frac{\Delta G_\ell}{kT}\right\} \qquad (2.52)$$

is the so called dissociation constant given by the difference $\Delta G_\ell$ of the Gibbs energy between the molecule and its constituents. $k_{\mathcal{M}_\ell}$ is a function of the temperature alone.

In a closed system without nuclear reactions, the total number of atoms of every element is conserved. This can be expressed by

$$\sum_{\ell=1}^{L} \nu_{\ell,j} \overset{\circ}{p}_{\mathcal{M}_\ell} = \sum_{\ell=1}^{L} \nu_{\ell,j} k_{\mathcal{M}_\ell} \prod_{j=1}^{J} \left(\overset{\circ}{p}_j\right)^{\nu_{\ell,j}} = \varepsilon_j n_{<H>} kT \qquad (2.53)$$

where $\varepsilon_j$ is the abundance of the element $E^{(j)}$ relative to hydrogen.

Eq. (2.53) is a set of $J$ equations for the partial pressures of the atoms of the $J$ elements. It can be solved for a given $T$ and $n_{<H>}$. An appropriate method for the solutions of this system is the Newton-Raphson iteration [Gail and Sedlmayr 1987a], [Tsuji 1973]. With the knowledge of the partial pressures of the elements one can obtain the partial pressures of the molecules by means of Eq. (2.51).

Since the solution of Eq. (2.53) still requires a considerable amount of computer time, it is advantageous to look for further simplifications. In a first approach it suffices to consider only the molecules having large concentrations. Neglecting the less important molecules does not significantly affect the solution[8].

An important simplification is guided by the fact, that the important element abundances and molecule concentrations differ by orders of magnitude. With a sound experience about the concentrations one expects it is possible to decouple the equations and solve them analytically without any time consuming iteration.

The necessary basis for decoupling the equations can be obtained from [Gail and Sedlmayr 1987a]. In this paper the authors have developed and discussed an equilibrium chemistry consisting of the elements $H$, $O$, $C$, $N$, $Fe$, $Si$, $Mg$, $S$, and $Al$ which are the most abundant elements in typical mixtures in astrophysics [Lang 1974]. Only the noble gases $He$, and $Ar$ have been neglected since they are chemically inactive.

In stars of population $I$ the element abundances are arranged in groups. Elements in each group have comparable abundances, while the groups are separated by about one order of magnitude in abundance (with the exception of nitrogen). These groups are (in brackets, the logarithm of the abundance relative to hydrogen is given (solar abundances after [Allen 1973])):

1. $H(0)$

2. $He(-1.00)$

3. $O(-3.18)$, $C(-3.48)$, $N(-4.04)$

4. $Fe(-4.40)$, $Si(-4.48)$, $Mg(-4.58)$, $S(-4.80)$

5. $Al(-5.39)$

---

[8]This is different in a non equilibrium chemistry, since elements and molecules with low abundances may have a large influence on the solution being transition states and catalysts in important reaction path ways.

For the simplification of the chemistry it is important to distinguish the different cases for the abundance ratio of carbon to oxygen. Due to the extraordinary strong bond in the $CO$ molecule ($11.2eV$), it is one of the first molecules formed after which it cannot be destroyed (the only important exception is the destruction by UV photons from the interstellar UV radiation field in the outer optically thin regions of the envelope). This has important consequences for the chemistry. In carbon stars (in which carbon is more abundant than oxygen), almost all oxygen is contained in the $CO$ molecule and therefore chemically blocked. Since the same is true for nitrogen (blocking itself by being bound in the $N_2$ molecule), the remaining carbon is by far the most abundant element (after hydrogen, of course). Therefore the chemistry in carbon stars is dominated by hydrocarbons and other carbon bearing species and is almost free of oxygen. In M-type stars (in which oxygen is more abundant than carbon), all carbon is contained in $CO$, and the chemistry is dominated by water vapor and molecules containing less abundant elements like $Mg$, $Si$, and $Fe$.

In [Gail and Sedlmayr 1987a] an equilibrium chemistry containing 141 diatomic and polyatomic molecules has been studied and the most important molecules of every element have been determined for both the carbon-rich and the oxygen-rich case. These molecules are

- *in the oxygen-rich case*
  $H_2, N_2, CO, OH, H_2O, SiH, SiO, H_2S, HS, SO, SiS, MgH, MgO, MgOH, MgS, FeO, Fe(OH)_2, AlOH, AlO_2H, Al_2O, AlH$

- *in the carbon-rich case*
  $H_2, N_2, CO, C_2, C_2H, C_2H2, CH_4, HCN, MgH, SiS, CS, SiC_2, Si_2C, HS, H_2S, FeS, AlH$.

Due to the very large differences in the abundances of the elements and in the concentrations of molecules it is now (as an example) possible to solve the dissociation equilibrium between the $H$ atom and the $H_2$ molecule separately (i.e. only one equation of type (2.53)) since the amounts of hydrogen being incorporated in molecules other than $H_2$ is always negligible compared to the hydrogen in $H$ and $H_2$. Having solved this equilibrium, the partial pressure of atomic hydrogen is known and the dissociation equilibrium of another, less abundant element including molecules containing hydrogen can be solved with the given value for the partial pressure of hydrogen[9].

---

[9]Of course this type of calculation violates the conservation of mass in the chemical system since the amount of hydrogen contained in less abundant molecules was neglected in the element balance of hydrogen. However, the errors remain small if the equations for the elements are solved in the order of decreasing element abundances. The errors could be further reduced by a second and third iteration of the system of chemical equations where the amounts of material contained in molecules of low concentration can be taken into account.

Using the results of [Gail and Sedlmayr 1987a] the following hierarchy of dissociation equilibria has been developed:

| The oxygen-rich case | The carbon-rich case |
|---|---|
| 1. Equilibrium for $H$, $H_2$. ($\Rightarrow \mathring{p}_H$) | 1. Equilibrium for $H$, $H_2$. ($\Rightarrow \mathring{p}_H$) |
| 2. Equilibrium for $N$, $N_2$. ($\Rightarrow \mathring{p}_N$) | 2. Equilibrium for $N$, $N_2$. ($\Rightarrow \mathring{p}_N$) |
| 3. All $C$ in $CO$ | 3. All $O$ in $CO$ |
| 4. Equilibrium for $O$, $OH$, $H_2O$ ($\Rightarrow \mathring{p}_O$) | 4. Equilibrium for $C$, $C_2$, $C_2H$, $C_2H_2$, $CH_4$, $HCN$ ($\Rightarrow \mathring{p}_C$) |
| 5. Equilibrium for $Si$, $SiH$, $SiO$, $S$, $H_2S$, $HS$, $SO$, $SiS$ ($\Rightarrow \mathring{p}_{Si}, \mathring{p}_S$) | 5. Equilibrium for $Si$, $S$, $SiS$, $CS$, $SiC_2$, $Si_2C$, $HS$, $H_2S$ ($\Rightarrow \mathring{p}_{Si}, \mathring{p}_S$) |
| 6. Equilibrium for $Mg$, $MgH$, $MgO$, $MgOH$, $MgS$ ($\Rightarrow \mathring{p}_{Mg}$) | 6. Equilibrium for $Mg$, $MgH$ ($\Rightarrow \mathring{p}_{Mg}$) |
| 7. Equilibrium for $Fe$, $FeO$, $Fe(OH)_2$ ($\Rightarrow \mathring{p}_{Fe}$) | 7. Equilibrium for $Fe$, $FeS$ ($\Rightarrow \mathring{p}_{Fe}$) |
| 8. Equilibrium for $Al$, $AlOH$, $AlOH_2$, $Al_2O$ ($\Rightarrow \mathring{p}_{Al}$) | 8. Equilibrium for $Al$, $AlH$ ($\Rightarrow \mathring{p}_{Al}$) |

This hierarchy of equations leads to quadratic equations for the partial pressures of the atoms. The only exception here is the calculation of the silicate-sulfur equilibrium in the carbon-rich case, where a cubic equation has to be solved. This is due to the fact that molecule $Si_2C$ contains two silicon atoms. Therefore Eq. (2.53) leads to a quadratic equation for the partial pressure of silicon. If solved simultaneously with the equation for sulfur a cubic equation results. All equations can be solved analytically[10].

The solution can be improved by a fix point iteration: In the next step, the calculated amounts of the most abundant elements contained in molecules of lower abundance can be taken into account. However, the experience with this algorithm shows, that for a large range in temperature ($400K \leq T \leq 2500K$) and in the hydrogen density ($1cm^{-3} \leq n_{<H>} \leq 10^{15}cm^{-3}$) the results without iteration are consistent with the results of the much larger chemistry of [Gail and Sedlmayr 1987a] to within a factor of about two. The advantage of this kind of code is not only the fast execution time. Since no iteration is required, it is furthermore perfectly designed for vectorization on computer facilities.

---

[10]Since the quadratic equations of the form $p^2 + ap + b = 0$ always have two solutions $p_{1/2} = -a/2 \pm \sqrt{a^2/4 - b}$ it is important to know, which of the solution is the correct one. It is quite simple to show, that in our case, always the larger solution $p_+$ (that one with the "+" sign) is the physical one. If $a$ is positive, the solution $p_+$ is thus calculated as a difference. Since the the partial pressures of the atoms become very small at low temperatures, the solution $p_+$ becomes a small difference of two large numbers; a fact that might lead to large numerical errors. These errors can be avoided if first the smaller solution $p_-$ is calculated by a sum. Then, the larger solution can be obtained by the application of the theorem of Vieta: $p_+ = b/p_-$ without numerical risks.

For the calculation of the dissociation constants analytical approximations have been used for the thermodynamic data of the molecules given in [Stull and Prophet 1971]. I received these fits from Dr. H.-P. Gail (Heidelberg).

### 2.5.1 Applicability of an Equilibrium Chemistry

A simple equilibrium chemistry may only be applied to a model calculation, if the expected deviations from this equilibrium have no crucial consequences for the results. Large non-equilibrium effects in red giants have to be expected, if either a rapid expansion of the medium leads to a frozen chemistry or if UV radiation stronger than that of the usual photospheric radiation field (e.g. the interstellar UV radiation field or radiation originating from a chromosphere) are present ([Goeres et al. 1988, Clegg et al. 1983]) .

It is well known that late type giants later than about $M5$ do not have active chromospheres. Therefore the UV radiation is only the photospheric field in the inner region of the envelope. Since the stars considered in this work are characterized by effective dust formation, the interstellar UV radiation field will be effectively absorbed in the outer regions of the envelope and will not affect the inner regions, where dust formation takes place.

The velocity fields are especially important in the outer regions of the envelope where the densities are rather small and the chemical timescale exceeds the expansion timescale.

Therefore an equilibrium chemistry seems to be acceptable in our context.

## 2.6 Dust Chemistry

It has already been discussed that the chemical composition of the gas phase depends critically on the carbon to oxygen ratio. This has considerable impact on the dust condensing in the atmosphere.

### 2.6.1 Dust in C-Stars

There is hardly any debate about the chemical composition of the dust in the envelopes of carbon giants. Due to the overabundance of carbon relative to oxygen, the chemistry in the envelope is dominated by carbon bearing molecules. Furthermore carbon has a solid phase that is stable at high temperatures (about $1700K$). Therefore it seems to be clear that the dust essentially consists of carbon with either graphite or amorphous structure. Some features in the infrared spectra of these objects point to the existence of a few other dust species like $SiC$ [Treffers and Cohen 1974] and, possibly, $MgS$ [Goebel and Moseley 1985]. However, due to the low abundances of both $Si$ and $Mg$, these species are probably more impurities in essentially pure carbon grains and they will be neglected in the model calculations in Chap. 3. Only the condensation of pure carbon grains is considered.

Since only the carbon bearing molecules are of interest for the dust formation, a further reduction of the gas phase chemistry is possible: for the model calculations in Chap. 3 only the steps 1, 3, and 4 of the hierarchy of equations on page 47 have been solved. The equilibrium for $N$ was not solved and the molecular species $HCN$ has been neglected. This changes the densities of the other carbon bearing molecules by less than 5% compared to the values obtained if $N$ is included.

The nucleation process is believed to follow a path through polyaromatic hydrocarbons [Keller 1987, Frenklach and Feigelson 1989, Gail and Sedlmayr 1987a]. Unfortunately, no quantitative theory was available, when these model calculations were performed. Therefore the nucleation rates are calculated from classical nucleation theory according to the equations given in [Gail et al. 1984]. The surface tension of the dust material was assumed to be $1400 erg/cm^2$.

The relevant growth reactions, the assigned reaction probabilities, and the changes in volume are [Gail and Sedlmayr 1988]:

| Reaction | $\alpha$ | $V_\iota [cm^3]$ |
|---|---|---|
| $C_N + C_2H_2 \to C_{N+2} + H_2$ | 0.34 | $1.76 \cdot 10^{-23}$ |
| $C_N + C_2H \to C_{N+2} + H$ | 0.34 | $1.76 \cdot 10^{-23}$ |

The changes in volume have simply been calculated from the monomer volume of a carbon atom in graphite which is given by $V_{mono} = 8.78 \cdot 10^{-24} cm^3$ [Weast et al. 1988]. An equilibrium chemistry in the gas phase, equal gas and dust temperatures, and negligible drift velocities are assumed. Then the generalized supersaturation ratios are simply given by

$$S_\iota = S_0^{n_\iota} \qquad (2.54)$$

[Gail and Sedlmayr 1988] where $n_\iota$ is the number of carbon atoms in the molecule responsible for the growth reaction $\iota$ and $S_0$ is the supersaturation ratio of the carbon atom given by

$$S_0 = \frac{p_{C_1}}{\overset{\circ}{p}_{C_1}}. \qquad (2.55)$$

The vapor pressure $\overset{\circ}{p}_{C_1}$ of the carbon atom can be calculated from the thermodynamical data of the atom and the bulk material given in [Stull and Prophet 1971].

| Name | Formula | Mol.wt [a.u.] | Density [$gcm^{-3}$] | Melting point [$°K$] | Boiling point [$°K$] |
|---|---|---|---|---|---|
| pure elements | | | | | |
| Aluminium | $Al$ | 26.98 | 2.70 | 933 | 2740 |
| Diamond | $C$ | 12.01 | 3.51 | tr Graphite | ...... |
| Graphite | $C$ | 12.01 | 2.25 | s 3925 vac 1773 | ...... |
| Iron | $Fe$ | 55.85 | 7.86 | 1808 | 3023 |
| Magnesium | $Mg$ | 24.31 | 1.74 | 921 | 1380 |
| Silicon | $Si$ | 28.09 | 2.32-2.34 | 1683 | 2628 |
| carbide | | | | | |
| Al carbide | $Al_4C_3$ | 143.96 | 2.36 | stab. to 1673 | d 2473 |
| Fe carbide | $Fe_3C$ | 179.55 | 7.69 | 2110 | ...... |
| Si carbide | $SiC$ | 40.10 | 3.22 | s ~2973 | ...... |
| oxide | | | | | |
| Corundum | $Al_2O_3$ | 101.96 | 3.97 | 2288 | 3253 |
| Wuestite | $FeO$ | 71.85 | 5.70 | 1642 | ...... |
| Hermatitie | $Fe_2O_3$ | 159.69 | 5.24 | 1838 | ...... |
| Magnetite | $Fe_3O_4$ | 231.54 | 5.18 | 1867 | ...... |
| Spine | $MgAl_2O_4$ | 142.27 | 3.60 | 2408 | ...... |
| Ferrite | $MgFe_2O_4$ | 200.00 | 4.44-4.60 | 2023 | ...... |
| Periclase | $MgO$ | 40.30 | 3.58 | 3125 | 3873 |
| silicate, silicide | | | | | |
| Quarz etc. | $SiO_2$ | 60.08 | 2.64-2.66 | 1883 | 2503 (2863) |
| Opal | $SiO_2 \cdot xH_2O$ | ...... | 2.17-2.20 | >1873 | ...... |
| Si monoxide | $SiO$ | 44.08 | 2.13 | >1975 | 2153 |
| Gruenerite | $FeSiO_3$ | 131.93 | 3.50 | 1419 | ...... |
| Fayalite | $Fe_2SiO_4$ | 203.78 | 4.34 | (?) 1776 | ...... |
| Clinoenstatite | $MgSiO_3$ | 100.39 | 3.19 | d 1830 | ...... |
| Forsterite | $Mg_2SiO_4$ | 140.69 | 3.21 | 2183 | ...... |
| sulfide, sulfate | | | | | |
| Al sulfide | $Al_2S_3$ | 150.14 | 2.02 | 1373 | s 1773 ($N_2$) |
| Pyrite | $FeS_2$ | 119.97 | 5.00 | 1444 | ...... |
| Mg sulfate | $MgSO_4$ | 120.36 | 2.66 | d 1397 | ...... |
| Mg sulfide | $MgS$ | 56.37 | 2.84 | d >2273 | ...... |
| Si sulfide | $SiS_2$ | 92.21 | 2.02 | s 1363 | white heat |
| Si m.sulfide | $SiS$ | 60.15 | 1.85 | s 1213 | ...... |
| others | | | | | |
| Cyanite | $Al_2O_3 \cdot SiO_2$ | 162.05 | 3.25 | 1818 tr Mullite | >1818 |
| Mullite | $3Al_2O_3 \cdot 2SiO_2$ | 426.05 | 3.16 | 2193 | ...... |
| Mg silicide | $Mg_2Si$ | 76.70 | 1.94 | 1375 | ...... |

Table 2.1:

High temperature condensates consisting of the elements $C$, $O$, $S$, $Mg$, $Al$, $Si$ and $Fe$. s (sublimes), d (decomposes), tr to (transformes to), vac (in vacuum). Source: [Weast *et al.* 1988]

A reasonable fit to this data is taken from [Gail and Sedlmayr 1988]:

$$\log \overset{\circ}{p}_{C_1} = -\frac{86300}{T} + 32.89. \qquad (2.56)$$

### 2.6.2 Dust in M-Stars

The situation is a bit more confusing in M-stars. Since all carbon is locked in $CO$, the most abundant element is oxygen which has no solid phase at the conditions in the circumstellar shell. As it has been discussed by [Gail and Sedlmayr 1986], less abundant elements have to be incorporated into the grains. The most abundant elements which have stable phases above $500K$ have to be selected. A look on Table 2.1 shows that there is a large number of different condensates which fulfill these conditions.

The elements in question are $Fe$, $Si$, $Mg$, and $Al$. Though $S$ in the laboratory may form several high temperature condensates, it does not seem to be incorporated into grains in the interstellar medium [Jenkins 1987]. This is probably due to the fact that oxygen nearly always is present in astrophysical environments. An examination of the equilibrium constants of several molecules bearing either oxygen or sulfur shows that the oxygen bearing molecule usually is more stable. Since $O$ is much more abundant in astrophysical situations, it will replace the sulfur in most molecules and condensates (an exception to this rule is possible in carbon rich environments where all oxygen is blocked in $CO$). Therefore $S$ is neglected in the calculations in Chap. 4. Furthermore the element $Al$ will be neglected, since it is much less abundant than $Si$, $Mg$, and $Fe$.

In Chap. 4 a test calculation is presented where the formalism for heterogeneous dust formation presented in Chap. 1 is applied to the formation of grains consisting of a mixture of $Si$, $Mg$, $Fe$, and $O$.

#### The Equations for the Dust

For every dust species condensing in the envelope, a complete set of the Eqs. (1.39) has to be solved in the model calculation.

## 2.7 Absorption Coefficients for Gas and Dust

For the absorption coefficients of the gas, tables provided by M. Scholz have been used. These tables provide the *Rosseland mean* of the absorption coefficient as a function of gas pressure and temperature. The absorbers considered are described in [Scholz and Tsuji 1984].

For the solution of the radiative transfer problem in the approximation given by [Lucy 1971, Lucy 1976], the *flux mean* of the absorption coefficient is required. For the calculation of this mean the Eddington flux of the radiation field has to be known (which is only the case if the radiative transfer is explicitly solved). An appropriate approximation would be the *Planck mean*, but unfortunately a table providing this data was not available. Therefore the available *Rosseland mean* had to be used[11].

The absorption coefficients for the dust generally can be calculated from the *Mie* theory. In the calculations presented here the *small particle limit* of the Mie theory has been used which is applicable if the particle radii are small compared to the wavelength of the light divided by $2\pi$. If this condition is valid, the absorption cross section of the particle is proportional to its volume.

Since the total volume occupied by grains per unit gas volume is given by $\mathcal{L}_3$, the absorption coefficient may be written as

$$\kappa_{\text{abs}}(\lambda) = \frac{6\pi}{\lambda} \text{Im}\left\{\frac{m^2-1}{m^2+2}\right\} \mathcal{L}_3 \qquad (2.57)$$

(c.f. [Wickramasinghe 1972]). The scattering coefficient in this approximation always is small compared to the absorption coefficient and may be neglected.

The Planck mean for the absorption coefficient usually is fitted by a power law of the form [Lucy 1976]

$$\kappa_{\text{abs}} = \frac{3}{4}\mathcal{L}_3 Q_0 T^\beta \qquad (2.58)$$

where $Q_0$ and $\beta$ have to be fitted to measurements of the optical properties of the material. For the calculations in Chap. 3 the values for amorphous carbon given by [Gail and Sedlmayr 1985] have been used:

$$\begin{aligned} Q_0 &= 6.7 \\ \beta &= 1.0. \end{aligned} \qquad (2.59)$$

---

[11]This point is being worked out in another PhD thesis in our group: [Winters 1992].

## 2.8 Parameters, Boundary Conditions and Numerical Solution

The set of equations is dependent on the following stellar parameters:

| | |
|---|---|
| stellar mass | : $M_*$ |
| luminosity | : $L_*$ |
| effective temperature | : $T_*$ |
| element abundances | : $\varepsilon_i$ |

Except for the carbon abundance the solar element abundances [Allen 1973] are assumed. Therefore, the relevant abundance parameter for the calculations is the *overabundance of carbon relative to oxygen* $\varepsilon_C/\varepsilon_O$.

There is one further parameter which is present in the equations: The mass loss rate $\dot{M}$. However, it will now be shown that the system of equations contains one excess equation that allows to eliminate $\dot{M}$.

The equations one has to solve in order to compute a stationary model of dust forming winds are ordinary differential equations. They are subject to the following boundary conditions:

1. At the inner boundary which is defined to be the stellar photosphere, no dust is present and, therefore, all moments of the distribution function vanish:

$$\mathcal{L}_j = 0 \qquad j = 0, 1, \ldots \tag{2.60}$$

   If the radiative transfer problem is solved by Lucy's approximation, at the photosphere the condition

$$\tau_L(R_*) = \frac{2}{3} \tag{2.61}$$

   must hold.

2. At the sonic point, where the hydrodynamic velocity $v$ equals the isothermal speed of sound $c_s$, the equation of motion is subject to the condition

$$\frac{2c_s^2}{r_s} - \frac{dc_s^2}{dr_s} - g(r_s) + g_{\rm rad}(r_s) = 0. \tag{2.62}$$

3. At infinity the diluted optical depth $\tau_L$ must vanish:

$$\tau_L(\infty) = 0. \tag{2.63}$$

54                    CHAPTER 2.  A MODEL OF DUST FORMING WINDSIt is an important property of the set of equations, that the equation for $\tau_L$ has two boundary conditions. Therefore the effective temperature is not a free parameter of the set of equations but a result of the solution of the radiative transfer problem. One may, however, turn the tables, fix the effective temperature and make another parameter dependent on the others, for example the mass loss rate. Thereby one obtains a formulation of the system allowing for the self-consistent determination of the mass loss rate for given stellar parameters $L_*$, $T_*$, and $M_*$ (besides the element abundances). The fruits of this step are described in Sect. 3.2.

Since the boundary conditions are given at different sites, the set of equations cannot be solved simply by an outward integration. Such problems usually are essentially treated by two different methods.

1. *Multiple shooting method.*
   The application of this method to the wind problem has been described in detail by [Gail and Sedlmayr 1985, Gail and Sedlmayr 1987b]. Here the problem of the boundary conditions is solved with a shooting method for as many parameters as boundary conditions are defined outside the inner boundary of the model. In [Gail and Sedlmayr 1987b] the hydrodynamic velocity $v(R_*)$ was iterated in order to meet the condition for the equation of motion at the sonic point and the luminosity of the star was iterated in order to meet the outer boundary of the diluted optical depth. Here this method is used for the model of an M-star. In this model due to the experimental stage of the description of grain growth and composition, a method is required where changes in the model may easily be incorporated. Since the multiple shooting method only requires a forward integration of ordinary differential equations, it is perfectly designed for this work. For these calculations an improved version of the program of Dr. H.-P. Gail has been used. In this code the formulation of the theory of dust formation in the form described in Chap. 1 and the chemistry discussed in Sect. 2.5 has been implemented.

2. *Implicit code.*
   The application of this code to the wind problem has been described in detail in [Dominik 1987]. It uses a Henyey-type method in order to solve the set of equations. The differential equations are not integrated forward but the discretized equations at every point are solved by a Newton-Raphson method. For two reasons this code is not designed for the experimental stage of model construction: Firstly it is necessary to have a complete starting model which is not too far from the final solution. And, secondly, change or addition of any equation requires some effort since all derivatives have to be evaluated and a new starting model has to be produced. However, this type of code is very fast and stable. Therefore it is adequate if a large number of models have to be calculated for a fixed input of physics. An implicit code has been developed and used for the self-consistent models of dust driven winds around carbon stars discussed in Chap. 3.

# Chapter 3

# Dust Driven Winds Around Carbon Stars

The complete solution of the system equations described in the last section invokes the concept of a *dust driven wind*, i.e. a wind solution where the radiation pressure on dust grains is the dominant hydrodynamic force responsible for the large scale motion of the material.

Models including radiation pressure on dust grains have previously been constructed by other authors. A list not claiming completeness includes [Salpeter 1974], [Kwok 1975], [Lucy 1976], [Menietti and Fix 1978], [Deguchi 1980], [Tielens 1983], [Berruyer and Frisch 1983], [Kozasa *et al.* 1984], and [Gail and Sedlmayr 1987b]. However, in most works the process of dust formation was treated only schematicly.

Besides being the first detailed and time dependent treatment of dust formation coupled with the solution of hydrodynamic equations, [Gail and Sedlmayr 1987b] was a step forward in so far as it allows for a self-consistent calculation of the mass loss rate. In this chapter the detailed structure of such a model will be discussed. The description has been enhanced by including the discussion of the grain size distribution which allows a more accurate monitoring of the dust formation process. In the second part of the chapter, some results of the calculation of an extended model grid are discussed.

## 3.1 The Structure of a Typical Model

In this section the structure of a typical model of a dust-driven wind around a carbon star will be discussed. As discussed in the previous chapter only the formation of amorphous carbon is considered and the absorption coefficients are calculated by the small particle limit of the Mie theory in the approximation given in Eq.(2.58). The parameters of the model are

$$\begin{aligned} L_* &= 1.49 \cdot 10^4 L_\odot \\ T_* &= 2000 K \\ M_* &= 1 M_\odot \\ \varepsilon_C / \varepsilon_O &= 2 \end{aligned}$$

All other chemical abundances are assumed to be solar abundances taken from [Allen 1973]. A model with these parameters leads to a mass loss rate of

$$\dot{M} = 1.0 \cdot 10^{-5} M_\odot/yr \qquad (3.1)$$

which is a typical value for the observed mass loss rates in carbon giants (e.g. [Knapp et al. 1982]).

### 3.1.1  The Hydrodynamic Structure

Figure 3.1 shows the radial hydrodynamic structure of the wind. The wind solution starts at very low velocities (around $10 cm/s$). In this region, no dust is present and radiation pressure only plays a minor role. This can be seen from the quantity $\alpha$ which denotes the radiative acceleration in units of the gravitational acceleration at the same site. $\alpha$ is of the order of $10^{-4}$ near the photosphere and differs from zero only due to the absorption of the gas phase.

With increasing distance from the star, the temperature decreases. At a distance of about $1.4 R_*$, corresponding to a temperature of about $1500 K$, the onset of dust formation takes place. The absorption coefficient per unit mass almost jumps by 4 orders of magnitude, marking a rather sharp borderline of the dust shell. The radiative acceleration is increased by the same factor and the material is accelerated away from the star.

At a distance of about $1.6 R_*$, enough dust has formed for the radiative acceleration to exceed the gravitational acceleration. Shortly before that, the velocity of the wind exceeds the velocity of sound (the wind passes the sonic point). Since the process of dust formation has not jet stopped, the radiative acceleration still increases and the wind is further accelerated up to a asymptotic speed of $25.5 km/s$.

### 3.1.2  Dust Formation in the Wind

For the discussion of dust formation the shell is divided into three regions of different significance for dust formation:

|     |                     |                                |                              |
| --- | ------------------- | ------------------------------ | ---------------------------- |
| I.  | Inner region        | $R_* \leq r \leq 1.4 R_*$      | (dust free atmosphere)       |
| II. | Sonic point region  | $1.4 R_* \leq r \leq 2.0 R_*$  | (condensation zone)          |
| III.| Outer region        | $2.0 R_* \leq r$               | (dust shell)                 |

In Fig. 3.2 the radial course of the quantities characterizing the process of dust formation is shown. An important quantity is the final grain size $a_\infty$ a grain will reach at large distances from the star. It is of course a function of the site in the wind, where the grain originates.

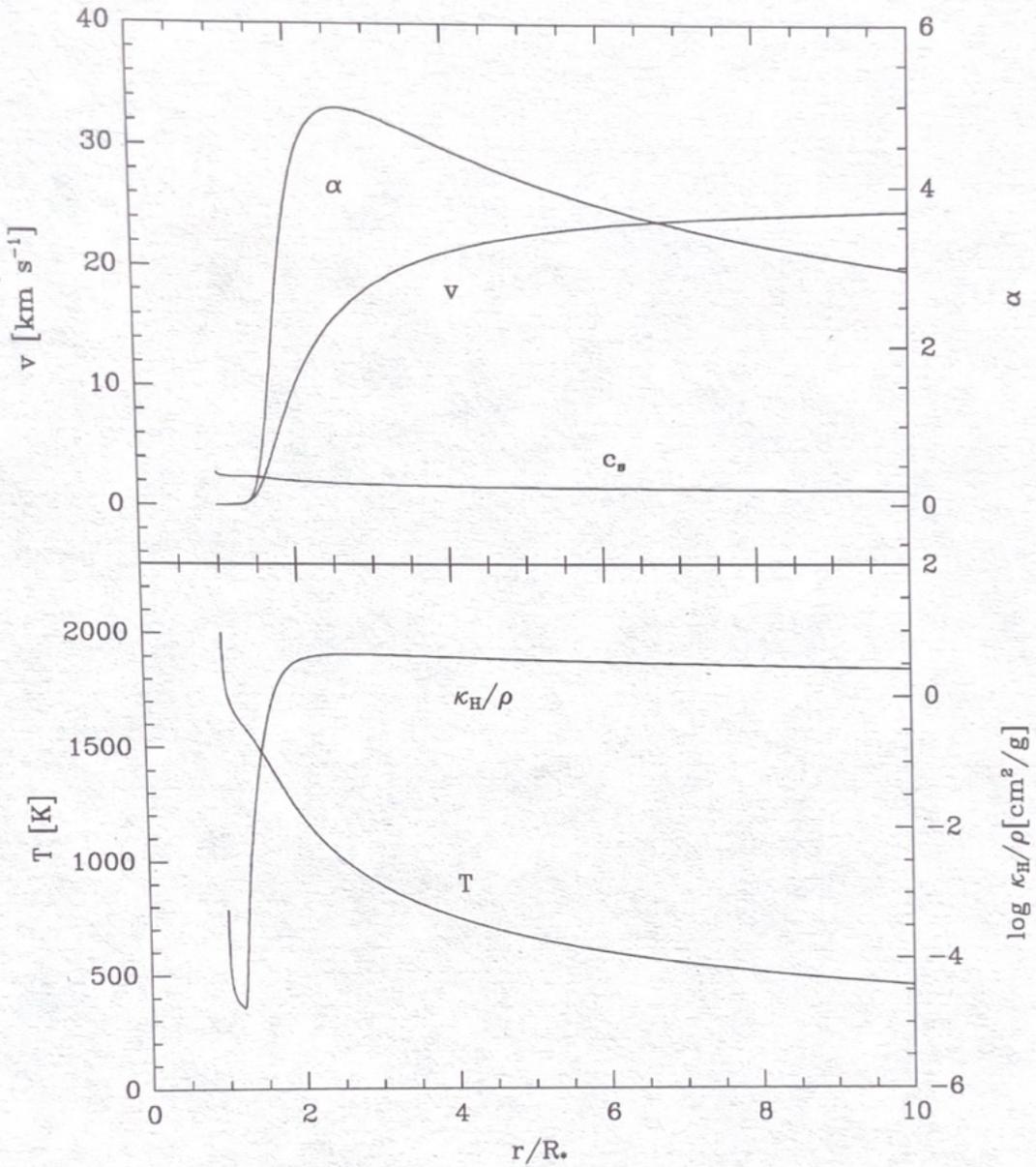

Figure 3.1:
Radial structure of the dust driven wind model. The plotted quantities are: hydrodynamic velocity $v$, speed of sound $c_s$, radiative acceleration in units of the gravitational acceleration $\alpha$, gas temperature $T$ (equal to the dust temperature), and absorption coefficient $\kappa_H$ of both gas and dust.

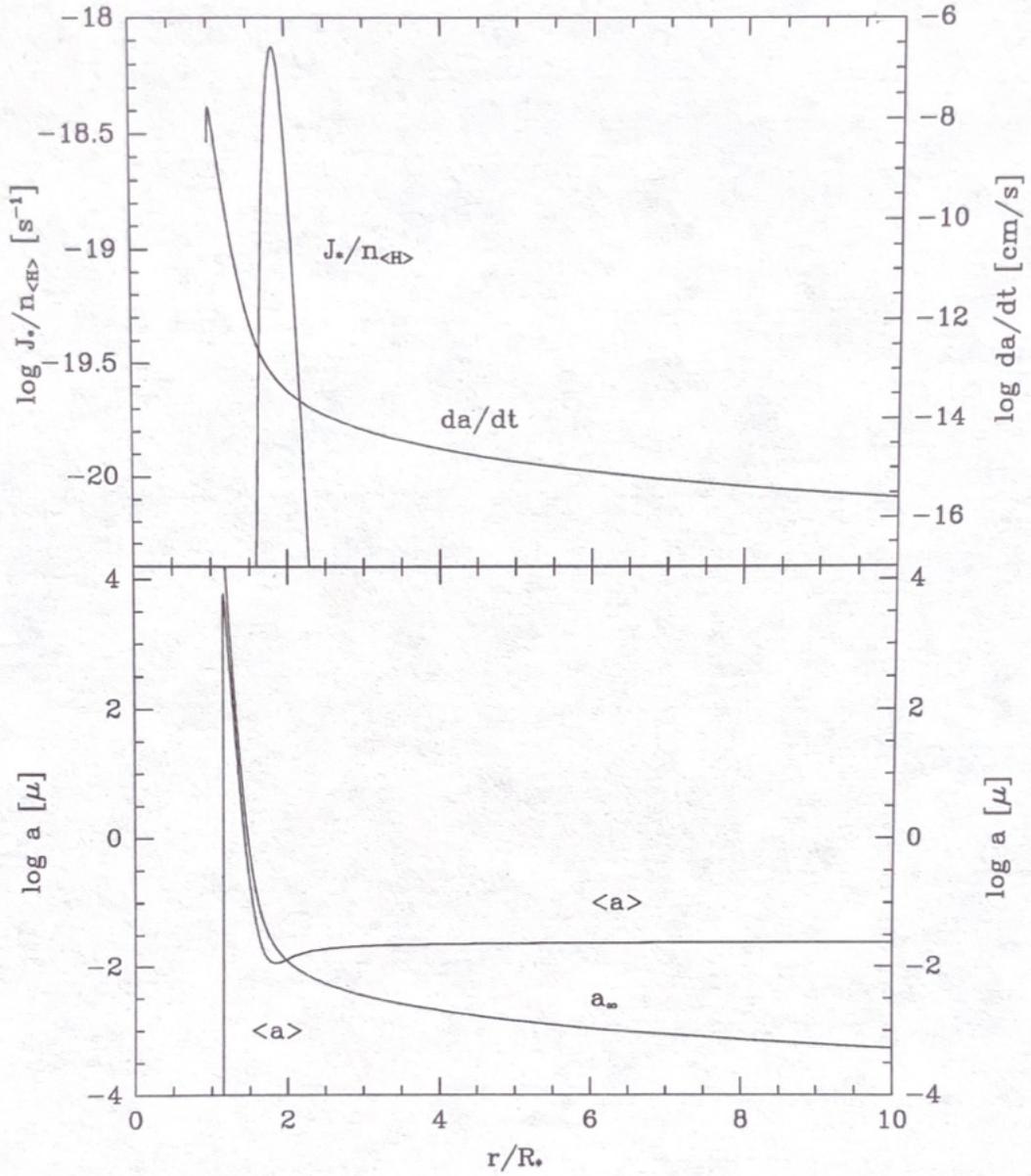

Figure 3.2:

Radial structure of the quantities characterizing the process of dust formation. The plotted quantities are: nucleation rate $J_*$ normalized to $n_{<H>}$, growth velocity of the particle radius $da/dt$, mean grain radius $<a>$, and final size $a_\infty$ of a dust particle originating at $r$.

### Inner Region

In the inner region only tiny amounts of dust are formed. The growth velocity of the particle radius is in the order of $10^{-10} cm/s$ (i.e. every 100 seconds the dust grains grow by one monomer layer). The hydrodynamic velocities are very small in that region. A dust grain stays there for about $10^9 s$ and consequently may reach a size of $100 \mu m$. Since the growth velocity is rapidly decreasing with increasing distance from the star, the additional growth in the outer regions of the shell is small and the final size $a_\infty$ is of the same order of magnitude. Hence, the growth of these particles is a merely local phenomenon. However, the nucleation rate in that region is very small ($\approx 10^{-30}$ nuclei per second and H-atom) and only a small amount of condensing material is contained in these grains.

The mean grain size in this region always is slightly larger than the final grain size of the particles originating from here. This is due to the larger particles formed at smaller $r$ that still dominate the distribution function.

The existence of these very large particles is a rather artificial consequence of two assumptions made in the model:

- Stationary Model.
  In fact, most late type giants are long period variable stars with a period of about $10^8 s$. This period should be an upper limit for the time a dust particle may grow under constant conditions. This reduces the maximum particle size by about one order of magnitude.

- Negligible drift motion.
  Though the drift in this region is small compared to the sound velocity, it can still be larger than the hydrodynamic velocity of the wind. This has two consequences: The drift increases the number of collisions per time interval between the grain and the gas atoms and it decreases the time the grain stays in the high density region near the photoshere. The competition between both effects leads to larger grains if the nuclei are formed outside the sonic point and to smaller grains if the nuclei are formed inside the sonic point (for a detailed discussion see [Dominik et al. 1989]). Since the very large grains are formed inside the sonic point, their size is overestimated.

### Sonic Point Region

With increasing $r$, $a_\infty$ is decreases rapidly and the growth process becomes less localized. There are two causes for the rapid decrease of $a_\infty$: (1) The growth velocity decelerates due to the lower densities. Near the sonic point the growth velocity is about $10^{-13} cm/s$. (2) The hydrodynamic velocity increases and, consequently, the time of residence in the region where the particle nuclei are formed and where the growth conditions are rather favorable, is small. Particles originating from the sonic point reach final sizes of $0.3\mu m$. Parallel to the decrease of $a_\infty$, the nucleation rate $\mathcal{J}_*$ dramatically increases. $\mathcal{J}_*$ peaks at $r = 1.8R_*$ with the value of $10^{-18.1}$ clusters per second and H-atom. Since $a_\infty(1.8R_*) = 0.03\mu m$ the maximum of the size distribution function has to be expected at that size.

Around the sonic point the relation between the mean and the final particle size is reversed. The large number of small particles produced by the peak of $\mathcal{J}_*$ now dominates the size distribution.

Though here the growth conditions are less favorable than at $r = 1.4R_*$, the massive production of particles with subsequent growth leads to a relevant consumption of carbon atoms in the gas. The dust grains effectively absorb the radiation from the star and the resulting acceleration pushes the wind through the sonic point.

**Outer Region**

In the outer region the nucleation rate rapidly decreases and only few new particles are formed. On the other hand the existing particles still grow which leads to a further increase of the radiative acceleration. When dust formation finally stops, nearly 50% of the condensable material in the gas phase (condensable here means all carbon except for that blocked in $CO$) is condensed into grains. Not all the material can condense since the large hydrodynamic velocity (more than $10km/s$) rapidly dilutes the material and collisions between the grains and the molecules become very rare.

For the same reasons, the ratio between mean and final particle size is again reversed and the mean particle size is larger than that of a grain originating in the outer envelope.

### 3.1.3  The Size Distribution

With Eq. (1.61) the size distribution of the dust particles at different sites in the shell can be calculated. In Fig. 3.3 the size distribution at $r = 1.4R_*$, $r = r_s = 1.6R_*$, $r = 4R_*$, and $r = 20R_*$ has been plotted. The final size distribution (equal to that at $20R_*$) is an almost perfect power law in the large size interval from about $0.01\mu m$ up to the largest grain size present. The index of the spectrum in this model is equal to $-4.7$:

$$f(a) \sim a^{-4.7}. \tag{3.2}$$

The other curves represent the evolution of the size distribution in the wind. As argued before, the growth of the large particles is merely a local phenomenon, due to the combined effect of large densities and large timescales inside the sonic point. Therefore, the large particle end of the size distribution is formed first and already reaches its final shape at $1.4R_*$. This seems to be rather surprising, since the growth of the particles can be interpreted as a translation of the size distribution function in the space of grain radii [Gauger et al. 1990b]. This also is true here of course, but cannot be seen due to the logarithmic scale of the $a$-axis. The effect of grain growth can be seen in the changes of the size distribution at $4R_*$ to that at $20R_*$. At $4R_*$, the distribution has reached its final shape for the particle sizes larger than about $0.01\mu m$. The smaller particles are more abundant here than in the final distribution. This is due to the fact that the nucleation rate already is negligible at $4R_*$. The only possible effect is the growth of already existing particles, which shifts the size distribution to larger sizes.

Nevertheless, since the growth conditions change rapidly in the wind one can argue that the size distribution in a stationary dust driven wind is built up from the large particle end.

### 3.1.4 The Power Law

The size distribution in a dust driven wind may be fitted by a power law over a large range in size. This feature is not limited to the model discussed. The same is true for models with a wide parameter range. According to Eq. (1.61) this requires a very specific relation between the nucleation rate $\mathcal{J}_*$ and the growth velocity $\chi$ as a function of the final size of a dust grain. It especially requires that the steep increase of $J_*$ and the decrease of $a_\infty$ occur simultaneously. One can distinguish between three different cases in the relation of $a_\infty$ and $J_*$:

(a) The nucleation rate reaches its maximum at still very large values of $a_\infty$. In this case only big particles are formed which probably consume all condensable material.

(b) The nucleation rate becomes large in the outer regions where $a_\infty$ already is small and does not change much. In this case, a narrow distribution function centered at small particle sizes will be the result.

(c) The increase of $J_*$ and the decrease of $a_\infty$ occur simultaneously.

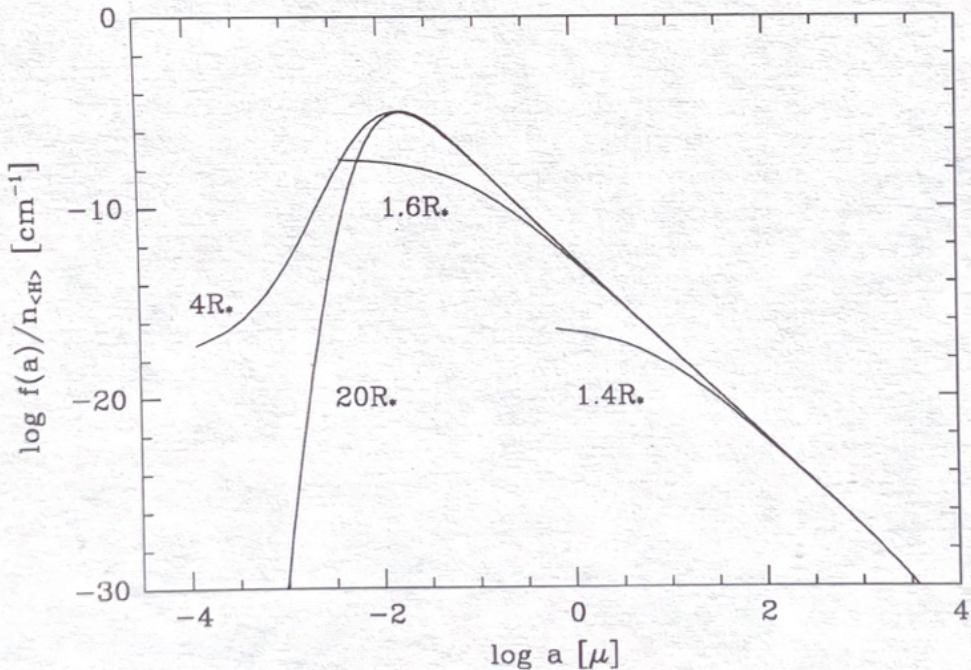

Figure 3.3:
The particle size distribution at four different sites in the wind. The logarithmic diagram shows that $f(a)$ follows a power law. The labels indicate the distance from the star.

Case (c) is the situation found in the dust driven wind models. In general, case (c) seems to be rather improbable since only slight changes in the position of the peek of $J_*$ would lead to either case (a) or (b). However, case (c) is the standard situation which can bee understood as follows:

**The Position of the Decrease of $a_\infty$ in the Shell**

As argued above, the growth of the particles to their final size is merely a local phenomenon due to the rapidly changing conditions. Therefore it is possible to estimate the final size of a particle from the local densities and timescales. Assuming that the concentrations of the chemical species are almost constant, the final size of the particle originating at the site $r$ in the wind is proportional to the growth velocity of the particle radius $da/dt$ and to the timescale $\Delta t$ in which this velocity changes. This timescale simply is

$$\Delta t \sim \left(\frac{\partial \ln \chi}{\partial t}\right)^{-1} = v^{-1}\left(\frac{\partial \ln \chi}{\partial r}\right)^{-1}. \qquad (3.3)$$

The growth velocity (compare Eq. (1.43)) is proportional to the densities at $r$, thus proportional to $(r^2v)^{-1}$. Therefore one finds

$$a_\infty \sim \frac{\Delta t}{r^2 v} = \frac{\left(\frac{\partial \ln \chi}{\partial r}\right)^{-1}}{r^2 v^2}. \qquad (3.4)$$

Almost everywhere in the shell the variations of both the radial distance $r$ and the scale height of $\chi$ are small compared to the changes in $v$. Therefore, the rapid decrease of $a_\infty$ is essentially due to the increase of $v$ at about the sonic point.

**The Position of the Increase of $J_*$**

A dust driven wind is a wind in which the radiation pressure on dust grains is responsible for the rapid acceleration of the wind material at about the subsonic to supersonic transition. Therefore this transition is shifted to the site at which the onset of dust formation takes place and the increase of $J_*$ (which is essentially a function of temperature), therefore necessarily occurs at the sonic point.

In this way the increase of $J_*$ and the decrease of the final grain size $a_\infty$ always occur simultaneously and stretch the size distribution to it's power law shape.

## 3.2 Dust Driven Mass Loss in the Hertzsprung-Russel-Diagram

It has already been discussed that the set of equations for the dust driven wind allows for a self-consistent determination of the mass loss rate as a function of the stellar parameters $M_*$, $L_*$, $T_*$, and the element abundances (in our case especially $\varepsilon_C/\varepsilon_O$ ). This feature of the model calculations is a far-reaching one: In calculations of stellar evolution it is often necessary to include mass loss rates for the star in certain evolutionary stages. For hot giants this is the case during every moment of their evolution. Low mass stars suffer heavy mass loss during their giant stages.

For the evolutionary calculations one usually depends on fit formulae to observed mass loss rates. A widely used formula is that of [Reimers 1978] which has been fitted to the observations of mass loss rates from G type giants. A formula fitted to a much wider set of observational data is that of [deJager et al. 1988]. However, all these formulae are not based on a physical mechanism driving the mass loss and, therefore, are not capable of treating the individuality of stars. Considerable improvement of the reliability of stellar evolutionary calculations including mass loss therefore requires a deeper understanding of the physics of the mass loss process itself and the determination of mass loss rates from model calculations of the stars considered.

As discussed in Sect. 2.1 radiation pressure on dust grains seems to have a decisive influence on the mass loss. Since other effects like the dissipation of waves tend to increase the mass loss rates [Gauger et al. 1990a], a lower limit for the mass loss rates to be expected in late type giants can be obtained from the study of wind models in which radiation pressure on gas and dust is (apart from the gas pressure gradient) the only driving force for the wind. The dependence of the mass loss rates calculated from these models on the stellar parameters will be discussed in the rest of this chapter.

### 3.2.1 The Mass Loss Rate

A dust driven wind solution does not exist for all possible parameters sets. First of all, the star has to be sufficiently cool and extended in order to allow for dust formation in its outermost layers. Furthermore, radiation pressure on dust grains can only be the dominating hydrodynamic force if both the luminosity of the star and the absorption coefficient of the dust (which is proportional to the fraction of condensable material that actually has been converted to grains) are high. Therefore, there exists a minimum star luminosity at about $10^3 L_\odot$ if one assumes complete condensation of all condensable material [Gail and Sedlmayr 1987c]. Since dust formation is effective only at high gas densities (which are proportional to the mass loss rate), there also exists a minimum mass loss rate for the wind to be dust driven [Gail and Sedlmayr 1987c]. If the condensation is not complete (due to low gas densities), the minimal luminosity is shifted to higher values.

In Fig. 3.4a a dividing line for purely dust driven winds in the HRD has been plotted which results from numerical calculations for a large series of models like the one discussed above (for more details see [Dominik et al. 1990]). It is located around $10^4 L_\odot$. Above this line purely dust driven winds are possible, below this line they are not. Note that the precise location of this line is dependent on the mass of the star and on the element abundances. In this plot a mass of $1 M_\odot$ and an overabundance of $\varepsilon_C/\varepsilon_O = 2$ have been used.

The self-consistently determined mass loss rates resulting from the model calculations are plotted in Fig. 3.4b as contour lines which are, of course, present only inside the allowed area. The minimum mass loss rate is about $10^{-7} M_\odot/yr$. Mass loss rates as high as $10^{-4} M_\odot/yr$ can be reached for sufficiently high luminosities. The value of the mass loss rate increases with increasing luminosity and decreasing effective temperature of the star. This is interesting for the investigation of stellar evolution of AGB stars: The bold line in Fig. 3.4b shows a schematic evolutionary track of such a star along the Hayashi line up to the tip of the AGB and at constant luminosity towards higher effective temperatures (e.g. [Schönberner 1981b]. Note that the effective temperature of the model is defined differently in our calculations than in stellar evolution calculations and, therefore, the low temperatures at the sketched track are not necessarily in contradiction to the usual evolutionary tracks). It can be seen that the mass loss rate will steadily increase as the evolution of the

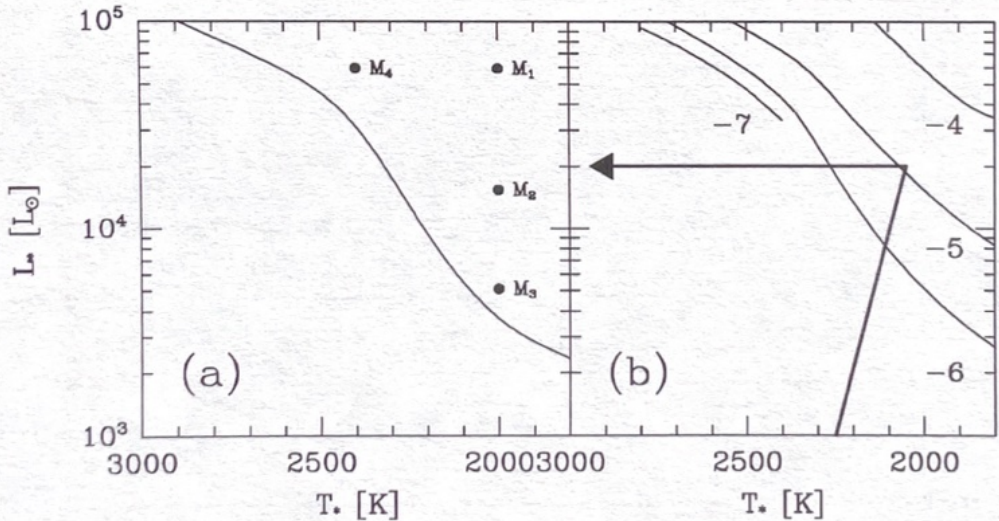

Figure 3.4:

HR-diagram for the dust driven winds around carbon stars. In the left diagram, the limiting line for dust driven winds is shown. The marks $M_1 \ldots M_4$ denote the positions of the four models used in the discussion on the dependence of the mass loss rate on the stellar mass and the chemical abundances. The right diagram shows contour lines of constant mass loss rate and a hypothetical evolutionary track of an AGB star.

star progresses until the star becomes a central star of a planetary nebula. This already suggests that dust driven winds might be an important mass loss mechanism during the transition from AGB stars to planetary nebulae.

The dependence of the mass loss rates on the overabundance of carbon relative to oxygen (i.e. on the abundance of the condensing material) is shown for four models in Fig. 3.5. In order to construct this diagram, four pairs of $(L_*, T_*)$, i.e. four points in the HR-diagram have been chosen. These points are marked in Fig. 3.4b and have been chosen in order to cover the area in the HRD where dust driven winds are possible. For these $(L_*, T_*)$ the overabundance of carbon relative to oxygen has been varied in a parameter range from 1 to 4. For values of $\varepsilon_C/\varepsilon_O$ very close to unity, almost all carbon is locked up in CO and only little carbon dust may condense which does not suffice to drive a wind. However, if $\varepsilon_C/\varepsilon_O$ exceeds a value of about 1.4, the mass loss rate hardly depends on the precise value of $\varepsilon_C/\varepsilon_O$ any more. Thus, the dust driven mass loss rate does not depend critically on the amount of dust formed in the envelope. If the radiative acceleration becomes sufficiently large to drive a wind, the mass loss rate is essentially independent of the chemical abundances.
The main effect of the large overabundances is an increase of the terminal velocity of the model: Though at the sonic point an almost fixed amount of dust is required for the subsonic-supersonic transition, a high carbon abundance increases the amount of dust formed *outside* the sonic point. Here a larger radiative acceleration cannot increase the mass loss rate but will further accelerate the wind. The model calculations show an almost linear relation between the terminal velocity and the initial amount of condensable material [Dominik *et al.* 1990].

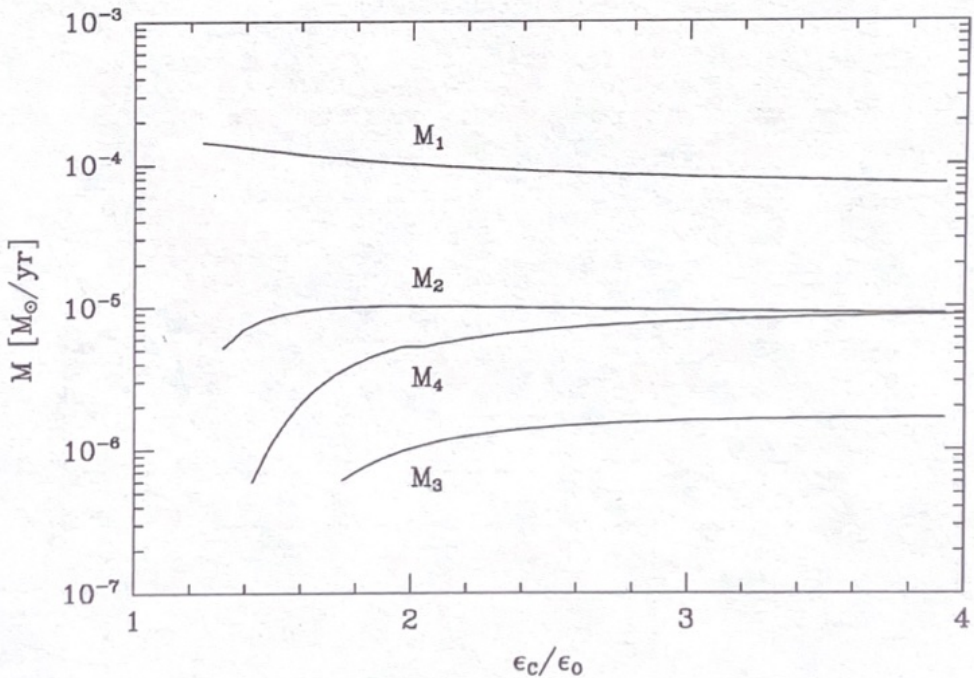

Figure 3.5:
Dependence of the mass loss rate on the overabundance of carbon relative to oxygen for the four models marked in Fig. 3.4a.

Figure 3.6 shows the dependence of the mass loss rate on the stellar mass. Here the mass of the models has been varied from a lower boundary of $0.5 M_\odot$ (which seems to be a good value of the mass of the white dwarf inside the red giant [Schönberner 1981a]) up to the highest value where a dust driven wind solution is possible. At this mass, the dividing line in the HRD meets the respective point $(L_*, T_*)$. An exception has been made for the model $M_1$. Here the calculations have been stopped at $M_* = 3 M_\odot$, although the limit is higher. From the figure it can be obtained directly that $\dot{M}$ increases dramatically with decreasing mass. Reducing $M_*$ to 50% of is original value may increase the mass loss rate by a factor of 3 to 100!

Until now it is not possible to follow the evolution of an AGB star with a treatment of dust driven mass loss, mainly due to the problems of the stellar evolution codes with the complicated processes in the interior of these stars. However, the results described above give rise to the following estimate: Let us assume that a low mass star ($M_* \simeq 1 \ldots 3 M_\odot$) evolves along the asymptotic giant branch. The stellar mass loss is not considered until its evolutionary track crosses the limiting line for purely

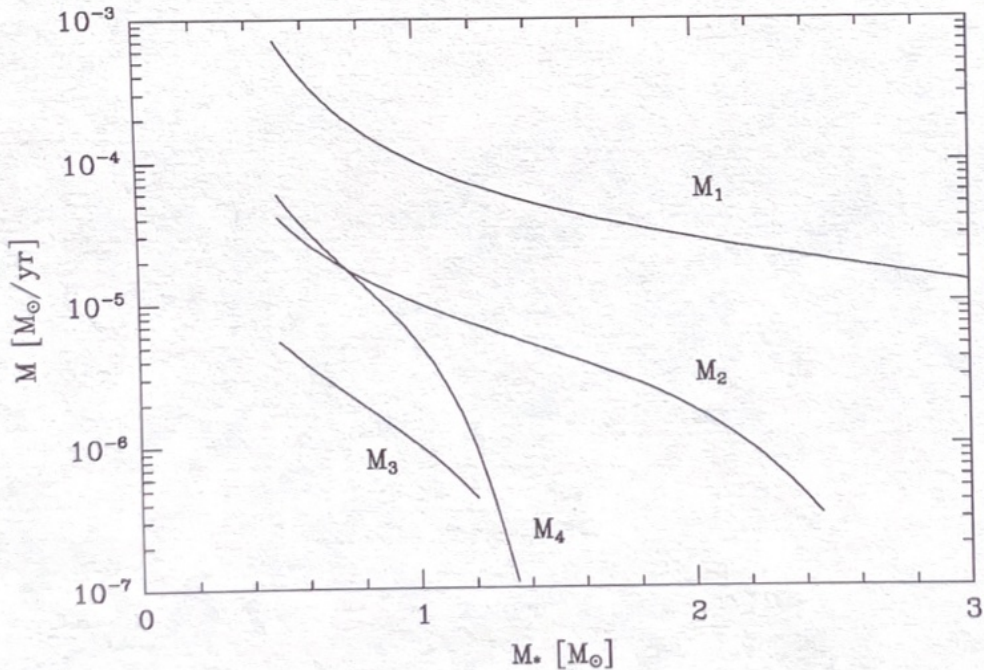

Figure 3.6:
Dependence of the mass loss rate on the stellar mass for the four models marked in Fig. 3.4a.

68                CHAPTER 3.  DUST DRIVEN WINDS AROUND CARBON STARSdust driven winds (note again that the corresponding luminosity depends on both mass and effective temperature of the star). Let us further assume that, from that moment on, the star does not change its luminosity and effective temperature. This can be interpreted by assuming that the typical timescale for the mass loss is smaller than the timescales for the evolution of the stars' interior. Then any change of the mass loss rate is due to the decreasing stellar mass only. The time evolution of the stellar mass $M_*(t)$ may then be obtained by simply integrating the corresponding $\dot{M}(M_*)$ curve in Fig. 3.6.

The results of this calculation are plotted in Fig. 3.7. It shows that for all four models the mass loss in fact is a self-accelerating process. All models loose their envelope mass within a typical timescale of $10^6$ yrs. However, a large part of the mass is lost in much shorter times of about $10^5$ yrs – $10^4$ yrs just before the white dwarf at the core becomes visible. The models $M_1$, $M_2$, $M_4$ which have luminosities above $10^4 L_\odot$ can reach very high mass loss rates. Their "evolution" looks very much like a mass loss instability and all three models loose at least $0.5 M_\odot$ during the last $10^4$ yrs. This behavior is very close to the specification of the so called "superwind" given by [Renzini 1981] in order to explain the formation of a planetary nebula during the late phases of AGB evolution.

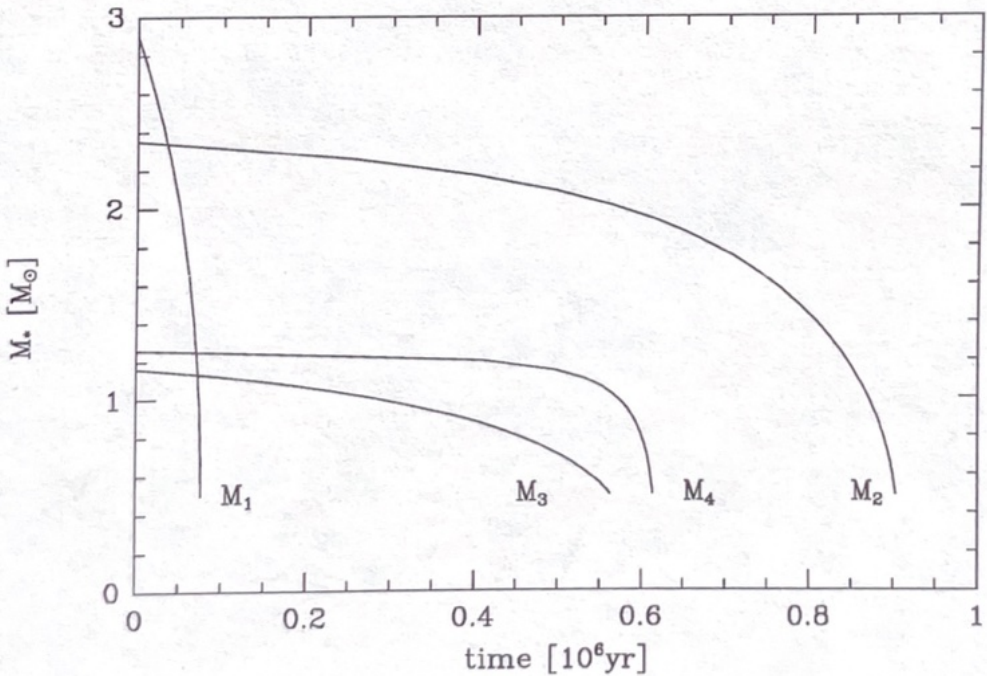

Figure 3.7:
Theoretical time evolution of the stellar mass with the assumption of constant $L_*$ and $T_*$. $(L_*, T_*)$ is given by the marks in Fig. 3.4, respectively. For $M_2$, $M_3$, $M_4$, $t = 0$ is the onset of a dust driven wind when the star meets the dividing line for the possibility of a purely dust driven wind. M1 has been shifted by an appropriate value in time.

Again this suggests that dust driven mass loss is a possible mechanism driving this transition.

For calculations of the evolution of low mass stars on the AGB it is useful to construct a fit formula for mass loss rates due to dust driven winds as a function of the stellar parameters. The model calculations can quite accurately be fitted by

$$\dot{M}[M_\odot/yr] = 10^{-3.7-0.93 M_*[M_\odot]} \cdot \left\{ \log L_*[L_\odot] - \frac{T_*}{600K} \right\}^{2.59 M_*[M_\odot]+1.71} \quad (3.5)$$

which is correct to within a factor of five for mass loss rates larger than $10^{-6} M_\odot/yr$ and, therefore, covers the region of interest. It should be noted that this formula has been constructed for C-stars where only the formation of carbon grains has been considered. Adding other dust species should increase the resulting mass loss rates.

Concerning the use of this formula in calculations of stellar evolution it should be emphasized that it is necessary to use the correct definition of $T_*$ (similar to that used in this calculations). This is important since the fit formula is very sensitive to the value of $T_*$. However, a consistent treatment requires an evaluation of the whole outer layer integration in stellar evolution calculations with the dynamical equations. For a discussion see [Dominik et al. 1990].

### 3.2.2 Comparison with the Formula of deJager

It is interesting to compare the mass loss rates given by the fit formula (3.5) with that given by the empirical formula of [deJager et al. 1988]. The linear approximation of this formula is

$$\log \dot{M}[M_\odot/yr] = 1.955 \log L_*[L_\odot] - 3.54 \log T_* - 0.99. \quad (3.6)$$

The results of both formulae for stars with one solar mass are plotted in Fig. 3.8.

It is important to note that the formula of [deJager et al. 1988] is a purely empirical formula that is fitted equally to the mass loss rates in the entire HRD. The empirically fitted mass loss rates show a slow increase at higher luminosities and lower temperatures, but the temperature dependence is very small.

On the contrary the mass loss rates derived from the dust driven winds describe a special physical mechanism for the mass loss. Here the rates show a much more drastic increase with both decreasing temperature and increasing luminosity. Furthermore these mass loss rates are negligible outside a certain region.

A possible interpretation for the discrepancy between the two diagrams would be that the empirical mass loss rates for most stars are driven by a mechanism different from that discussed in this chapter. The dust driven mass loss rates become important only in the uppermost corner of the HR diagram where the stars stay only for a small time. In this area, however, the mass loss rates are higher by up to a factor of 20 and therefore might be responsible for a large fraction of the total mass loss.

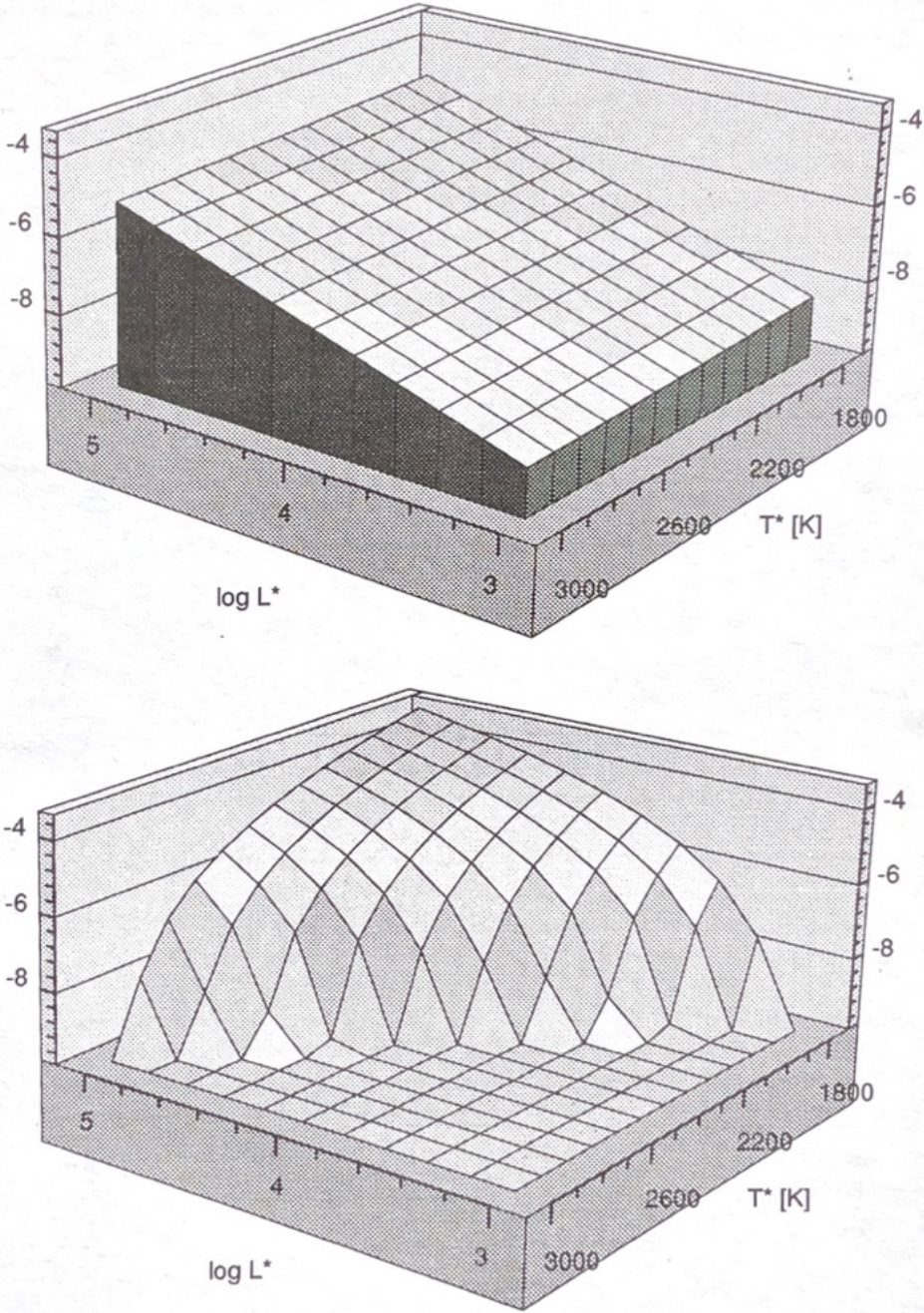

Figure 3.8:
Comparison of the mass loss rate given by the empirical formula by [deJager *et al.* 1988] (upper diagram) and the dust driven mass loss rates given by Eq. (3.5) (lower diagram). The ordinate shows $\log \dot{M}[M_\odot/yr]$.

Interestingly enough [deJager *et al.* 1988] need a correction factor to their formula in order to explain the unusually high mass loss rates of carbon stars. They suggest 11 as the value for this factor. Thus it seems to be a natural assumption that the mass loss rate at least in carbon stars is strongly influenced by the massive dust formation in their envelope.

# Chapter 4

# A Test Calculation for the Formation of Composite Grains in the Environment of an M-Star

In this chapter the theoretical concept developed in Chap. 1 shall be applied to the formation of composite dust grains in the envelope of an oxygen rich red giant.

## 4.1 Other Descriptions of the Formation of Composite Grains

Up to now there have been some different attempts to describe the dust formation process in oxygen rich environments where the grains are composed of several chemical elements.

[Gail and Sedlmayr 1986] have considered the formation of dust in M-stars stressing the question of the primary condensates that are responsible for the formation of the grain nuclei. Their result was that the primary condensates should be $MgS$, $SiO$, $Fe$, and eventually $MgO$. The growth of these particles was not discussed in detail. Instead, the formation of pure iron grains and pure magnesium oxide grains was considered. In the case of silicate, the formation of $MgFeSiO_4$ was assumed as long as magnesium and iron were left in the gas phase. The moment method used by these authors is the basis for the work presented here.

Several papers dealing with the formation of dust grains in M-stars originate from the Japanese group of Yamamoto, Hasegawa, Kozasa and co-workers. They usually consider the formation of pure $MgSiO_3$, $Al_2O_3$, $CaTiO$ and some more. In the most recent version of their theory [Kozasa and Hasegawa 1987] define a generalized supersaturation ratio (essentially) by the quotient of the right and left hand side is Eq. (1.16). With that supersaturation ratio the equations of the classical nucleation theory are used in order to calculate the nucleation rate. For the calculation of the growth velocity define a *key species* which is the species with the smallest collisional rate. This species triggers the growth process. Furthermore they assume a constant monomer of the dust material determining the composition of the con-densing material[1]. However, the direct application of the classical nucleation theory to the formation of dust grains where to monomer does not exist in the gas phase (which is true for almost all the condensates considered) is at least controversial (e.g. [Sedlmayr 1989b, Donn and Nuth 1985]).
All physical conditions (temperature, densities, etc.) are assumed to either be independent or linear functions of time. The equations of nucleation, growth, and depletion are then integrated analytically. This is a good approximation only in systems where the dust formation itself has a negligible influence on the structure of the whole system. In a situation like a dust driven wind where the whole hydrodynamic and thermodynamic structure is critically triggered by dust formation, the equations cannot be solved in that way in order to obtain realistic results.

An interesting paper is [Sharp and Huebner 1990]. In this work the authors undergo chemical equilibrium calculations including the existence of a large variety of solid compounds. All compounds are assumed to be in pure stoichiometric form and the equilibrium is calculated by a global minimization of the Gibbs energy of the whole system. The results of this calculation are the equilibrium densities of all relevant molecules in the gas phase. However, there are several reasons that these calculations cannot be applied to our situation:

- The authors considerer a complete equilibrium situation which is the limit for large timescales. It is, however, well known that the process of dust formation in the envelope of red giants is essentially a non equilibrium process occurring at temperatures lower than those to be expected from equilibrium situations.

- Only pure condensates are considered. In our case the condensates will certainly not be pure but a mixture of varying composition.

- The equilibrium calculations give no information on the time dependent behavior of the system. There is no possibility to monitor the number and size of dust grains formed.

The test calculation following here shall show that the *Ansatz* described in Chap. 1 is a step forward towards a realistic and quantitative description of the grains formed in the winds of red giants.

## 4.2  The Situation for Dust Formation in M-Stars

It has already been mentioned (Sect. 2.6), that the question of the condensates in M-stars cannot be answered as clearly as in the case of carbon stars. Since almost all carbon is locked up in the $CO$ molecule, the five most abundant elements ($H$, $He$, $O$, $C$, and $N$) cannot form stable solid particles. Therefore it is necessary to incorporate less abundant elements into the dust grains. The candidates next in abundance are the elements $Mg$, $Si$, $S$, $Fe$, and $Al$. Observations point to the fact, that sulfur is not incorporated in dust grains to a high degree - presumably an effect of the larger stability of the oxygen bearing compounds compared to the sulfur bearing compounds. Since in almost all astrophysical situations oxygen is present and more abundant than sulfur is, it is probably capable of drive out the sulfur of solid particles condensing in such environments.

---

[1]Only one remark in their paper points to the possibility that they sometimes calculate a nonstoichiometric composition, but in the results this is never discussed.

Due to the high abundance of oxygen it is very probable that oxygen is incorporated in the developing solids. In Table 2.1 it can be seen that the oxygen bearing solids are among the most stable ones.

$Al$ has the lowest abundance of the five elements. Since compounds of aluminium and oxygen are stable at high temperatures it might be a candidate for a species forming grain nuclei. However, due to its low abundance it is of minor importance for the growth process of the grains. Therefore $Al$ will not be considered in the following test calculation of grain growth. The formation of solid particles consisting only of $O$, $Mg$, $Si$, and $Fe$ will be discussed.

## 4.3 The Stability of Different Solid Compounds and the Composition of the Condensing Material

It has been discussed in Sect. 1.3.2 that it is possible to calculate the equilibrium densities over a certain solid compound for all molecules consisting of the elements incorporated in the solid compound. For example, if one considers the compound $MgSiO_3$, the equilibrium densities of the atoms $Mg$, $Si$, and $O$, and the molecules $O_2$, $MgO$, $SiO$, $SiO_2$, etc. can be calculated. At least for these atoms and molecules it is possible to evaluate the generalized supersaturation ratios (cf. Eq. (1.10)). If all these supersaturation ratios are greater than unity, the solid may be considered as stable as far as reactions with these molecules are concerned.

However, in astrophysical situations, a large fraction of the elements of interest is contained in molecular compounds containing hydrogen and other elements. For these molecules the equilibrium densities cannot be calculated the same way. For an exact treatment, the reaction probabilities for both growth and chemical sputtering reactions involving these molecules must be known. For the model calculation here it will be assumed that these molecules contribute only to the growth of the grains and that the chemical sputtering reactions are improbable.

Furthermore the probabilities $\alpha_\iota$ for all reactions contributing to the growth of the grain are assumed to be unity.

With these assumptions one can evaluate the growth rates of all molecules of interest and the destruction rates due to spontaneous evaporation. Adding up the rates one gets the element balances as they are defined in Eq. (1.66). If these element balances are *all* pointing towards growth, the compound will be called *stable in the actual* *environment*. The growth rate of this compound is then given by the net element fluxes on the surface of the solid reduced to the stoichiometric relations for the compound

In general it is not necessary that only one special compound be stable in a given situation. Actually, since dust formation occurs in a cooling gas flow, there will exist a hierarchy of possible dust materials that successively become stable. Since a large variety of element mixtures is stable at high temperatures it has to be expected that the different materials do not condense separately but will form a common grain material without a clear stoichiometric composition. The composition that forms only has to be compatible with certain boundary conditions due to the stability of the different materials. These boundary conditions are that the stoichiometric relations of the solid do not leave the settings given by the composition of the stable materials.

It is reasonable to assume that for every element the maximum flux compatible with at least one stable solid compound is realized. This assumption is used in the following calculations.

The last quantity one has to evaluate is the volume of the material condensing on the grains. Generally this quantity will depend on the detailed properties of the actual grain surface. However, as a first shot one can assume that every atom has a fixed volume in all possible compounds. These volumes have to be fitted in order to match the monomer volumes of all considered compounds as closely as possible. Then the *change in volume* $V_i$ of the different reactions can simply be calculated by

$$V_i = \sum_{j=1}^{J} V_{E(j)} \tag{4.1}$$

where $V_{E(j)}$ is the fit volume of the atom $j$.

To sum up: In order to calculate the growth velocity the following procedure is applied:

1. Calculate the equilibrium densities of as much molecules as possible over a variety of solid compounds for a given dust temperature.

2. Calculate the generalized supersaturation ratios for the molecules.

3. Calculate the hypothetical element fluxes on the surfaces of the different grain materials.

4. Check if for at least one of the grain materials all element fluxes are positive. If this is the case, the growth of an existing grain nucleus is possible.

5. Reduce the element fluxes to the values consistent with the corresponding stoichiometric composition of the compound.

6. Determine for each element the maximum flux on the surface of a grain compatible with one stable solid compound.

7. Calculate the growth velocity $\chi$ with these fluxes and the appropriate values for the element volumes.

## 4.4 Test Calculation of Dust Formation in a Stellar Wind

### 4.4.1 Assumptions for the Wind Model

In order to discuss the effects of dust formation independently of a complicated hydrodynamic structure a very simple model of the stellar wind will be used. It is assumed that the wind has a constant outflow velocity of $1 km/s$. This rather small velocity corresponds to the assumption, that the dust formation occurs inside the sonic point where the hydrodynamic velocity is small.

$$v = 1 km/s \tag{4.2}$$

The star is assumed to be an oxygen rich red giant with the following parameters:

| | |
|---|---|
| luminosity | $10^4 L_\odot$ |
| effective temperature | $2500 K$ |
| mass loss rate | $10^{-6} M_\odot/yr$ |
| chemical abundances | solar abundances [Allen 1973] |

The temperature structure in the wind is assumed to be given by radiation dilution:

$$T(r) = \frac{T_*}{\sqrt{2}} \left\{ 1 - \sqrt{1 - \frac{R_*^2}{r^2}} \right\}. \tag{4.3}$$

Furthermore it is assumed that the gas temperature and the dust temperature are equal.

Concerning the nucleation rates this calculation follows [Gail and Sedlmayr 1986], only the nucleation rates due to the molecules $MgO$ and $MgS$ have been neglected. These two turned out to be important in the mentioned work due to an overestimate of the equilibrium densities of both molecules. This overestimate results from an error in the chemical constants of the chemistry. If calculated with the correct equilibrium densities and from the surface tension of the bulk material the nucleation rates of $MgS$ and $MgO$ can be neglected[2].

---

[2] However, both $MgS$ and $MgO$ are still in the discussion of possible nucleating species since they form ionic clusters that are stable at rather high temperatures. Therefore, eventhough no large nucleation rates can be expected, these clusters may well be the first to become stable. These processes, however, are still under investigation, e.g. [Köhler 1992].

Therefore the only nucleating species considered here are $Fe$ and $SiO$ in the form described in [Gail et al. 1984, Gail and Sedlmayr 1986]. The physical constants required for the nucleating species also have been adopted from these papers.

### 4.4.2 Equilibrium Molecule Densities Over Some Solid Compounds

For the calculation of the equilibrium densities the Gibbs formation energies of the different grain materials have to be known. The term $\Delta G/kT$ usually is fitted by

$$\frac{\Delta G_s}{kT} = \frac{a}{T} + b \qquad (4.4)$$

where $a$ and $b$ are constants. These constants have been fitted for all high temperature condensates consisting only of $O$, $Si$, $Mg$, and $Fe$ and where the thermodynamic data was available in [Chase et al. 1985]. The fit coefficients are given in Table 4.1.

| Compound | $a$ [K] | $b$ | $V_{\mathrm{mono}}[10^{-23}cm^3]$ | molecules contributing to growth |
|---|---|---|---|---|
| $Si$ | -54669 | 17.88 | 2.0157 | $Si$ |
| $Fe$ | -50168 | 17.62 | 1.1882 | $Fe$ |
| $Mg$ | -17528 | 13.14 | 2.3359 | $Mg$ |
| $FeO$ | -113603 | 33.08 | 2.1079 | $Fe, O, FeO, O_2, OH, H_2O, Fe(OH)_2$ |
| $MgO$ | -121249 | 33.98 | 1.8827 | $Mg, O, MgO, O_2, OH, H_2O, MgOH$ |
| $SiO_2$ | -225943 | 54.63 | 3.7916 | $Si, O, SiO, O_2, OH, H_2O$ |
| $Fe_2O_3$ | -291604 | 89.53 | 5.0964 | $Fe, O, FeO, O_2, OH, H_2O, Fe(OH)_2$ |
| $Fe_3O_4$ | -407716 | 122.14 | 7.4749 | $Fe, O, FeO, O_2, OH, H_2O, Fe(OH)_2$ |
| $Mg_2Si$ | -99326 | 44.79 | 6.6112 | $Si, Mg$ |
| $MgSiO_3$ | -352292 | 89.40 | 5.2627 | $Si, Mg, O, SiO, MgO, O_2, OH, H_2O, MgOH$ |
| $Mg_2SiO_4$ | -476738 | 123.18 | 7.3296 | $Si, Mg, O, SiO, MgO, O_2, OH, H_2O, MgOH$ |

Table 4.1:

Data for the different dust materials: Fit coefficients $a$ and $b$ for the expression $\Delta G_s/kT$ as given in Eq. (4.4), monomer volume and THE molecules that are assumed to contibute to the growth. Data taken from [Weast et al. 1988] and [Chase et al. 1985].

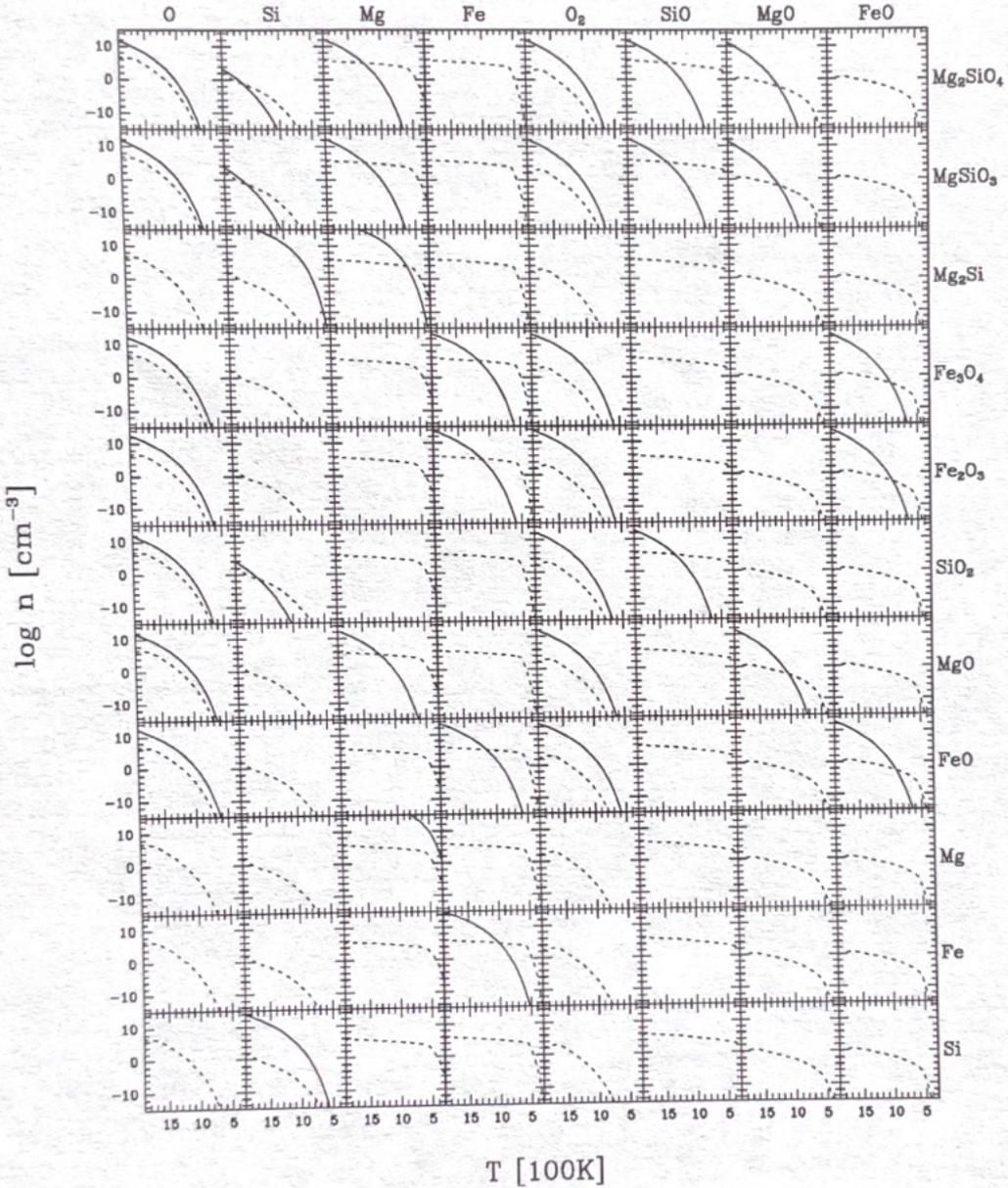

Figure 4.1:

Equilibrium densities of some molecules (columns) over 11 grain materials (rows). Solid lines: equilibrium densities over the compounds. Dashed lines: densities in the gas phase as they result from an equilibrium chemistry in the model discussed in this chapter without depletion by dust formation.

With Eqs. (1.16) and (1.19) and the procedure described in Sect. 1.3.2 the equilibrium densities of the molecules consisting only of elements included in the respective solid can now be calculated. The results are shown in Fig. 4.1. The solid lines in the plot represent the equilibrium densities of the molecules over the solids. They are present only in those diagrams in which the molecule consists only of elements contained in the solid. The dashed lines represent the molecule densities in the gas phase as a result of the equilibrium chemistry in the wind model discussed in this chapter. In the figure it can be seen directly, whether the generalized supersaturation ratio is greater or less than unity for a given molecule:

- If the dashed line is below the solid line, the real density of the molecule in the gas is less than the equilibrium density of the molecule over the solid. Therefore, this molecule will be evaporated from the solid *more frequently* than it will stick to the solid. Thus, the reactions with this molecule will work in the direction of grain *destruction*.

- If the dashed line is above the solid line, the real density of the molecule in the gas phase is greater than the equilibrium density of the molecule over the solid. Therefore, this molecule will be evaporated from the solid *less frequently* than it will stick to the solid. Thus, the reactions with this molecule will work in the direction of grain *growth*.

Though the effects of other molecules (e.g. the hydrogen bearing molecules) cannot be seen in the figure, it suggests that the solids $Si$, $Mg$, and $Mg_2Si$ are never stable in the wind. For all other compounds at least some supersaturation ratios become greater than unity at sufficiently small temperatures. This result also holds if the contributions of other molecules are taken into account as will be shown below.

### The Atom Volumes in the Solids

To obtain the changes in volume due to every reaction the atom volumes of the four relevant elements $O$, $Mg$, $Si$, and $Fe$ have been fitted so that they additively reproduce the monomer volumes in Table 4.1 as closely as possible. Since the solids $Si$, $Mg$, and $Mg_2Si$ are not stable they are neglected in this fit. The results are given in the following table. The fit leads to a maximum error of 4% in the monomer volume of the solids.

| Element | Volume [$10^{-24} cm^3$] |
|---|---|
| $O$ | 9.342 |
| $Si$ | 17.86 |
| $Mg$ | 8.727 |
| $Fe$ | 12.11 |

### 4.4.3 Gas Phase Chemistry

In Fig. 4.2 the densities of the molecules most important for the dust formation have been plotted. All densities have been normalized to the total density of carbon $n_{<C>}$ which in the reduced chemistry described in Sect. 2.5 always equals the density of the $CO$ molecule. The densities have been plotted against the gas temperature since in our model this is the most important quantity (because of the adopted very simple hydrodynamic model). For clarity, the plot has been split into four diagrams, each showing the most important molecules bearing one of the four considered dust forming elements.

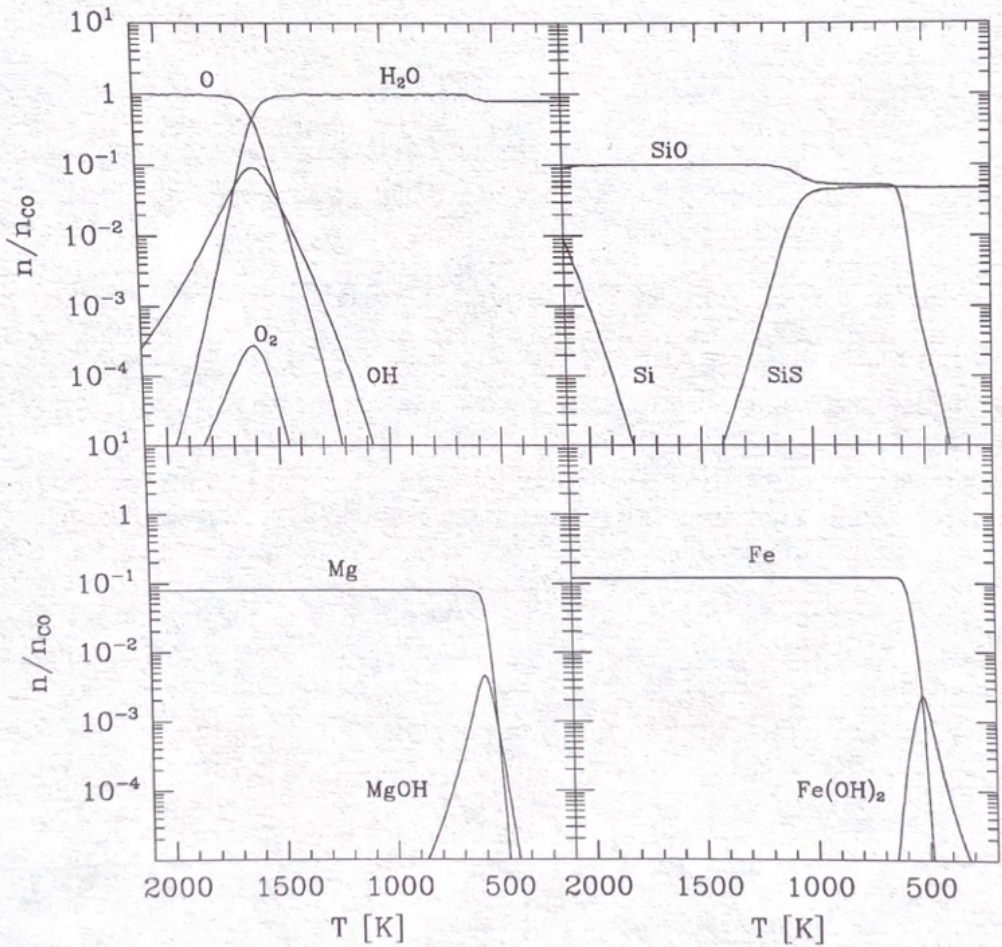

Figure 4.2:

Chemical composition of the gas phase as a function of temperature. The densities are normalized to the total density of the C atom and thus to the density of the CO molecule. Only the molecules relevant for dust formation are plotted.

$O$: Oxygen is essentially present in atomic form at higher temperatures and as $H_2O$ at lower temperatures. The $OH$ molecule and the $O_2$ molecule play only minor roles. The small decrease in the density of $H_2O$ around $600K$ is due to the consumption of oxygen by the onset of dust formation.

$Si$: Silicon is essentially present as $SiO$ at higher temperatures. Around $1200K$, the $SiS$ molecule also becomes important and binds as much silicon as there is sulfur present. At the onset of dust formation around $600K$, the silicon contained in $SiO$ is completely consumed while the silicon combined with $S$ is assumed to stay in the gas phase.

$Mg$: Magnesium is present in atomic form almost in the complete temperature range. The only other important molecule is $MgOH$ which starts to change roles with $Mg$ at around $600K$. However, since dust formation starts simultaneously, magnesium is consumed very efficiently and both $Mg$ and $MgOH$ disappear from the gas phase.

$Fe$: Iron also is present in atomic form almost in the compleste temperature range. Here the only other important molecule is $Fe(OH)_2$. When $Fe(OH)_2$ becomes important, dust formation onsets and consumes most of the available $Fe$.

### 4.4.4  Stability Hierarchy of the Grain Materials and the Condensing Composition

In Fig. 4.3 the most important results of this test calculation are put together. In the upper diagram the stability limits for the different grain materials are shown. The arrows indicate the highest temperature in the wind where all element balances for the respective compound are positive for the first time (c.f. Sect. 4.3). The first of the considered compounds to become stable is $Mg_2SiO_4$, soon followed by $MgSiO_3$, $SiO_2$, and $MgO$. The iron bearing compounds successively become stable. From the assumption of the maximum possible element flux the chemical composition of the material that would condense on existing dust grain nuclei can be calculated. The results of this calculation are plotted in the bottom diagram of Fig. 4.3. At a given temperature the part by number of a certain element in the condensate is proportional to the height of the respective area. For example, the composition at some temperatures can be written according to Eq. (1.67) as:

| temperature | composition | | | |
|---|---|---|---|---|
| [K] | $Mg$ | $Fe$ | $Si$ | $O$ |
| 1700 | 0.203 | 0.000 | 0.199 | 0.598 |
| 1500 | 0.182 | 0.111 | 0.176 | 0.531 |
| 1000 | 0.219 | 0.218 | 0.126 | 0.437 |
| 500  | 0.043 | 0.326 | 0.142 | 0.489 |

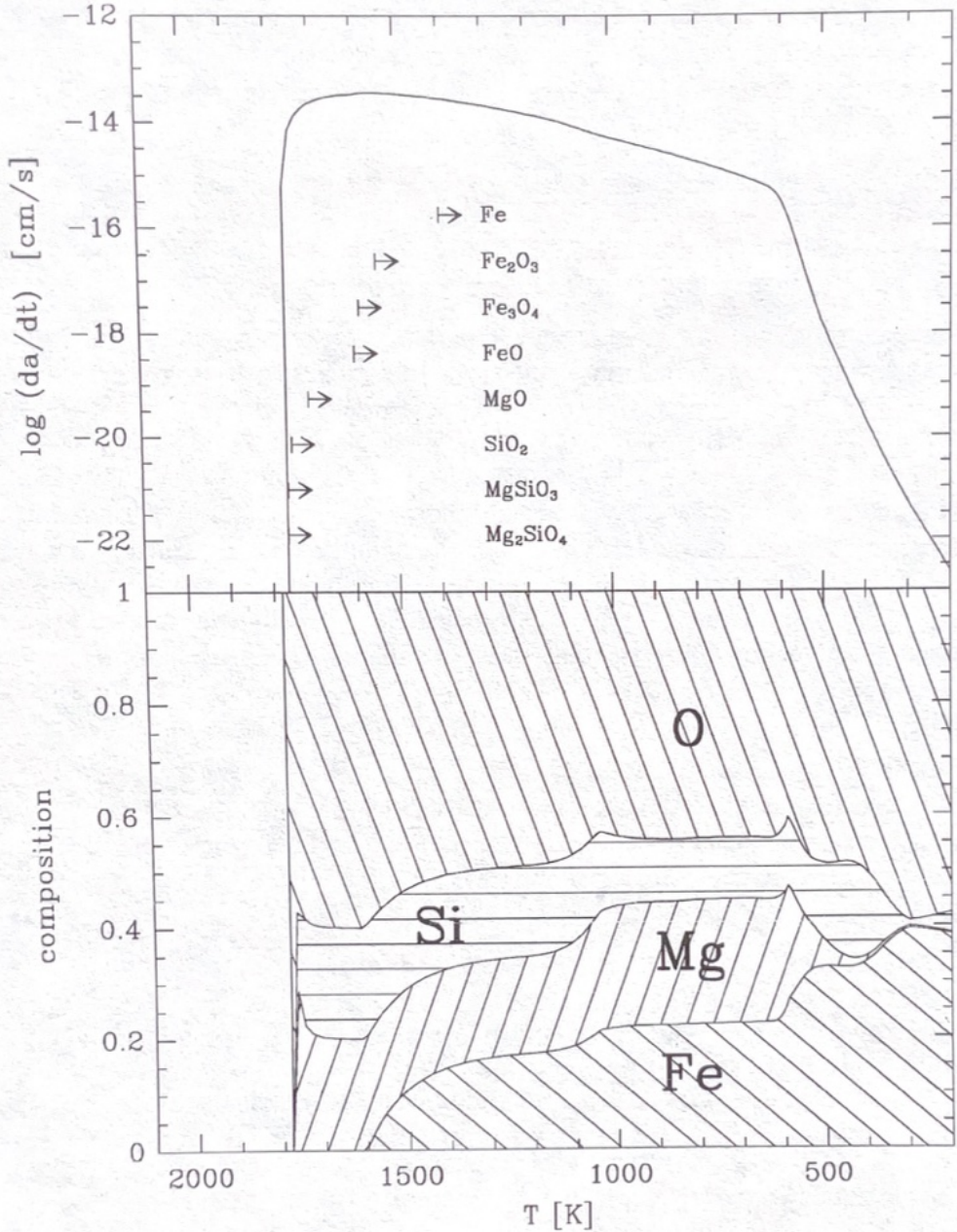

Figure 4.3:

Dust formation in the wind of the M-star model. Upper diagram: The arrows show the stability limits of the different grain materials. The line shows the (hypothetical) growth velocity of the particle radius as a function of temperature (distance from the star) in the wind. Lower diagram: Chemical composition of the material condensing on existing grain nuclei. The fraction of each element is proportional to the height of the respective area.

In this model, the condensing composition does not include iron at temperatures above $1600K$. However, this should not be taken too literally, since not all possible stable compounds have been taken into account. For example, a compound which is stable at very high temperatures is $MgFeSiO_4$, but this has not been included into the calculations since no thermodynamic data for this compound was found in [Chase et al. 1985].

A further point that should be kept in mind is, that growth of dust grains can only occur if the grain nuclei have already been formed. If the nucleation theory (as it does in our model) produces nuclei only at temperatures below $1000K$, the composition at higher temperatures is degraded to a hypothetical one.

The composition shown in Fig. 4.3 is not directly the radial structure of the formed dust grains, since it has to be transformed with the growth velocity $da/dt$ which is also plotted in Fig. 4.3 (upper diagram). This growth velocity is at first negative since none of the considered compounds are stable. Then, when $Mg_2SiO_4$ becomes stable, the growth velocity quickly increases up to a value of about $10^{-13.5} cms^{-1}$. From this point on the growth velocity essentially remains constant. This is due to the fact that in the simple hydrodynamic model adopted here the outflow velocity is constant and, therefore, only moderate variations in the gas density do occur. The slight decrease of the growth velocity thus is essentially due to the dilution in the wind. At about $600K$, a steep decrease follows. Here the dust formation is terminated and most of the condensable material is consumed. The molecules bearing iron and magnesium have almost vanished from the gas and no relevant growth can occur.

For the same reason, the composition of the condensing material below $500K$ is merely theoretical since no relevant growth can occur in this region. The calculated composition is the ratio of very small numbers.

### 4.4.5  Nucleation Rate and Degree of Condensation

The nucleation as it occurs in the model is plotted in Fig. 4.4. The most important nucleating species is iron with a pronounced peak in the nucleation rate around $600K$. The increase of $J_*$ is essentially a temperature effect, the decrease is due to the massive consumption of iron in the dust formation process. Below $600K$ silicate also shows some nucleation but the peak is much smaller than that of iron and does not play any role in this calculation.

In the top diagram of Fig. 4.4 the degrees of condensation for the element $O$, $Si$, $Mg$, and $Fe$ are shown. Iron and magnesium show complete condensation in quite a narrow region between 700 and $500K$. Complete condensation (differently from the C-star model in Sect. 3) is possible due to the small hydrodynamic velocity $v$. The dilution is not as fast as in the case of the dust driven wind and the growth reactions have enough time in order to consume all the material.

The $Si$ and $O$ condensation is not complete. This has different reasons.

In the case of oxygen this is due to the fact that only a certain fraction of oxygen can be incorporated into the grains in order to form stable condensates. The condensate with the maximum fraction of oxygen in our model is $SiO_2$. Thus no more than 66.6% of the material can be oxygen. Of course this condition is always fulfilled as can be seen in Fig. 4.3.

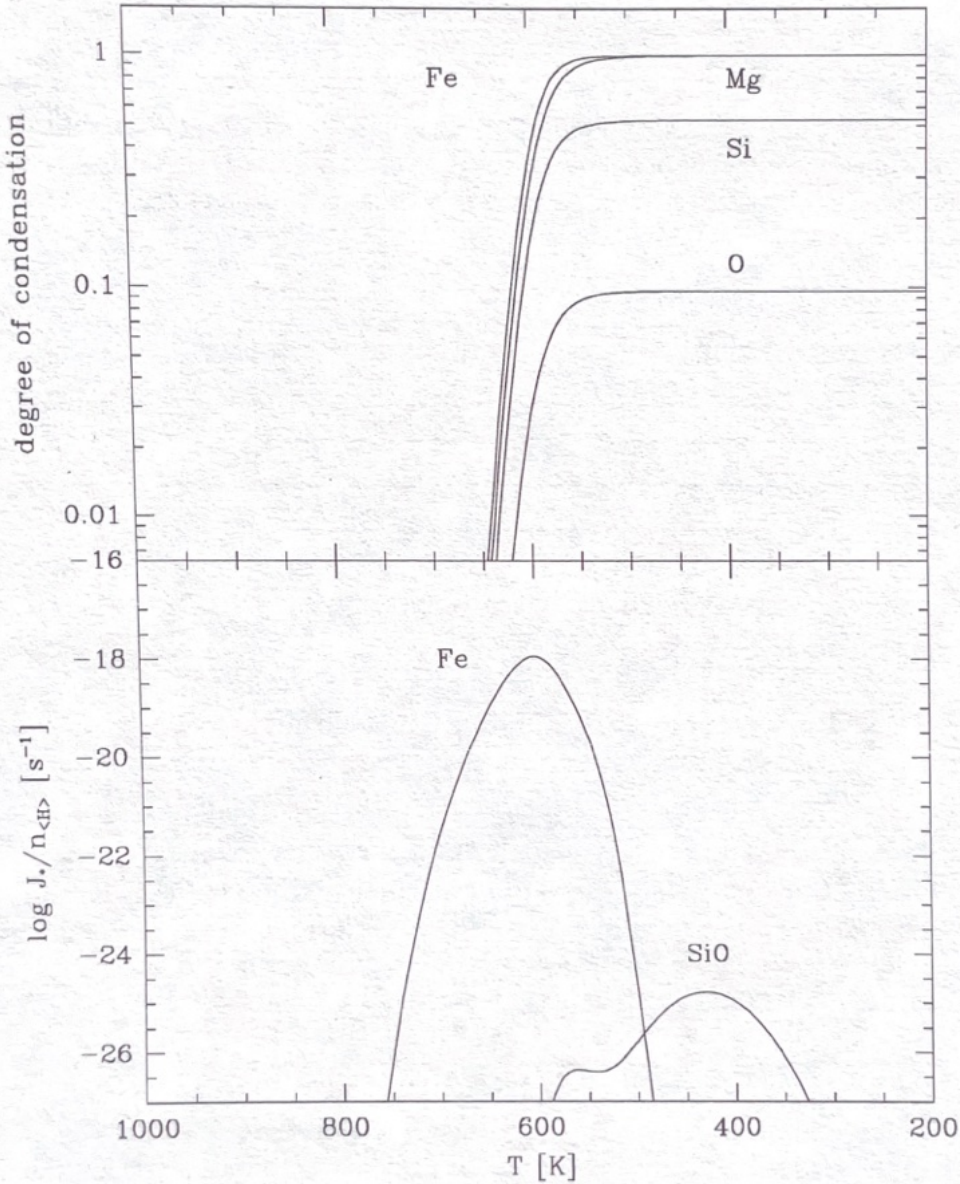

Figure 4.4:
Nucleation rate and degree of condensation. Lower diagram: The nucleation rates of $Fe$ and $SiO$. Upper diagram: The fraction of the elements condensed into grains.

In the case of silicon its smaller than unity degree of condensation is due to the fact that $SiS$ has not been included in the growth mechanism. Irrespective of the reliability of this assumption it shows that the incorporation of certain elements into the grains may be hindered by the existence of stable molecules that need large activation energies in order to be broken up. The most important example for this mechanism is of course the blocking of carbon in $CO$ that hinders the incorporation of carbon into the grains.

Another important point is that the condensation process including composite grains is much more efficient than that of only pure grains of, say, $Fe$ and $SiO$. In the following table the final degrees of condensation are shown for the model discussed here and a similar model where only the formation of pure grains has been considered.

| element | degree of condensation | |
|---|---|---|
| | composite | pure |
| $O$ | 0.097 | 0.026 |
| $Si$ | 0.532 | 0.521 |
| $Mg$ | 1.000 | 0.000 |
| $Fe$ | 1.000 | 0.652 |

### 4.4.6 The Composition of an Individual Dust Grain

In Chap. 1 it has been shown that the internal structure of the chemical composition of a dust grain can easily be calculated from the time evolution of the condensing composition and from the growth velocity (c.f. Eq. (1.71)). The result of this calculation for a dust grain of final size $10^{-6} \mu m$ is shown in Fig. 4.5.

The nucleus of this grain was produced at the temperature of $633K$, consequently the composition is that of the condensing material below that temperature. The grains growth has almost finished at $500K$ where the growth velocity (c.f. Fig. 4.3) is quite small. Therefore the total interior of the grain covers only a small fraction of the growth region indicated in Fig. 4.3 and the interior structure is rather constant. Only in the outermost part of the grain does the depletion of magnesium below $500K$ become visible as a reduction of the magnesium fraction in the composition. The composition of the condensing material below $400K$ plays no role since the growth velocity is already negligibly small.

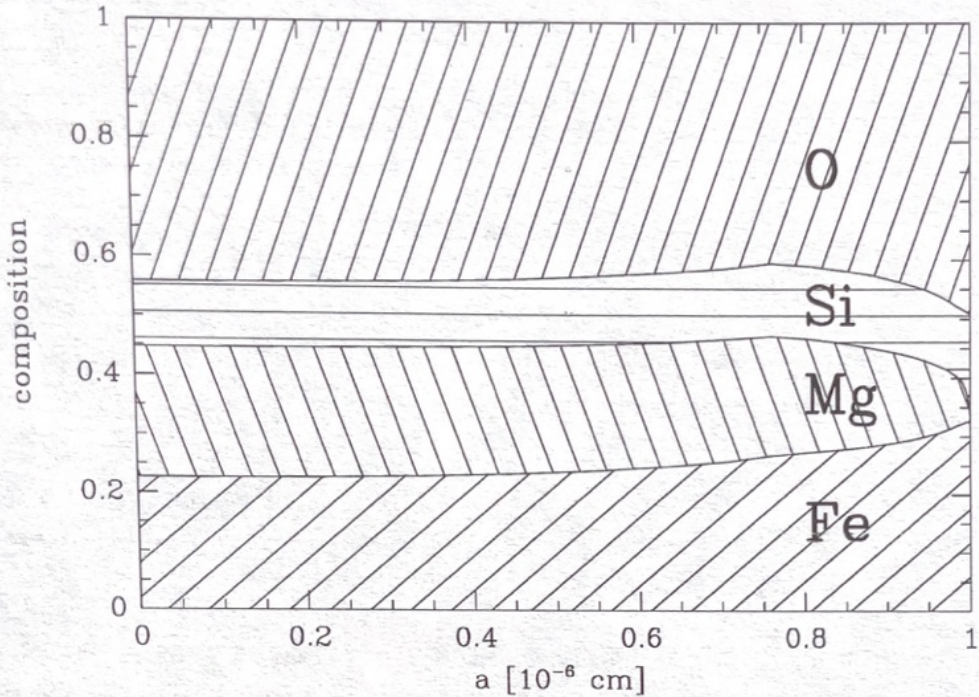

Figure 4.5:
Chemical composition of a dust grain of final size $10^{-6}$cm. The fraction of each element is proportional to height of the respective area.

## 4.5 Open Questions

A test calculation for the formation of composite grains has been performed and discussed in this chapter. It should be kept in mind that there are still several open questions and problems in the calculation.

The formalism devlloped in Chap. 1 allows for tracking the grain growth if the required physical constants (the rate coefficients for the different reactions between grains and molecules) are known. This, however, is still a great uncertainty and can be solved only by a detailed study of the possible reactions.

The formalism still leaves open the detailed structure of the crystals formed. It should be clear from the arguments above that the resulting solid compound will not have a definite stoichiometric composition. Descriptions of dust formation which claim such a composition have to be taken with care.

One can think of two different ways which both can lead to the realization of the chemical composition discussed above.

1. The different grain materials might form rather pure "islands" on the grain so that a certain percentage of the surface is covered by the first grain material, another by a second material and so on. This type of condensation requires, that either

   - the molecules colliding with the grain are able to migrate over the surface until they arrive at "their" specific material or
   - that the reaction probabilities on every island are zero for the molecules not consistent with the islands' material. In this case also different dust grain species could evolve each consisting of only one material.

2. The different materials form a completely amorphous structure. This should happen if the colliding molecules are not able to migrate freely over the surface but stick to the grain at the place of their impact.

An approach to this question might be taken from [Gail and Sedlmayr 1984]. Here the authors compare (for the case of pure carbon grains) the timescales for the particle growth to the typical atom hopping timescales on the surface. From these timescales a criterion may be derived deciding whether the atoms might reach an equilibrium position in the grain (therefore forming the most stable compound at the considered temperature) or whether they stay approximately at the position where they impacted the grain.

# Concluding Remarks

In this thesis essentially two subjects have been worked on: A description of composite dust grain formation and stationary dust driven winds.

The formation of composite dust grains is especially important in the envelopes of M-stars. Here several materials are possible condensates (this is different in C-stars where pure carbon clearly is the favorite material). A formalism has been developed which allows for tracking the growth of such composite dust grains. Furthermore a method has been introduced that enables both the calculation of the size distribution of the dust grains and the representation of the chemical composition of the grains as a function of depth below the surface. With this state of the theory of grain growth the tools are available that allow for an easy description of dust formation in various situations.

However, the input physics to this formalism still suffers from several problems. As already discussed, the detailed growth reactions are only poorly understood which hinders the correct determination of the crystal structure of the dust grains. And, also important, the calculations of the nucleation rates are to be improved. Working out the detailed chemical pathways that lead to dust formation is an indispensible condition for the really quantitative calculation of the dust formed in astrophysical systems. Only then the condensation temperatures can be determined correctly. Furthermore the actual rate of the formation of grain nuclei critically determines the size distribution formed in the condensation process. Only a detailed chemical understanding also allows the calculation of the byproducts of grain formation (e.g. small PAH's can be formed during the condensation of carbon grains [Keller 1987]) which can be important for the chemical evolution of the ISM.

The second part of the thesis was attended to the field of dust driven winds. It is an interesting property of these wind solutions that they only to some extent depend on above mentioned shortcomings of the theory of dust formation. This has essentially two reasons.

1. The feedback of the dust on the hydrodynamic and thermodynamic structure is due to the absorption coefficient of the dust. The absorption coefficient is mainly dependent on the amount of dust formed whereas the detailed size distribution of the dust is only of minor importance.

2. A dust driven winds is a self-tuning system. In order to accelerate a wind it is necessary to produce enough force in the envelope to surmount the gravitational acceleration. Thereby the exact value of the condensation temperature and of the amount of dust formed is not decisive since the interaction between dust formation and the hydrodynamics and thermodynamics of the wind always forces the dust to form at the "correct" position in the wind (see Sect. 3.1.4). This property of the winds is also reflected in the small dependence of the dust driven mass loss rate on the available amount of condensable material (see Fig. 3.5).

Nevertheless the concept of a purely dust driven wind requires that the formation of dust occurs rather near to the star. In the models of carbon stars discussed in Chap. 3 this condition was fulfilled. However, due to the low nucleation temperatures it has not been possible up to now to construct the same type of wind solution in M-type stars. Possibly this will change when the nucleation rates acting at higher temperatures become available.

Purely dust driven winds are a rather extreme situation that can only apply at the top of the AGB. Only here the photospheric temperatures are low enough and the stellar luminosities are sufficiently high. Other stars which also show dust in their envelopes will have a combined driving mechanism for the wind (see Sect. 2.1). The consistent description of dust formation including other possible driving mechanisms is an important project.

A second one is the connection to observations of late type giants. Consistent models describing both the mass loss mechanism and the produced spectrum of the object are the key for a better understanding of this important phase of stellar evolution. Sufficiently accurate modeling of dynamical atmospheres also holds the key to the masses and to other important parameters of red giants which are still very uncertain.

Another interesting field of work is the study of the evolutionary effects of the extremely high mass loss rates during the top AGB phase. Once a viable understanding of the mass loss mechanisms is achieved, it will be possible to include self-consistently determined mass loss rates in the calculations of stellar evolution which will give these calculations a new significance.

# Appendix I

# A Simple Estimate for Dust Formation

Dust formation in astrophysical environments generally is a non equilibrium process. If a system develops towards conditions at which some solid compounds are stable, this is no guaranty that the formation of solid particles in fact takes place. Therefore the transition region in which dust formation occurs is not completely determined by the thermostatic conditions like gas temperature, gas density, and concentrations of the different chemical species. In order to describe the transition region an additional parameter (the typical timescale for an evolution of the system in a non equilibrium state towards its equilibrium state) must be specified.

This timescale is of crucial importance.

On one hand one may have a system, in which dust formation, though thermodynamically possible, is very slow due to a low nucleation rate and/or a low growth rate. However, this system may form large amounts of dust if it remains at these condition for a sufficiently long time.
On the other hand one may have a system in which dust formation is rather fast (corresponding to a large nucleation rate and a large growth rate). Nevertheless the formation of dust can be effectively suppressed if these favorable conditions last only for a very short time.

To estimate these effects the following simple model is used: Let us consider a gas volume in which the gas density, the temperature, and the chemical concentrations are constant with time. Now one may ask for the time necessary to produce significant amounts of dust (e.g. to convert 10% of the total condensable material into dust grains). As long as the depletion of the condensable material is less than about 10%, dust formation does not have a strong influence on the chemical composition of the gas phase. Hence, the assumption of constant molecular concentrations approximately holds during the considered phase of dust formation.

Under this conditions both the nucleation rate $\mathcal{J}_\ell$ and the growth rate $\chi$ are independent of time. Then the moment Eqs. (1.39) can be integrated analytically. Starting without dust at time $t = 0$ one obtains

$$\mathcal{L}_0(t) = \int_0^t \mathcal{J}_\ell dt' = \mathcal{J}_\ell t \tag{I.1}$$

$$\mathcal{L}_j(t) = \int_0^t \mathcal{J}_\ell V_\ell^{\frac{j}{d}} + \frac{j}{d}\chi\mathcal{L}_{j-1}(t')dt'$$

$$= \mathcal{J}_\ell \sum_{l=0}^j \frac{j!}{(l+1)!(j-l)!} \left(\frac{\chi}{d}\right)^l V_\ell^{\frac{j-l}{d}} t^{l+1} \qquad j \geq 1. \tag{I.2}$$

An analysis of the terms in the sum shows that the one with $l = j$ dominates as soon as $t \cdot \chi \gg d \cdot V_\ell^{\frac{1}{d}}$, i.e. as soon as the volume added to the grain by growth is considerably larger than $V_\ell$. For the purpose of the estimate the sum may be approximated by this term:

$$\mathcal{L}_j(t) \simeq \frac{1}{j+1} \left(\frac{\chi}{d}\right)^j \mathcal{J}_\ell t^{j+1} \qquad j \geq 0. \tag{I.3}$$

Let $\varepsilon_{\text{cond}}$ be the *abundance of the dust forming element* defined by the *maximum possible concentration* (relative to the total density of hydrogen $n_{<H>}$, see Eq. (1.17)) of the dust forming molecule or atom (e.g. if the formation of graphite is considered one has $\varepsilon_{\text{cond}} = \varepsilon_C$. If the formation of a pure silicate is considered $\varepsilon_{\text{cond}} = \min\{\varepsilon_{Si}, \varepsilon_O\}$ and so on.).

We restrict the discussion to homogeneous dust formation. Then, with Eq. (1.24) the degree of condensation f may be defined by

$$f(t) = \frac{\mathcal{L}_d}{\varepsilon_{\text{cond}} n_{<H>} V_{\text{mono}}} \tag{I.4}$$

where $V_{\text{mono}}$ is the volume of the monomer in the solid state.

It is now simple to calculate the *condensation time* $\tilde{\tau}_{\text{cond}}(f_0)$ which is defined as the time at which a certain degree of condensation $f_0$ is reached.

$$f_0 \stackrel{!}{=} f(\tilde{\tau}_{\text{cond}}) = \frac{\mathcal{L}_d(\tilde{\tau}_{\text{cond}})}{\varepsilon_{\text{cond}} n_{<H>} V_{\text{mono}}}$$

$$= \frac{\frac{1}{d+1} \left(\frac{\chi}{d}\right)^d \mathcal{J}_\ell \tilde{\tau}_{\text{cond}}^{d+1}}{\varepsilon_{\text{cond}} n_{<H>} V_{\text{mono}}}. \tag{I.5}$$

Solving Eq. (I.5) for $\tilde{\tau}_{\text{cond}}$ yields

$$\tilde{\tau}_{\text{cond}} = \left\{ \frac{f_0 \varepsilon_{\text{cond}} n_{<H>} V_{\text{mono}} (d+1) d^d}{\mathcal{J}_\ell \chi^d} \right\}^{\frac{1}{d+1}}. \tag{I.6}$$

After this time a fraction of $f_0$ of the condensable material has been converted to dust grains.

One should note that the condensation time calculated this way not always is a realistic timescale for dust formation. If the critical cluster is identical with the monomer in the gas phase, the degree of condensation $f_0$ can be reached by nucleation without any relevant growth of the dust grains: every monomer that collides with another monomer then is considered to be a dust grain. A part of $f_0/2$ of dimers is enough for the desired degree of condensation. These dimers do not behave like macroscopic particles and the result that effective dust formation occurs is somewhat misleading. Therefore it is demanded that the condensation time has to be longer than the time needed for the growth of a particle by some molecular layers (here five layers are chosen). This time is denoted by $\tau_{\text{cond},min}$ and is given by

$$\tau_{\text{cond},min} = \frac{\chi}{5 d V_{\text{mono}}^{\frac{1}{d}}}. \tag{I.7}$$

Then, the condensation time $\tau_{\text{cond}}$ is the maximum of $\tilde{\tau}_{\text{cond}}$ and $\tau_{\text{cond},min}$:

$$\tau_{\text{cond}} = \max \left\{ \tilde{\tau}_{\text{cond}}, \tau_{\text{cond},min} \right\}. \tag{I.8}$$

Furthermore it is possible to calculate the mean grain size after the time $\tau_{\text{cond}}$ from Eq. (1.22):

$$<a>(\tau_{\text{cond}}) = \frac{\xi_a^{(d)} \mathcal{L}_1(\tau_{\text{cond}})}{\mathcal{L}_0(\tau_{\text{cond}})}$$

$$= \xi_a^{(d)} \frac{1}{2} \frac{\chi}{d} \tau_{\text{cond}}. \tag{I.9}$$

Figures I.1, I.2, and I.3 show contour plots for the *nucleation rate* and the *condensation time* for the dust forming species $SiO$, $Fe$, and $C$. In all three cases classical nucleation theory has been used in order to calculate the nucleation rate $\mathcal{J}_\ell$ which is assumed to equal the stationary nucleation rate $\mathcal{J}_*$. In the case of $SiO$ and $Fe$ condensation, a solar element mixture has been used. In the case of $C$ condensation the abundance of carbon is two times the abundance of oxygen. The chemical composition has been calculated by the reduced equilibrium chemistry described in Sect. 2.5. The application of these diagrams is discussed with the example of the $SiO$ condensation.

Let us consider a volume of gas in an astrophysical environment evolving from high to low gas temperatures (e.g. the ejected matter of a supernova, the wind of a giant star, a molecular cloud forming from the interstellar medium). The dotted line in the plot shows the line at which the supersaturation ratio equals unity. Below this line, solid particles are stable and will form after a sufficiently long time. However, the solid line nearest to the $S = 1$ line in the plot of $\tau_{cond}$ denotes a time of $10^{18}$ seconds, approximately equal to the age of the universe. Therefore a gas volume characterized by $(n_{<H>}, T)$ on this line may have formed considerable amounts of dust only if the conditions for dust formation have been at least equally favorable since the Big Bang. Above this curve, no dust can have formed by pure $SiO$ condensation during that time. For situations relevant for astrophysics, the condensation temperature is always lower than 1500 K, mostly lower than 1000 K. Near the border line the variations of $\tau_{cond}$ with $n_{<H>}$ and especially with $T$ are large. This is essentially due to the fast variation of the nucleation rate which on its part is due to the changing critical cluster size.

At temperatures around $300 K$ the condensation time becomes almost independent of the temperature. This is due to the fact that, at these temperatures, the monomer becomes the critical cluster. All clusters larger than the monomer are more stable and the phase transition becomes possible at almost all densities. The characteristic timescale for this transition is governed by the collision frequency between $SiO$ molecules. Thus the condensation time in this region is proportional to the density.

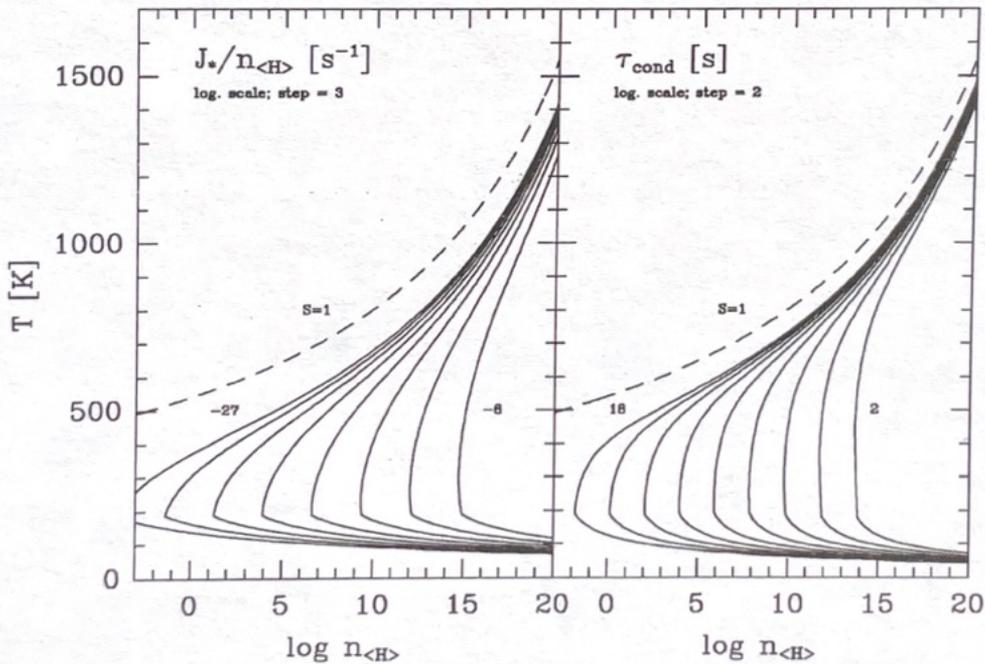

Figure I.1:
The nucleation rate $J_*$ and the condensation time $\tau_{cond}$ for the dust forming species $SiO$ in an oxygen-rich element mixture with cosmic abundances.

At temperatures even lower dust formation become impossible again since the collision frequency between $SiO$ molecules is also proportional to $\sqrt{T}$. Thus the collision frequency approaches zero as T does, even for higher densities.

Typical properties of astrophysical plasmas under different conditions are collected in Table (I.1). In winds, novae, and supernovae the timescale $\tau_{typ}$ of the system is either the cooling or the expansion timescale of the system at temperatures typical for $SiO$ condensation. In molecular clouds and $HII$ regions the approximate life times of the clouds have been used. The values for temperatures and densities of the interstellar medium have been adopted from [Duley and Williams 1984]. Typical life times of these object have been taken from [Osterbrok 1974, Spitzer 1968].

For $\tau_{typ}$ in novae and supernovae, the typical timescales for dust formation after the explosion have been used (50 days for novae and 1 year for supernovae). The densities in these objects have been obtained from the assumption of a uniform distribution of the ejected matter ($\approx 10 M_\odot$ in supernovae and $\leq 10^{-4} M_\odot$ in novae) after the typical time in a volume $V_{eject} = (4\pi/3) R_{eject}^3$. $R_{eject}$ is calculated by $R_{eject} = v_{exp} \tau_{typ}$ with expansion velocities $v_{exp}$ of 500 km/s in novae and 10000 km/s in supernovae.

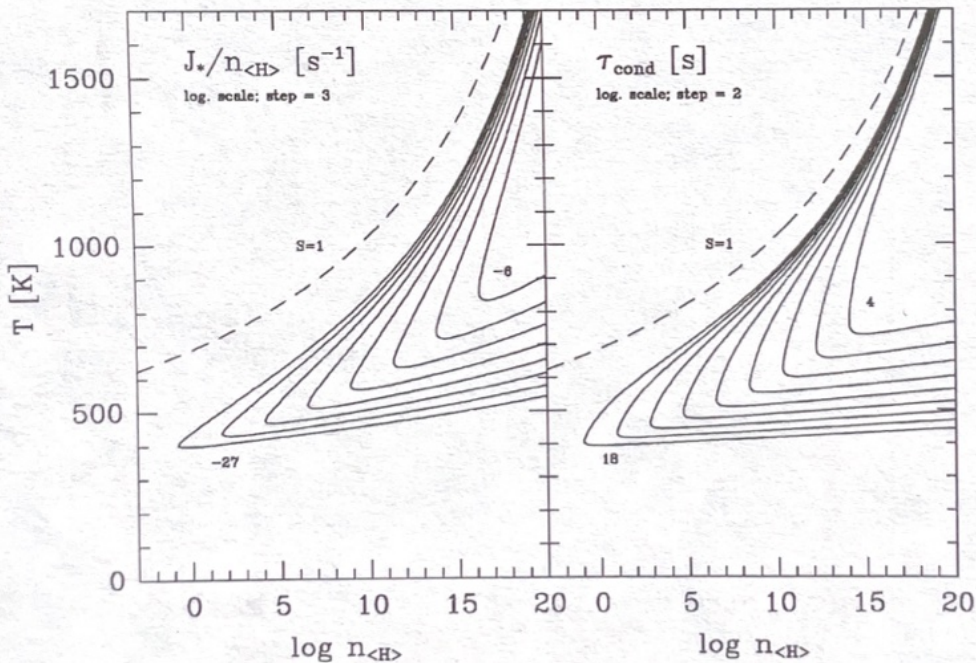

Figure I.2:

The nucleation rate $J_*$ and the condensation time $\tau_{cond}$ for the dust forming species $Fe$ in an oxygen-rich element mixture with cosmic abundances.

For the calculation of densities and timescales in massive winds the following numbers have been adopted: Mass loss rates of $10^{-5} M_\odot/yr$, a radius of dust formation of $2 \cdot 10^{14} cm$ in both cases, and expansion velocities of 10 km/s for cool stars and 1000 km/s for hot stars.

Comparing the numbers in Table I.1 with the curves of constant $\tau_{cond}$ in Fig. I.1 it can easily be seen that in all cool objects in the interstellar medium the collisional rates are too small for effective dust formation due to low gas densities or low temperatures or both.

Dust can be formed by stellar matter returning to the interstellar medium (massive winds, novae, and supernovae). Only here the densities are high enough and the temperatures are not too low for the collisions necessary for the gas-dust transition to occur.

This result is a rather clear outcome of the estimate.

However, it should be kept in mind that there are some shortcomings. The calculation assumes a completely passive role of the dust component and does not consider any feedback of the dust on the system. In this aspect it is nearly on the same level of approximation as the analytical calculations of [Draine and Salpeter 1977, Yamamoto and Hasegawa 1977]. However, in systems were the dust component

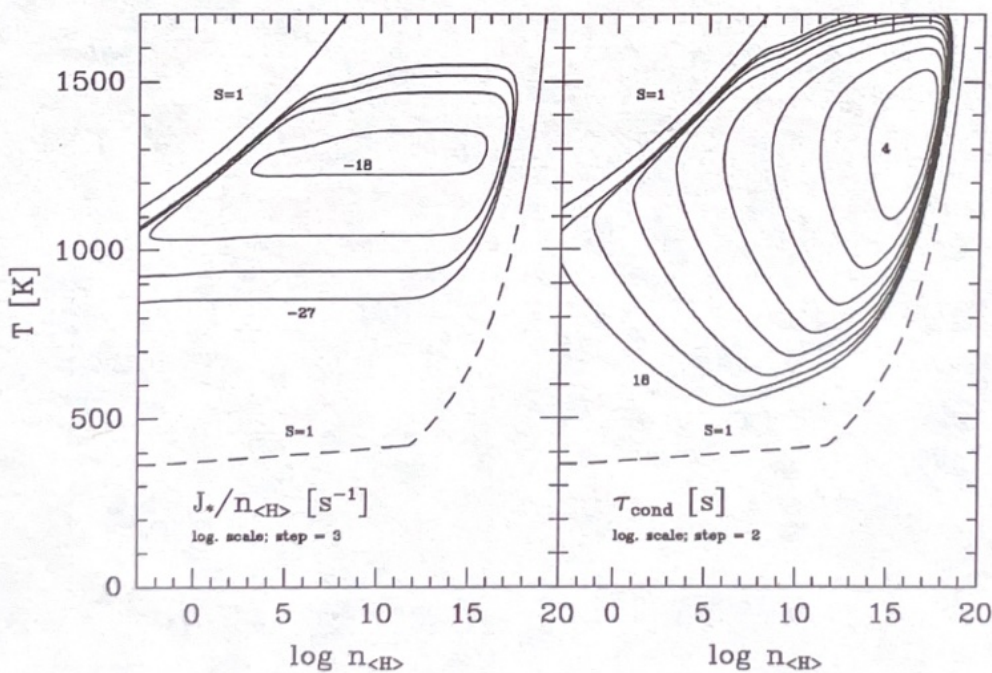

Figure I.3:
The nucleation rate $J_*$ and the condensation time $\tau_{cond}$ for the dust forming species $C$ in an carbon-rich element mixture with cosmic abundances except for $\varepsilon_C = 2\varepsilon_O$.

| Object | Temperature [°K] | log $n_{<H>}$ $n_{<H>}[cm^{-3}]$ | log $\tau_{\text{cond}}$ [s] | log $\tau_{\text{typ}}$ [s] | SiO cond? |
|---|---|---|---|---|---|
| *interstellar medium* | | | | | |
| H II region | 10000 | 2 − 3 | ∞ | ≈ 12 | no |
| Intercloud medium | 10000 | ≈ −1 | ∞ | ? | no |
| Diffuse clouds | 100 | ≈ 2 | > 18 | ? | no |
| Dark clouds | 10 − 20 | ≈ 4 | > 18 | ? | no |
| Molecular clouds | 50 | ≈ 6 | > 18 | ≈ 15 | no |
| Compact HII reg. | 100 − 1000 | 3 − 4 | ≈ 13 | ≈ 11 | no |
| *explosive ejection of matter* | | | | | |
| Novae | from 10000 down | ≤ 8 − 10 | ≈ 6 − 8 | 6 − 7 | ? |
| Supernovae | from 10000 down | 7 − 9 | ≈ 7 − 9 | 7 − 8 | yes |
| *massive winds of giant stars* | | | | | |
| Cool winds | from 2000 down | 7 − 10 | ≈ 7 − 9 | ≈ 8 | yes |
| Hot winds | from 20000 down | 6 − 8 | ≈ 8 − 10 | ≈ 6 | no |

Table I.1:
The condensation timescale in different astrophysical situations. $n_{<H>}$ total density of hydrogen in the gas, $\tau_{\text{typ}}$ typical hydrodynamic timescale in the system, $\tau_{\text{cond}}$ condensation timescale, $SiO$ cond?: is the homogeneous condensation of $SiO$ possible?

---

plays a dominant role, the diagrams plotted in this chapter are no longer applicable the same easy way. In a dust driven wind (compare Chap. 3) the dust in fact controls the hydrodynamic and thermodynamic timescales of the system. A dust driven wind is a system which does not cross the critical $p - T$ region at a more or less constant timescale but rapidly changes the relevant timescales. Then, there is no unique crossing point of the wind trajectory and the actual timescale. Instead, the wind trajectory meets several lines of condensation time with precisely the correct timescale. If one thinks of the plots in this chapter as a three dimensional plot $\tau_{\text{cond}}(n_{<H>}, T)$, the wind trajectory does not cross the surface at a definite point but follows the surface for some distance. Then relevant amounts of dust are formed at considerably different timescales (connected with different timescales for grain growth). In this way the broad size distribution function obtained in Chap. 3 is formed.

A poet once said, "The whole universe is in a glass of wine." We will probably never know in what sense he meant that, for poets do not write to be understood. But it is true that if we look at a glass of wine closely enough we see the entire universe. There are things of physics: the twisting liquid which evaporates depending on the wind and weather, the reflections in the glass, and our imagination adds the atoms. The glass is a destillation of the earth's rocks, and in its composition we see the secrets of the universe's age, and the evolution of stars. What strange array of chemicals are in the wine? How did they come to be? There are the ferments, the enzymes, the substrates, and the products. There in wine is found the great generalisation: All life is fermentation. Nobody can discover the chemistry of the wine without discovering, as did Louis Pasteur, the cause of much disease. How vivid is the claret, pressing its existence into the conciousness that watches it! If our small minds, for some convenience, divide this glass of wine, this universe, into parts – physics, biology, geology, astronomy, psychology, and so on – remember that nature does not know it! So let us put it all back together, not forgetting ultimately what it is for. Let it give us one more final pleasure: drink it and forget it all!

(R. P. Feynman)

106                                                                    BIBLIOGRAPHY[Schönberner 1981b] D. Schönberner, 1981. Late Stages of Stellar Evolution: II. Mass Loss and the Transition of Asyptotic Giant Branch Stars into Hot Remanents. *Astron. Astrophys.*, **272**, 708–714.

[Sedlmayr 1989a] E. Sedlmayr, 1989. Dust Formation in Stellar Winds. In M.O. Mennessier and A. Omont, editors, *From Miras to Planetary Nebulae: Which Path for Stellar Evolution*, pages 179–185, Edition Frontiéres, 91192 Gif sur Yvettes Cedex.

[Sedlmayr 1989b] E. Sedlmayr, 1989. Dust Formation in Stellar Winds. In *Interstellar Dust*, pages 467–477, Reidel, Dordrecht.

[Sharp and Huebner 1990] C.M. Sharp and W.F. Huebner, 1990. Molecular Equilibrium with Condensation. *Astrophys. J.(Letters)*, **72**, 417–431.

[Shu *et al.* 1987] F.H. Shu, F.C. Adams, and S. Lizano, 1987. Star Formation in Molecular Clouds: Observation and Theory . *Ann. Rev. Astron. Astrophys.*, **25**, 23–81.

[Spitzer 1968] L. Spitzer, 1968. *Diffuse Matter in Space.* Whiley, New York.

[Stull and Prophet 1971] D.R. Stull and H. Prophet, 1971. *JANAF Thermochemical Tables. Second Edition.* National Bureau of Standards.

[Taylor *et al.* 1991] S.D. Taylor, D.A. Williams, and A. Bennet, 1991. Electric Charge on Grains in Diffuse Clouds. *Mon. Not. R. Astron. Soc.*, **248**, 148–152.

[Tielens 1983] A.G.G.M. Tielens, 1983. Stationary Flows in the Circumstellar Envelopes of M-Stars. *Astrophys. J.*, **271**, 702–716.

[Treffers and Cohen 1974] R. Treffers and M. Cohen, 1974. High Resolution Spectra of Cool Stars in the 10 and 20 Mikron bands. *Astrophys. J.*, **188**, 545–552.

[Tsuji 1973] T. Tsuji, 1973. Molecular Abundances in Stellar Atmospheres. II. *Astron. Astrophys.*, **23**, 411–431.

[Veen and Olofsson 1989] W.E.C.J. van der Veen and H. Olofsson, 1989. Determinations of Mass Loss Rates. In M.O. Mennessier and A. Omont, editors, *From Miras to Planetary Nebulae: Which Path for Stellar Evolution*, pages 139–157, Edition Frontiéres, 91192 Gif sur Yvettes Cedex.

[Veen and Rogers 1989] W.E.C.J. van der Veen and M. Rogers, 1989. A Comparison between CO-, OH-, and IR- Mass Loss Rates of Evolved Stars. *Astron. Astrophys.*, **226**, 183.

[Weast *et al.* 1988] R.C. Weast, M.J. Astle, and W.H. Beyer, editors, 1988. *CRC Handbook of Chemistry and Physics.* CRC Press, Boca Raton, Florida, 69 edition.

# Lebenslauf

| | |
|---|---|
| Name | Carsten Dominik |
| Geburtsdatum | 11. September 1961 |
| Geburtsort | Berlin-Schöneberg |
| Eltern | Renate Dominik, geb. Frik (Buchhändlerin) |
| | Hans Dominik (Ingenieur) |
| Staatsanghörigkeit | deutsch |
| Familienstand | seit 15.3.1991 verheiratet mit Susanne Dominik, geb. Reitis, 1 Kind |

## Schulbesuch

| | |
|---|---|
| 1968–1974 | Christoph-Ruden-Grundschule in Berlin-Buckow |
| 1974–1980 | Albrecht-Dürer-Gymnasium in Berlin-Neukölln |
| Juni 1980 | Abitur (Allgemeine Hochschulreife) |

## Studium

| | |
|---|---|
| 1980–1987 | Studium der Physik am Fachbereich 4 der Technischen Universität Berlin |
| Nov. '83 – März '87 | Tutor am Fachbereich Mathematik |
| Apr. '85 – März '87 | Stipendiat der Studienstiftung des Deutschen Volkes |
| Oktober 1987 | Diplom in Physik, Diplomarbeit am Institut für Astronomie und Astrophysik über das Thema: *Selbstkonsistente Modelle staubgetriebener Winde bei C-Sternen.* |
| seit 12. 10. 1987 | Wissenschaftlicher Mitarbeiter bei Prof. Dr. E. Sedlmayr (Inst. für Astronomie und Astrophysik, TU Berlin) |

Berlin, den 5. Februar 1992